\documentclass[11pt]{article}
\usepackage{gbmacros}
\usepackage{times,mathpazo,charter,graphicx,tocloft}
\usepackage[bookmarks=false,colorlinks=true,linkcolor=blue,citecolor=red]{hyperref}
\usepackage{color}
\usepackage{subcaption}

\def\maketitle{\par\noindent{\LARGE\bf\sffamily\thetitle}\\[1.6ex]
{\large\theauthor}\\[0.6ex]
\textit{\theaddress}\\[0.2ex]
{\small\today}\par\vglue1.4\bigskipamount}
\def\title#1{\def\thetitle{#1}}
\def\author#1{\def\theauthor{#1}}
\def\address#1{\def\theaddress{#1}}
\allowdisplaybreaks

\advance\textwidth 12mm
\advance\hoffset -6mm
\advance\textheight 18mm
\advance\voffset -10mm

\makeatletter
\graphicspath{{figs/}}
\def\d{\mathrm{d}}
\def\sech{\mathop{\rm sech}\nolimits}
\def\diag{\mathop{\rm diag}\nolimits}
\def\tr{\mathop{\rm tr}\nolimits}

\def\Real{\mathbb{R}}
\def\Complex{\mathbb{C}}
\def\Re{\mathop{\rm Re}\nolimits}
\def\Im{\mathop{\rm Im}\nolimits}
\def\arg{\mathop{\rm arg}\nolimits}
\def\pvint{\int\kern-0.94em-\kern0.2em}
\let\^=\hat
\let\==\bar
\let\~=\tilde
\def\@#1{{\mathbf{#1}}}

\let\ge=\geqslant
\def\d{\mathrm{d}}
\def\e{\mathrm{e}}

\makeatother

\def\be{\begin{equation}}
\def\ee{\end{equation}}
\def\bse{\begin{subequations}}
\def\ese{\end{subequations}}

\def\02{\boldsymbol{0}_{2 \times 2}}

\def\bfsigma{{\boldsymbol\sigma}}
\def\bfQ{\mathbf{Q}}
\def\Kosqr{\kappa_o^2}
\def\Ko{\kappa_o}
\def\M{\mathbf{M}}
\def\N{\mathbf{N}}
\def\I{{\mathrm{I}}}
\def\II{{\mathrm{II}}}
\def\III{{\mathrm{III}}}
\def\IV{{\mathrm{IV}}}

\begin{document}
\title{Solitons and soliton interactions in repulsive spinor\\[0.4ex]
Bose-Einstein condensates with non-zero background}
\author{Asela Abeya$^1$, Barbara Prinari$^1$, Gino Biondini$^1$ and P.G. Kevrekidis$^2$}
\address{$^1$Department of Mathematics, State University of New York at Buffalo, Buffalo, NY 14260 \\
$^2$Department of Mathematics and Statistics, University of Massachusetts Amherst, Amherst, MA 01003}
\maketitle
\begin{quote}
\textbf{Abstract.}~
We characterize the soliton solutions and their interactions for a system of coupled evolution equations of nonlinear Schr\"odinger (NLS) type
that models the dynamics in one-dimensional repulsive Bose-Einstein condensates with spin one, taking advantage of the representation of such model as a special reduction of a $2\times2$ matrix NLS system. Specifically, we study in detail the case in which solutions tend to a non-zero background at space infinities. First we derive a compact representation for the multi-soliton solutions in the system using the Inverse Scattering Transform (IST). We introduce the notion of canonical form of a solution, corresponding to the case when the background as $x\to\infty$ is proportional to the identity. We show that solutions for which the asymptotic behavior at infinity is not proportional to the identity, referred to as being in non-canonical form, can be reduced to canonical form by unitary transformations that preserve the symmetric nature of the solution (physically corresponding to complex rotations of the quantization axes). Then we give a complete characterization of the two families of one-soliton solutions arising in this problem, corresponding to ferromagnetic and to polar states of the system, and we discuss how the physical parameters of the solitons for each family are related to the spectral data in the IST.  We also show that any ferromagnetic one-soliton solution in canonical form can be reduced to a single dark soliton of the scalar NLS equation, and any polar one-soliton solution in canonical form is unitarily equivalent to a pair of oppositely polarized displaced scalar dark solitons up to a rotation of the quantization axes.
Finally, we discuss two-soliton interactions and we present a complete classification of the possible scenarios that can arise depending on whether either soliton is of ferromagnetic or polar type.
\end{quote}

\medskip

\section{Introduction}

Over the past two decades, the platform of atomic Bose-Einstein condensates (BECs) has emerged as a ripe one for exploring numerous aspects of nonlinear
phenomena~\cite{becbook1,becbook2,Panos_book2015}.
More recently,  within this framework, the realm of multicomponent systems has been gaining considerable
traction~\cite{Ueda,Stamper_Kurn}.
This is a topic that has been of considerable interest not only in atomic physics, but also in optics and in nonlinear waves, more
generally~\cite{Panos_survey2016}.
Indeed, this setting provides a natural testbed for the exploration, both theoretically and experimentally, of various intriguing structures, such as dark-bright solitons, or domain walls, as well as for instabilities such as phase separation that cannot arise in the simpler, single-component settings.

Multicomponent BECs, more concretely, may be composed by two or more atomic gases, and may have the form of various (homonuclear or even heteronuclear) mixtures~\cite{Ueda,Stamper_Kurn}.
Unlike what happens in multicomponent nonlinear optics~\cite{Manakov74,kivshar_agr}, where (typically) Kerr-type
nonlinearities depend on the squared moduli of the components, the equations describing spinor condensates exhibit nonlinear terms
reflecting the $SU(2)$ symmetry of the spins: the spin-exchange interactions that are the sources of the spin-mixing within condensates
deviate from the above mentioned intensity-coupled nonlinearity,
{when  more than two components are involved.}
%, and
%they have no analogue in conventional nonlinear optics.

Spinor BECs have been realized by employing optical trapping techniques, which allow for the confinement of atoms regardless of
their hyperfine spin state~\cite{Ueda,Stamper_Kurn}.
Spinor BECs formed by atoms with spin $F$ are described by a macroscopic wave function with $2F + 1$
components.
Indeed, experimental works summarized in the above reviews have
considered both $F=1$, $3$-component states and $F=2$, 5-component
ones; indeed, even spin-3 cases in Cr have been considered~\cite{liyu}.
%These systems give rise to various phenomena that are not present in single-component BECs,
%such as formation of spin domains, spin textures, and topological
%states.
Indeed, exploiting the non-zero hyperfine spin of
the gas as an additional accessible degree of freedom, various experimental studies demonstrated new fundamental phenomena (e.g.,
paramagnet-to-ferromagnet, and polar-to-antiferromagnetic phase transitions, Dirac monopoles, quantum
knots, condensation of magnon excitations, etc), as well as various types of solitonic structures (e.g.,
bright and dark solitons, topological states, polar-core spin vortices,
and topological Wigner crystals of half-solitons), {as summarized
in~\cite{Panos_survey2016}.
The ability of recent, state-of-the-art
experiments to capture
numerous
among these exotic states, including knots~\cite{m20}, merons and
skyrmions~\cite{m18}, and monopoles~\cite{m19} only adds to the appeal
of this rich setting.}
%\cite{Panos_book2008,Panos_book2015,Panos_survey2016}.

Atoms in $F=1$ spinor BECs can be described by the three-component
macroscopic condensate vector wave function $(\Phi_1,\Phi_0,\Phi_{-1})^T$,
where
each of the $\Phi_j(x,t)$
is a scalar wave function describing
atoms with magnetic spin quantum number $j$.
In a mean-field approximation,
$\Phi_j$ is shown to satisfy the following system of PDEs~\cite{Ueda,Stamper_Kurn}:
\bse
\label{e:spinor}
\begin{gather}
i \hbar \partialderiv{\Phi_{\pm1}}t + \frac{\hbar^2}{2m}\partialderiv[2]{\Phi_{\pm1}}x
= (\bar c_o+\bar c_2)(|\Phi_{\pm1}|^2+|\Phi_0|^2)\Phi_{\pm1}
+ (\bar c_o-\bar c_2)|\Phi_{\mp1}|^2\Phi_{\pm1} + \bar c_2\Phi_{\mp1}^*\Phi_0^2\,,
\\
i \hbar \partialderiv{\Phi_{0}}t + \frac{\hbar^2}{2m}\partialderiv[2]{\Phi_{0}}x
= (\bar c_o+\bar c_2)(|\Phi_{1}|^2+|\Phi_{-1}|^2)\Phi_{0} + \bar c_o |\Phi_0|^2\Phi_0 + 2\bar c_2\Phi_{0}^*\Phi_1\Phi_{-1}\,,
\end{gather}
\ese
where $\bar c_j$ are the coupling constants (related to the scattering lengths),
and asterisk denotes complex conjugate.
The above system admits special reductions which are integrable.
Specifically,
the case
$\bar{c}_2=0$ yields the three-component generalization of the Manakov system \cite{Manakov74},
whose properties and solutions were studied analytically in \cite{APT2004,BiondiniKrausPrinari,JPA2015v48p395202,JMP2015v56p071505}.
Conversely, the case
$\bar{c}_o=\bar{c}_2=\nu $
is a special reduction of
the matrix NLS (MNLS) equation,
which we write here in normalized, dimensionless form as
%\vspace*{-1ex}
\bse
\label{e:mNLSgen}
\begin{equation}
iQ_t+Q_{xx}-2\nu Q\,Q^\dagger\, Q=0\,,
\end{equation}
where subscripts $x$ and $t$ denote partial differentiation and the dagger denotes Hermitian conjugate,
when $Q(x,t)$ is a symmetric $2\times2$ matrix:
\begin{equation}
\label{e:MNLSreduction}
Q(x,t)=\begin{pmatrix} q_1 & q_0 \\ q_0 & q_{-1}\end{pmatrix}.
\end{equation}
\ese
Here, the values $\nu=\pm 1$ identify the
defocusing/focusing nonlinearity regimes, respectively,
%\panos{nonlinearity (respectively, in atomic realm,
%  repulsive/attractive
%  interparticle interaction) regimes,}
$q_j(x, t)$
are suitable normalizations of the scalar wave functions $\Phi_j(x,t)$
for $j = 0,\pm 1$,
$Q^{\dagger}$ is the Hermitian conjugate of $Q$; subscripts $x,t$ denote partial derivatives with respect to the spatial variable $x$ and the time variable $t$, respectively.
Indeed, the system \eqref{e:mNLSgen} was proposed as a model to describe hyperfine spin $F = $1 spinor BECs with either
repulsive interatomic interactions and anti-ferromagnetic spin-exchange interactions ($\nu=+1$), or attractive interatomic interactions and
ferromagnetic spin-exchange interactions ($\nu=-1$),
and the fields
$q_1,q_0,q_{-1}$ are related to the vacuum expectation values of the three components
of the quantum field operator in the three possible spin configurations $1,0,-1$ \cite{Wadati1,Wadati1.1}.
The system was subsequently extended to include repulsive mean field interactions and anti-ferromagnetic spin exchange interactions ($\nu=1$), as well as
finite background (i.e., $Q\to Q_\pm\ne 0$ as $x\to \pm \infty$) \cite{Wadati2,Wadati4,Wadati5,Wadati6,Ueda,QinMu,BP1,BP2},
and higher spin cases, e.g., spin $F = 2$ condensates when $Q(x,t)$ is a $4 \times 4$ complex, symmetric potential \cite{Wadati3,Gerdjikov1,Gerdjikov2}.
Solitons an soliton interactions in symmetric spaces were studied in \cite{KF,Gerdjikov3,Gerdjikov4,Gerdjikov5}.
%There are some open questions both on the explicit form of the solitons and on their interactions, which
%could be due to the fact that in [29-31] a Lax operator L \in so(5) is considered, while in the present work
%we are using an equivalent L operator in sp(4) \simeq so(5). 

While dark-dark (DD) and dark-bright (DB) solitons and soliton trains in 2-component BECs
have been studied theoretically and observed experimentally for over a decade \cite{Becker2008,Engels1,Engels2,Engels3,Engels4,Panos3,Panos5,Panos5'},
an extension to 3 components and spinor systems had not been rave been observed in experiments ealized in experiments until very recently.
In \cite{Panos6}, the existence of robust DBB and DDB solitons in a defocusing spinor $F = 1$ condensate of $^{87}\mathrm{Rb}$ atoms was reported. In general, the systems considered in the experiments are non-integrable, and as such researchers have often relied on perturbation-based techniques
of related integrable systems to study solitons and their evolution:
the theoretical predictions for the soliton solutions in integrable
cases are an extremely valuable tool for the investigation of the
non-integrable solitary waves in regimes that are not too far from the
integrable ones. For instance, in \cite{Panos6} the coupling
coefficients for ``symmetric'' spin-independent and ``antisymmetric''
spin-dependent interaction terms $\lambda_a$ and $\lambda_s$,
respectively, are such that $\lambda_a/|\lambda_s|\sim 10^{-2}$ is a
small parameter up to which the model equation can be considered a
small perturbation of a 3-component Manakov system; see
also~\cite{ostrovskaya}.
{While the role of the spin-dependent term is often central to
  the observed spinor dynamics, experimentally it is also possible to
  eliminate the impact of the relevant term and realize the genuine
  3-component Manakov model. Indeed, this was achieved
in a recent experimental work \cite{Panos_PRL2020}, where pairs of
3-component dark-bright-bright solitons} in a BEC were prepared using a method based on local spin rotations which simultaneously
imprint suitable phase and density distributions. This enabled the observation of the striking collisional properties of the emerging multi-component solitons, and the results showed a remarkable quantitative agreement with the analytical predictions of collision-induced polarization shifts in the repulsive 3-component Manakov model in \cite{JPA2015v48p395202,JMP2015v56p071505}.
Eq.~\eqref{e:mNLSgen} in the defocusing case is another, distinct integrable model which one can use as the basis
to obtain analytical predictions for the above mentioned experimental results.
Additionally, the spinor model may provide insight on domain-wall type
solutions which are of interest in their own
right~\cite{Panos2}, but which have no analog as exact
solutions of the Manakov system. {Very recently, additional
  solitonic excitations in the form of magnetic solitons have also
  been considered in the realm of spinor BECs~\cite{raman1,raman2}.}

This work is concerned with the study of the defocusing MNLS equation, namely \eqref{e:mNLSgen} with $\nu=1$, within the framework of the Inverse Scattering Transform (IST),
with the main goal of providing a complete spectral characterization of the physical parameters of its dark soliton solutions, and of the soliton interactions.
The results obtained in this work pave the way for a comparison with the above mentioned experimental observations of solitons, domain walls and
other coherent structures in $F=1$ spinor BECs. {Indeed, we
  envision this
  as a starting point for the potential future consideration numerically
  and theoretically
  of a homotopic continuation (in a
  parameter
  such as $\bar{c}_2/\bar{c}_0$) of
  the present solutions towards the $F=1$ physical limit. This would
  be
a potentially fruitful direction towards identifying novel solutions that
might be even experimentally observable.}

The paper is organized as follows.
In section~2 we briefly review the IST for the defocusing MNLS with non-zero boundary conditions as developed in \cite{BP4},
and we then use it to  derive a compact, explicit representation for the multi-soliton solutions of the system.
In this context, we refer to the canonical form of a solution when the background $Q_+$ is proportional to the
identity. Solutions for which $Q_+$ is not proportional to the identity are referred to as being in non-canonical form,
and we show they can be reduced to canonical form by unitary transformations that preserve the symmetric nature of the solution
$Q(x,t)$ (physically, complex rotations of the quantization axes).
In section~3 we study the one-soliton solutions: the nature of the solitons depends on whether the associated norming constants (polarization matrices) are rank-one matrices (giving rise to ferromagnetic solitons) or full rank (corresponding to polar solitons), and we discuss their canonical and non-canonical forms as related to the
boundary conditions, and the characterization of their physical properties in terms of scattering data for both ferromagnetic and polar states.
We also show that the invariance of the system \eqref{e:mNLSgen} under arbitrary unitary transformations allows one to reduce
any ferromagnetic one-soliton solution to a single dark soliton of the
scalar NLS equation, and any polar one-soliton solution to a pair of oppositely polarized displaced scalar dark solitons up to a rotation of the quantization axes,
similarly to what was found for the solitons, breathers and rogue waves of the focusing spinor system in \cite{BP1,BP2,pol1,pol2,pol3,BP3}.
In section~4 we investigate two-soliton solutions in the long-time asymptotics, and we determine how the polarization matrix of each soliton changes because of the interaction. Explicit formulas for the soliton interactions are obtained for all possible types of interacting solitons, namely ferromagnetic-ferromagnetic, polar-polar, and polar-ferromagnetic soliton interactions. Finally, section~5 contains some concluding remarks and some more technical aspects are considered in the appendix.

\section{The defocusing spinor NLS equation with NZBC and its multi-soliton solutions}

In this work we study the solutions of the defocusing spinor system~\eqref{e:spinor}
with nonzero background, i.e., with nonzero boundary conditions (NZBC) as $x\to\pm\infty$.
To this end, it is convenient to rewrite the corresponding matrix NLS system~\eqref{e:mNLSgen} as
\be
\label{e:MNLS}
iQ_t+Q_{xx}-2(Q Q^{\dagger}-\Kosqr I_2)Q=0\,.
\ee
where
$\Ko$ is a real positive constant,
$I_n$ is the $n \times n$ identity matrix
and
$Q(x,t)$ is the $2 \times 2$ symmetric matrix-valued potential in \eqref{e:MNLSreduction}, as before.
The term proportional to $\Kosqr I_2$ in~\eqref{e:MNLS}
can be removed by the simple gauge transformation
$Q(x,t) \mapsto Q(x,t)e^{2i\Kosqr t}$,
but it ensures that the background values of the potential are independent of time.
Namely, $Q(x,t)$ satisfies
the following constant NZBC:
\be
\label{e:NZBC}
Q(x,t) \to Q_{\pm} \quad \text{as} \quad x \to \pm \infty\,.
\ee
%and the matrix potential $Q(x,t)$ is chosen to be a symmetric matrix:
%\[
%Q(x,t) = \begin{pmatrix} q_1(x,t) & q_0(x,t) \\ q_0(x,t) & q_{-1}(x,t) \end{pmatrix}\,.
%\]
Furthermore, we assume that the boundary conditions
(i.e., the asymptotic values for the potential)
$Q_\pm$ %= \displaystyle\lim\nolimits_{x \to \pm \infty}Q(x,t)$
satisfy the constraint
\begin{equation}
\label{e:BCconstraint}
Q^{\dagger}_{\pm} Q_{\pm} = Q_{\pm} Q^{\dagger}_{\pm} = \Kosqr I_2\,.
\end{equation}
In terms of the individual entries of the matrices $Q_\pm$, \eqref{e:BCconstraint} corresponds to the following equivalent set of constraints:
\be
\label{e:BCconstraint_expl}
|q_{1,\pm}|^2=|q_{-1,\pm}|^2\,, \qquad |q_{0,\pm}|^2=\Kosqr -|q_{1,\pm}|^2=\Kosqr -|q_{-1,\pm}|^2\,, \qquad q_{1,\pm}q_{0,\pm}^*+q_{0,\pm}q_{-1,\pm}^*=0\,.
\ee
Note that the above conditions imply that $Q_\pm$ are both normal matrices, and unitary up to normalization. Besides the norm of the background, $\Ko$, the boundary condition $Q_+$ is then specified by three additional real parameters: the (common) amplitude of the diagonal entries and their two phases, with the amplitude and phase of the off-diagonal entry being completely determined by the last two conditions in \eqref{e:BCconstraint_expl}.

%%%%%%%%%%%%%%%%%%%%%%%%%%%%%%%%%%%%%%%%%%%%%%%%%%%%%%%%%%%%%%%%%%%%%%%%%%%%%%%%%%%%%%%%%%%%%
\subsection{Canonical versus non-canonical solutions and conserved quantities}
\label{s:canonical}

Recall that the MNLS equation is invariant under unitary transformations.
Namely, if $Q(x,t)$ is a solution of~\eqref{e:MNLS}, then $\tilde{Q}(x,t)=U\,Q(x,t)V$
is also a solution, with $U$ and $V$ arbitrary unitary matrices.
On the other hand, these transformations are admissible only if they preserve the symmetry of $Q(x,t)$, namely, if $\tilde{Q}(x,t)$ is symmetric whenever
$Q(x,t)$ is,
in which case, as we show in Appendix~\ref{s:symmetries},
the transformations correspond physically to complex rotations of the quantization axes.

We say that
a solution is in canonical form when $Q_+=I_2$, and in non-canonical form if $Q_+\ne I_2$.
As we show next, any solution can be reduced to canonical form by rescaling and
a suitable rotation of the quantization axes.
(Note that we singled out the matrix $Q_+$, but an equivalent definition could be given using $Q_-$.)
Since
the background matrix $Q_+$ is a normal matrix
(by virtue of~\eqref{e:BCconstraint}),
it is
unitarily diagonalizable.
Moreover, one can easily show that the eigenvalues of $Q_+$ are
$\Ko e^{i(\alpha_1 + \alpha_{-1}) \pm i\delta}$,
where $\alpha_{\pm1}$ are determined by the phases of the diagonal entries of $Q_+$,
namely,
$\alpha_1 = 2\arg q_{1,+}$
and $\alpha_{-1} = 2\arg q_{-1,+}$,
and
\be
\sin \delta=\sqrt{1- (|q_{+,1}|^2/\Kosqr)\,\cos(\alpha_1+\alpha_{-1})}\,.
\ee
Furthermore, the orthogonal eigenvectors of $Q_+$ can be chosen to be the
real vectors
\be
\label{e:eigenmatrixforQ+}
\displaystyle
v_{\pm}= \Big( a_\pm\,,\, \sqrt{\Kosqr-|q_{+,1}|^2} \, \Big)^T  \,,
\qquad
a_\pm = |q_{+,1}|\sin(\alpha_1-\alpha_{-1})\pm \sqrt{\Kosqr-|q_{+,1}|^2\cos^2(\alpha_1-\alpha_{-1})}\,.
\ee
We can therefore write the background as
$Q_+= \Ko V^T \Delta_+ V$, where
$\Delta_+ = e^{i(\alpha_1+\alpha_{-1}) + i\delta \sigma_3}$,
with $\sigma_3 = \diag(1,-1)$ the third Pauli matrix,
and
$V=(v_+/\| v_+\|\,,\,v_-/\|v_-\|)$ is the real matrix of orthonormal eigenvectors of $Q_+$.
Now consider the transformation
\be
\tilde{Q}(x,t)= \Delta_+^{-1/2} V Q(x,t) V^T \Delta_+^{-1/2}\,.
\label{e:QtoQtilde}
\ee
It is easy to show that $\tilde{Q}(x,t)$
is symmetric whenever $Q(x,t)$ is.
As a consequence, without loss of generality one can take the background to be $Q_+=\Ko I_2$
up to admissible unitary transformations, i.e.,
complex rotations of the quantization axes.
Finally, recall that the MNLS equation is scale invariant. Namely, if $Q(x, t)$ is a solution of \eqref{e:MNLS}, so is
$\tilde{Q}(x, t) = c \hat{Q}(cx, c^2t)$
for any constant $c \in \Real$.
Therefore, we can take $\Ko = 1$,
which implies that $\hat{Q}(x,t)$ is in canonical form.
%Finally, a further combination of unitary transformations $\Delta ^{-1/2}\tilde{Q}(x,t)\Delta^{-1/2}$
%not possible, in general, to reduce the solution to one with a background proportional to the identity by admissible unitary transformations/rotations of the quantization axes.

The complete integrability of the MNLS equation implies that the system~\eqref{e:MNLS} has an infinite number of conserved quantities in involution. Of particular
importance for describing the physical properties of the condensate are the total number of holes/particles $\overline{N}$, and the total spin $\@F$, which can be
expressed, respectively, as integrals over the spatial domain of the particle number density $\=n(x,t)$ and of the spin densities in the three components, $f_1,f_0,f_{-1}$, namely
%\marginpar{\bb What's the\\difference\\btw $n$ \&\ $\bar n$?\eb}
\bse
\label{e:densities}
\begin{gather}
\displaystyle
\label{e:ndensity}
\overline{N} = \int_\Real \=n(x,t)\,\d x\,,\qquad \=n(x,t) = \tr(Q_\pm^\dag Q_\pm)- \tr(Q^\dag Q)\,,\\
\label{e:fdensity}
\@F = \int_\Real \mathbf{f}(x,t)\,\d x\,,\qquad \mathbf{f}(x,t)\equiv \left(f_1,f_0,f_{-1} \right) := \tr(Q^\dag \boldsymbol{\sigma} Q)\,,
\end{gather}
\ese
where $\boldsymbol{\sigma} = (\sigma_1,\sigma_2,\sigma_3)$ are the Pauli matrices.
For future reference, we note that the particle number density is invariant under arbitrary unitary transformations and remains the same
if a complex rotation of the quantization axes is performed to reduce the background $Q_+$ to the identity.
The spin density, on the other hand, is not invariant under unitary transformations from the left. Indeed, one can
easily verify that multiplying $Q$ by an arbitrary unitary matrix from the right does not change the spin density, while multiplication
from the left results in the spin density changing covariantly.
More specifically, under the transformation \eqref{e:QtoQtilde},
which, reduces the solution to canonical form up to the rescaling of $\Ko$,
the spin density becomes
\bse
\be
\label{e:fbarunitrayrelation}
\~{\@f}(x,t) = S_{\theta,\delta}\,\@f(x,t)
\ee
%where $\~{\@f}$ is the spin density for $\~{Q}(x,t) := U Q(x,t) V$, with $U := e^{i \alpha \sigma_3}[\cos \theta I_2 -i \sin \theta\,\sigma_2]$ and $V$ is a unitary matrix, and the invertible matrix
where,
writing the orthogonal matrix $V$ as $V=\cos \theta I_2 -i \sin \theta\,\sigma_2$
($\sigma_2$ being the second Pauli matrix, see Appendix~\ref{s:symmetries}),
\be
\label{e:rotationmatrix}
\displaystyle
S_{\theta,\delta} = \begin{pmatrix}
\cos\delta\,\cos(2\theta) & -\sin\delta & \cos\delta\,\sin(2\theta)\\
\sin\delta\,\cos(2\theta) & \cos\delta & \sin\delta\,\sin(2\theta)\\
-\sin(2\theta) & 0 & \cos(2\theta)
\end{pmatrix}\,,
\ee
\ese
It is worth noticing that $S_{\theta,\delta}$ is indeed an orthogonal matrix, and the
transformation to canonical form amounts to a rotation of the quantization axes (again, see Appendix~\ref{s:symmetries}).

%%%%%%%%%%%%%%%%%%%%%%%%%%%%%%%%%%%%%%%%%%%%%%%%%%%%%%%%%%%%%%%%%%%%%%%%%%%%%%%%%%%%%%%%%%%%%
\subsection{Overview of the IST for the defocusing MNLS equation with NZBC}

In order to derive an expression for the multi-soliton solutions of the defocusing MNLS equation \eqref{e:mNLSgen},
and to fully characterize the physical parameters of the solitons in terms of spectral data,
it is convenient to first briefly review the IST for \eqref{e:mNLSgen} with NZBC
that was developed in \cite{BP4}.

As originally shown in \cite{Wadati1}, the MNLS equation \eqref{e:MNLS} for a $2 \times 2$ potential matrix $Q(x,t)$ is equivalent to the compatibility condition ($\phi_{xt}=\phi_{tx}$) of the following $4\times 4$ Lax pair:
%\begin{subequations}
\[
\label{e:Laxpair}
%	\label{e:scatprob}
\phi_x= \@U \phi\,, \qquad
%	\label{e:timeevol}
\phi_t= \@V \phi\,,
\]
%\end{subequations}
with
\bse
\begin{gather}
\label{e:BigUandV}
\@U(x,t,k)=-ik\bfsigma_3+\bfQ, \quad
\@V(x,t,k)=-2ik^2\bfsigma_3+2k\bfQ+i\bfsigma_3[\bfQ_x+\Kosqr I_4-\bfQ^2],
\\
\bfsigma_3=\begin{pmatrix} I_2 & 0_2 \\ 0_2 & -I_2 \end{pmatrix}, \quad \bfQ=\begin{pmatrix} 0_2 & Q \\ Q^\dag & 0_2 \end{pmatrix},
\end{gather}
\ese
where
$0_n$ is the $n \times n$ zero matrix.
As usual,
one refers to the first equation of the Lax pair \eqref{e:Laxpair} as the scattering problem.
The IST for the MNLS \eqref{e:MNLS} with non-zero boundary conditions was developed in \cite{Wadati4,BP1,BP4}.
Next we give a brief overview of the IST formulation following \cite{BP4},
which we will then use to obtain a formula for the  multi-soliton solution.
%as well the complete spectral characterization of the solitons and their interactions.

Importantly, note that the constraint \eqref{e:BCconstraint} on the boundary conditions plays the same role as the ``equal amplitude'' boundary condition in the scalar and vector NLS equations, and it ensures
that the asymptotic scattering problems as $x\to \pm \infty$ are equal and only have two branch points.
Indeed, taking into account \eqref{e:BCconstraint}, the asymptotic scattering problems
(which are obtained by replacing $\bfQ$ with $\bfQ_{\pm}$ in \eqref{e:BigUandV})
have eigenvalues $\pm i\lambda$ with $\lambda= (k^2- \Kosqr )^{1/2}$, and each eigenvalue has multiplicity 2.
As in the IST for the Manakov system \cite{FT1987},
is convenient to introduce uniformization variable $z$ defined by the conformal mapping
\[
\label{e:zdef}
z = k+\lambda,
\]
whose inverse transformation is
\[
\label{e:kldef}
k=\frac{1}{2}(z+ \Kosqr /z), \quad \lambda=\frac{1}{2}(z- \Kosqr /z).
\]
Consequently, $\operatorname{Im} \lambda >0$ corresponds to the region $\Complex^{+}$ in the $z$-plane, and $\operatorname{Im} \lambda <0$ corresponds to the region $\Complex^{-}$ in the $z$-plane.
The Jost solutions are defined as the simultaneous solutions of both parts of the Lax pair identified by the BCs:
\bse
	\begin{align}
	\label{e:phi}
	{\Phi}(x,t,z)\equiv (\varphi(x,t,z),\bar{\varphi}(x,t,z))=X_{-}(z)e^{-i\theta(x,t,z)\bfsigma_3}(1 + o(1)),
	\qquad x \to -\infty, \\
	\label{e:psi}
	{\Psi}(x,t,z)\equiv (\bar{\psi}(x,t,z),\psi(x,t,z))=X_{+}(z)e^{-i \theta(x,t,z)\bfsigma_3}(1 + o(1)),
	\qquad x \to \infty,
	\end{align}
\ese
where $\varphi(x,t,z)$, $\bar{\varphi}(x,t,z)$, $\bar{\psi}(x,t,z)$ and $\psi(x,t,z)$ are $4 \times 2$ matrices,
\begin{equation}
\label{e:theta}
\theta(x,t,z)=\lambda(z)(x + 2 k(z) t),
\end{equation}
and
\bse
\label{e:assympX}
	\begin{align}
	X_{\pm}(z)= I_4-\frac{i}{z}\bfsigma_3\bfQ_{\pm},\qquad \label{e:Xinverse}
	X_{\pm}^{-1}(z)=\frac{1}{\gamma(z)}\left(I_4+\frac{i}{z}\bfsigma_3\bfQ_{\pm}\right),
	\qquad z \in \Real\setminus\{0,\pm \Ko\},\\
	\label{e:gammadef}
	\det X_{\pm}(z)=\left(\frac{2 \lambda}{\lambda + k}\right)^2=(\gamma(z))^2, \quad \gamma(z)=1-\frac{\Kosqr }{z^2}.
	\end{align}
\ese
As usual, the continuous spectrum of the scattering problem corresponds to values of $(k,\lambda)$, or, equivalently, $z$, such that
	 all four eigenfunctions above are bounded for all $x\in \mathbb{R}$, which requires $\lambda(k) \in \mathbb{R}\setminus\left\{0\right\}$ and hence $k\in (\infty,-\Ko)\cup (\Ko,+\infty)$.
	In the $z$-plane, the continuous spectrum is $\Sigma:=\mathbb{R}\setminus \left\{\pm \Ko\right\}$.
%\newline
A complete set of modified analytic eigenfunctions with constant limit as $x \to \pm \infty$ can be defined as
\bse
	\begin{align}
	\label{e:modeigM}
	\M(x,t,z) \equiv (M(x,t,z),\bar{M}(x,t,z))={\Phi}(x,t,z)\,e^{i\theta(x,t,z)\bfsigma_3}, \\
	\label{e:modeigN}
	\N(x,t,z) \equiv (\bar{N}(x,t,z),N(x,t,z)) = {\Psi}(x,t,z)\,e^{i\theta(x,t,z)\bfsigma_3},
	\end{align}
\ese
One can express the modified eigenfunctions $M$, $\bar{M}$, $N$ and $\bar{N}$
	as solutions of suitable Volterra-type integral equations,
	and show that under some mild integrability conditions of $Q(x,t)-Q_\pm$ for $x\in(x_o,\pm \infty)$ and any fixed $t\ge 0$, the modified eigenfunctions $M(x,t,z)$ and $N(x,t,z)$ can be analytically extended to $\Complex^+$ in the $z$-plane. Similarly, the modified eigenfunctions $\bar{M}(x,t,z)$ and $\bar{N}(x,t,z)$ can be analytically extended to $\Complex^-$ in the $z$-plane.	

Because
$
\det {\Phi}(x,t,z)=\det {\Psi}(x,t,z)=\det X_{\pm}=(\gamma(z))^2$ for all $x,t,z \in \Real$,
${\Phi}$ and ${\Psi}$ are both fundamental solutions of the scattering problem. Hence there exists a proportionality matrix ${S}(z)$ between the two fundamental solutions, such that
\[
\label{e:scatcoef}
{\Phi}(x,t,z)={\Psi}(x,t,z){S}(z), \qquad {S}(z)=\begin{pmatrix} a(z) & \bar{b}(z) \\ b(z) & \bar{a}(z) \end{pmatrix}, \quad x,t\in \mathbb{R}, \quad z \in \Real\setminus\{\pm \Ko\},
\]
where $S(z)$ is referred to as the scattering coefficient matrix and $a,b,\bar{a},\bar{b}$ are $2 \times 2$ block matrices. Since $\det \Phi = \det \Psi$ we have $\det S(z)=1$ for $ z \in \Real\setminus\{\pm \Ko\}$. In turn,
	from \eqref{e:scatcoef} it also follows that:
\begin{subequations}
\label{e:scatdet}
	\begin{align}
	\label{e:scatdet1}
	\det a(z)=\text{Wr}(\varphi,\psi)/\text{Wr}(\bar{\psi},\psi)\equiv \det (\varphi,\psi)/\det \Psi=\det (\varphi,\psi)/(\gamma(z))^2, \\
	\label{e:scatdet2}
	\det  \bar{a}(z)=\text{Wr}(\bar{\psi},\bar{\varphi})/\text{Wr}(\bar{\psi},\psi)\equiv \det (\bar{\psi},\bar{\varphi})/\det \Psi=\det (\bar{\psi},\bar{\varphi})/(\gamma(z))^2,
	\end{align}
\end{subequations}
where $\text{Wr}(u,v)$ denotes the Wronskian determinant of $4\times 2$ vector functions $u$ and $v$.

The Jost eigenfunctions satisfy the following symmetry relations with respect to the involution $z \mapsto z^*$:
\[
\label{e:sym1}
\Phi^{\dagger}(x,t,z^*)\bfsigma_3\Phi(x,t,z)=\Psi^{\dagger}(x,t,z^*)\bfsigma_3\Psi(x,t,z)=\gamma(z)\bfsigma_3\,.
\]
We will use the following notation to denote the $2 \times 2$ blocks of the eigenfunction matrices $\Phi$ and $\Psi$:
\begin{equation}
\label{e:eigfunblocks}
\Phi(x,t,z)=\begin{pmatrix} \varphi_{\text{up}} & \bar{\varphi}_{\text{up}} \\ \varphi_{\text{dn}} & \bar{\varphi}_{\text{dn}} \end{pmatrix}, \quad \Psi(x,t,z)=\begin{pmatrix} \bar{\psi}_{\text{up}} & \psi_{\text{up}} \\ \bar{\psi}_{\text{dn}} & \psi_{\text{dn}} \end{pmatrix},
\end{equation}
so \eqref{e:sym1} can be written in block-matrix form
\bse
	\begin{align}
	\gamma(z) a(z)= \bar{\psi}_{\text{up}}^{\dagger}(z^*)\varphi_{\text{up}}(z)- \bar{\psi}_{\text{dn}}^{\dagger}(z^*) \varphi_{\text{dn}}(z), \\
	\gamma(z) \bar{a}(z)= {\psi}_{\text{dn}}^{\dagger}(z^*) \bar{\varphi}_{\text{dn}}(z)- {\psi}_{\text{up}}^{\dagger}(z^*)\bar{\varphi}_{\text{up}}(z), \\
	\gamma(z) b(z)= {\psi}_{\text{dn}}^{\dagger}(z^*) \varphi_{\text{dn}}(z)- {\psi}_{\text{up}}^{\dagger}(z^*)\varphi_{\text{up}}(z), \\
	\gamma(z) \bar{b}(z)= \bar{\psi}_{\text{up}}^{\dagger}(z^*) \bar{\varphi}_{\text{up}}(z)- \bar{\psi}_{\text{dn}}^{\dagger}(z^*)\bar{\varphi}_{\text{dn}}(z)\,,
	\end{align}
\ese
where the $x,t$ dependence of the eigenfunctions on the right-hand side has been omitted for shortness.
The above relations show that $a(z)$ can be analytically extended to $\Complex^+$, and $\bar{a}(z)$ can be analytically extended to $\Complex^{-}$.
Also, we obtain
\begin{equation}
S^{-1}(z)=\bfsigma_3 S^{\dagger}(z^*)\bfsigma_3, \quad S^{-1}(z)= \begin{pmatrix} \bar{c}(z) & d(z) \\ \bar{d}(z) & c(z) \end{pmatrix}\,,
\end{equation}
which provides symmetries for the scattering coefficients:
\begin{subequations}
	\begin{align}
	a^{\dagger}(z^*) a(z) - b^{\dagger}(z^*)  b(z) = I_2 , \qquad
	a^{\dagger}(z^*)  \bar{b}(z) - b^{\dagger}(z^*)  \bar{a}(z) = 0_2, \\
	\bar{b}^{\dagger}(z^*)  a(z) - \bar{a}^{\dagger}(z^*)  b(z) = 0_2, \qquad
	\bar{b}^{\dagger}(z^*)  \bar{b}(z) - \bar{a}^{\dagger}(z^*)  \bar{a}(z) = -I_2.
	\end{align}
\end{subequations}
and
\[
\label{e:sym1SinvSrelations}
	\bar{c}(z) =  a^{\dagger}(z^*), \quad
	d(z) = -  b^{\dagger}(z^*) , \quad
	\bar{d}(z) = -  \bar{b}^{\dagger}(z^*) , \quad
	c(z) =  \bar{a}^{\dagger}(z^*) .
\]
The scattering problem also admits a second involution: $z \mapsto \Kosqr/z$.
The corresponding symmetries for the eigenfunctions are given by:
\begin{equation}
\label{e:2symmefs}
\Phi(x,t,z)=-\frac{i}{z}\Phi(x,t, \Kosqr /z)\bfsigma_3\bfQ_{-}\,, \quad \Psi(x,t,z)=-\frac{i}{z}\Psi(x,t, \Kosqr /z)\bfsigma_3\bfQ_+, \quad z \in \Sigma\,.
\end{equation}
Explicitly, each of the $4 \times 2$ Jost eigenfunctions satisfies
\begin{subequations}
	\begin{align}
	\label{e:eigfunsym2a}
	\varphi(x,t,z)=\frac{i}{z}\bar{\varphi}(x,t,\Kosqr /z)Q_{-}^\dagger, \quad \bar{\varphi}(x,t,z)=-\frac{i}{z}\varphi(x,t,\Kosqr /z)Q_{-}\,, \\
	\label{e:eigfunsym2b}
	\bar{\psi}(x,t,z)=\frac{i}{z}\psi(x,t, \Kosqr /z)Q_+^\dagger, \quad \psi(x,t,z)=-\frac{i}{z}\bar{\psi}(x,t, \Kosqr /z)Q_{+}\,.
	\end{align}
\end{subequations}
These symmetries imply the following relations for the scattering data:
\begin{subequations}
	\begin{align}
	\label{e:sym2a,abar}
	a(\Kosqr /z)=\frac{1}{\Kosqr }Q_{+}\bar{a}(z)Q^\dag_-\,, \qquad
	%	\label{e:sym2abar}
	\bar{a}(\Kosqr /z) = \frac{1}{\Kosqr }Q^\dag_+ a(z) Q_{-}\,, \\
	\label{e:sym2b,bar}
	b(\Kosqr /z) = -\frac{1}{\Kosqr } Q^\dag_+\bar{b}(z)Q^\dag_-\,, \qquad
	%	\label{e:sym2bbar}
	\bar{b}( \Kosqr /z) = -\frac{1}{\Kosqr }Q_{+} b(z) Q_{-}\,.
	\end{align}
\end{subequations}
A third symmetry follows from the fact that we assume the potential $Q(x,t)$ to be a symmetric matrix. Correspondingly, the eigenfunctions satisfy the following symmetries:
\begin{equation}
\Phi^T(x,t,z)\bfsigma_2\Phi(x,t,z)=\Psi^T(x,t,z)\bfsigma_2\Psi(x,t,z)=\gamma(z)\bfsigma_2, \qquad \bfsigma_2 = \begin{pmatrix} 0_2 & -iI_2 \\ iI_2 & 0_2 \end{pmatrix}\,,
\end{equation}
where the superscript $^T$ denotes matrix transpose,
which implies that
\begin{equation}
\label{e:sym3Srel}
S^T(z)\bfsigma_2S(z)=\bfsigma_2, \quad z \in \Sigma\,,
\end{equation}
and consequently
\begin{equation}
c(z)=a^T(z), \quad \bar{c}(z)=\bar{a}^T(z), \quad d(z)=-\bar{b}^T(z), \quad \bar{d}(z) = -b^T(z).
\end{equation}
The discrete spectrum is the set of all values $z_j \in \mathbb{C}\setminus \Real$ where $\det a(z) = 0$ or $\det \=a(z) = 0$. Since the scattering operator is self-adjoint, $z_j \in C_o:= \{z \in \Complex : |z| = \Ko\}$. Moreover, the symmetries of the scattering data imply that $\det a(z) =0$ if and only if $\det \=a(z^*)=0$. Suppose that $\det a(z)$ has a finite number $J$ of zeros $z_1,\ldots,z_J$ in $C_0^{+} = C_0 \cap \{z \in \mathbb{C}:\operatorname{Im} > 0\}$ and, by symmetry, $\det \bar{a}(z)$ has a finite number $J$ of zeros $z_1^*,\ldots,z_J^*$ in $C_0^{-} = C_0 \cap \{z \in \mathbb{C}:\operatorname{Im} < 0\}$. Let us define
\[
\label{e:Pdef}
P(x,t,z)=(\varphi(x,t,z),\psi(x,t,z)), \quad \bar{P}(x,t,z)=(\bar{\psi}(x,t,z),\bar{\varphi}(x,t,z))\,.
\]
As we will discuss next, the nature of the discrete eigenvalue $z_j$ (or, equivalently, $z_j^*$) depends on the rank of the matrix $P(x,t,z_j)$ (equivalently, $\bar{P}(x,t,z_j^*)$).

\paragraph{Norming constants and residue conditions: Case~1, $\operatorname{rank}P(x,t,z_n)=3$.}
As shown in \cite{BP4}, the Wronskian representation \eqref{e:scatdet} in this case yields
\[
\label{e:rank3proportionality}
\varphi(x,t,z_n) \alpha(z_n) = \psi(x,t,z_n)c_n, \qquad
\bar{\varphi}(x,t,z_n^*) \bar{\alpha}(z_n) = \bar{\psi}(x,t,z_n^*)\bar{c}_n,
\]
where $c_n$ and $\=c_n$ are constant 2 $\times$ 2 rank-1 matrices, and $\alpha(z),\,\={\alpha}(z)$ denotes the adjugate (or cofactor) matrix of $a(z)$ and $\=a(z)$, respectively.
These provide the residue relations
\bse
\label{e:norminconstrank3}
\begin{gather}
\underset{z=z_n}{\operatorname{Res}}[M(x,t,z_n)a^{-1}(z)]=e^{2i\theta(x,t,z_n)}N(x,t,z_n)C_n, \qquad C_n=\frac{c_n}{(\det a)'(z_n)}\, \qquad \det C_n = 0,\\
\underset{z=z_n^*}{\operatorname{Res}} [\bar{M}(x,t,z)\bar{a}^{-1}(z)] = e^{-2i\theta(x,t,z_n^*)} \bar{N}(x,t,z_n^*) \bar{C}_n, \qquad \bar{C}_n=\frac{\bar{c}_n}{(\det \bar{a})'(z_n^*)}\,,\qquad \det \bar{C}_n = 0,
\end{gather}
\ese
where prime denotes the derivative with respect to $z$.

\paragraph{Norming constants and residue conditions: Case~2, $\operatorname{rank}P(x,t,z_n)=2$.}
It is possible, on the other hand, to have double zeros of $\det a(z)$ and $\det \=a(z)$ for which the matrices
$Ma^{-1}$ and $\bar{M}\=a^{-1}$ still have a simple pole for such a value of $z$.
When this happens, $a(z_n)=\bar{a}(z_n^*)=0_{2 \times 2}$ and $\operatorname{rank}P(x,t,z_n)=\operatorname{rank}\bar{P}(x,t,z_n^*)=2$.
In this scenario a stronger condition of proportionality between the eigenfunctions holds, namely:
\[
	\varphi(x,t,z_n)=\psi(x,t,z_n)b_n, \qquad
	\bar{\varphi}(x,t,z_n^*)=\bar{\psi}(x,t,z_n^*){\bar{b}}_n,	
\]
where $b_n,\bar{b}_n$ are constant, non-singular $2 \times 2$ matrices. In this case, the residue conditions read
\begin{subequations}
\label{e:normingconstantrank2}
	\begin{align}
	\underset{z=z_n}{\operatorname{Res}}[M(x,t,z)a^{-1}(z)]=e^{2i\theta(x,t,z_n)}N(x,t,z_n)C_n, \qquad C_n=\frac{2b_n\alpha'(z_n)}{(\det a)''(z_n)}, \\
	\underset{z=z_n^*}{\operatorname{Res}}[\bar{M}(x,t,z)\bar{a}^{-1}(z)]=e^{-2i\theta(x,t,z_n^*)}\bar{N}(x,t,z_n^*)\bar{C}_n, \qquad \bar{C}_n=\frac{2\bar{b}_n\bar{\alpha}'(z_n^*)}{(\det \bar{a})''(z_n^*)}.
	\end{align}
\end{subequations}
and although the residue conditions formally have the same expression as in the rank-1 case (cf \eqref{e:normingconstantrank2}), here the norming constants $C_n,\bar{C}_n$ need not be rank-1 matrices.

The asymptotic behaviors of the eigenfunctions and the scattering data as $z \to \infty$ and $z \to 0$ are needed in order to properly formulate the
inverse problem for the eigenfunctions, and subsequently reconstruct the potential matrix. They are given in \cite{BP4} as:
\begin{subequations}
	\begin{align}
	M(x,t,z)=\begin{pmatrix} I_2+\frac{i}{z} \int_{-\infty}^x [Q(x',t)Q^\dag(x',t)- \Kosqr  I_2]dx'+O(1/z^2) \\ \frac{i}{z} Q^\dag(x,t)+O(1/z^2) \end{pmatrix} \qquad z \to \infty, \, z \in \Complex^{+},\\
	\label{e:Mbaratinf}
	\bar{M}(x,t,z)=\begin{pmatrix} -\frac{i}{z}Q(x,t)+O(1/z^2) \\ I_2-\frac{i}{z} \int_{-\infty}^x [Q^\dag(x',t)Q(x',t)- \Kosqr  I_2]dx'+O(1/z^2) \end{pmatrix} \qquad z \to \infty, \, z\in \Complex^{-}, \\
	\label{e:Nbaratinf}
	\bar{N}(x,t,z)=\begin{pmatrix} I_2+\frac{i}{z} \int_x^{\infty} [Q(x',t)Q^\dag(x',t)- \Kosqr  I_2]dx'+O(1/z^2) \\ \frac{i}{z}Q^\dag(x,t)+O(1/z^2) \end{pmatrix} \qquad z \to \infty,\, z\in \Complex^{-}, \\
	\label{e:Natinf}
	N(x,t,z)=\begin{pmatrix} -\frac{i}{z}Q(x,t)+O(1/z^2) \\ I_2-\frac{i}{z} \int_x^{\infty} [Q^\dag(x',t)Q(x',t)- \Kosqr  I_2]dx'+O(1/z^2) \end{pmatrix} \qquad z \to \infty,z\in \Complex^{+},\\
	M(x,t,z)=\begin{pmatrix}  QQ^\dag_{-}/\Kosqr +O(z) \\ iQ^\dag_{-}/z+O(1) \end{pmatrix}\,, \qquad
N(x,t,z)=\begin{pmatrix} -iQ_{+}/z+O(1) \\ Q^\dag Q_{+}/\Kosqr +O(z) \end{pmatrix} \quad \text{as } z \to 0, \, z \in \Complex^{+}, \\  \bar{M}(x,t,z)=\begin{pmatrix} -iQ_{-}/z+O(1) \\  Q^\dag Q_{-}/\Kosqr +O(z) \end{pmatrix}\,, \qquad
\bar{N}(x,t,z)=\begin{pmatrix}  Q Q^{\dag}_{+}/\Kosqr +O(z) \\ i Q^{\dag}_{+}/z+O(1) \end{pmatrix}, \quad \text{as } z \to 0, \, z \in \Complex^{-},
	\end{align}
\end{subequations}
implying
\be
S(z)=I_2+O(1/z),\quad
z \to \infty,
\qquad
S(z)=\frac{1}{\Kosqr} \begin{pmatrix}  Q_{+}Q^\dag_{-} & 0_2 \\ 0_2 &  Q^\dag_{+}Q_{-} \end{pmatrix}+O(z),
\quad z \to 0,
\ee
with both limits taken along the real axis.
%\eject

The inverse problem can be formulated as a matrix Riemann-Hilbert problem (RHP) in terms of the uniformization variable:
\begin{equation}
\label{e:mu2}
\mu^{-}(x,t,z)=\mu^{+}(x,t,z)(I_4-G(x,t,z)),\quad z \in \Sigma,
\end{equation}
where the sectionally meromorphic matrices are
\[
\label{e:mu1}
\mu(x,t,z) =
\begin{cases}
(Ma^{-1},N),\qquad \Im z > 0\,,
\\
(\bar{N},\bar{M}\bar{a}^{-1}),\qquad \Im z < 0\,,
\end{cases}
\]
with $\mu^{\pm}(x,t,z)$ denoting the projection of $\mu(x,t,z)$ to the real $z-$axis from above/below, the jump matrix is
\begin{equation}
\label{e:mu3}
G(x,t,z)=\begin{pmatrix} 0_2 & -e^{-2i\theta(x,t,z)}\bar{\rho}(z) \\ e^{2i\theta(x,t,z)}\rho(z) & \rho(z)\bar{\rho}(z) \end{pmatrix},
\end{equation}
and the reflection coefficients are $\rho(z) = b(z)a^{-1}(z)$ and $\={\rho}(z) = \=b(z)\={a}^{-1}(z)$. The matrices $\mu^{\pm}(x,t,z)-I_2$ are $O(1/z)$ as $z \to \infty$. After regularization, to account for the pole at $z =0$ and at the discrete eigenvalues $\{z_j,z_j^*\}_{j =1}^N$, the RHP can be solved via Cauchy projectors, and the asymptotic behavior of the upper $2 \times 2$ block of $N(x,t,z)$ as $z \to \infty$
yields the reconstruction formula
\begin{equation}
\label{e:Qpotential}
Q(x,t)=Q_{+}+i\sum_{j=1}^{J} e^{-2i\theta(x,t,z_j^*)}\bar{N}_{\text{up}}(x,t,z_j^*)\bar{C}_j - \frac{1}{2 \pi} \int_{\Real} e^{-2i\theta(x,t,\zeta)}\bar{N}_{\text{up}}(x,t,\zeta)\bar{\rho}(\zeta)d\zeta.
\end{equation}
Using the reconstruction formula and the second symmetry we obtain the symmetry relations for the norming constant
\bse
\label{e:symmetryofnormigcnst}
\begin{gather}
\label{e:symmetryofnormigcnst1}
\=C_n = C_n^\dag, \qquad Q_+ C_n = e^{2 i \arg(z_n)} \=C_n Q_+^\dag  \\
\label{e:symmetryofnormigcnst2}
C_n^{T} = C_n, \qquad \={C}_n^{T} = \=C_n
\end{gather}
\ese
where $n= 1,...,J$.
We are interested in potentials $Q(x,t)$ where the reflection coefficient $\rho(z)$ is identically zero for $z \in \Real$, which implies that $\bar{\rho}(z)$ is also zero for $z \in \Real$. Under this assumption of reflectionless potentials, we have
\begin{equation}
\label{e:Qpotnoreflection}
Q(x,t)=Q_{+}+i\sum_{j=1}^{J} e^{-2i\theta(x,t,z_j^*)}\bar{N}_{\text{up}}(x,t,z_j^*)\bar{C}_j,
\end{equation}
with
\begin{subequations}
\label{e:efs_norefl}
	\begin{gather}
	\label{e:Nbarupnoreflection}
	\bar{N}_{\text{up}}(x,t,z_n^*)=I_2+\sum_{j=1}^{J} \frac{e^{2i\theta(x,t,z_j)}N_{\text{up}}(x,t,z_j)C_j}{z_n^*-z_j}, \\
	\label{e:Nupnoreflection}
	N_{\text{up}}(x,t,z_n)=-\frac{i}{z_n}Q_{+}+\sum_{j=1}^{J} \frac{e^{-2i\theta(x,t,z_j^*)}\bar{N}_{\text{up}}(x,t,z_j^*)\bar{C}_j}{z_n-z_j^*}.
	\end{gather}
\end{subequations}
Solving the linear system \eqref{e:efs_norefl} for the eigenfunctions and substituting into the reconstruction formula \eqref{e:Qpotential} yields
the $J$ soliton solution for the defocusing MNLS.

\subsection{Multi-soliton solutions}

Using the IST formalism above, we now derive an explicit formula for the general multi-soliton solution of~\eqref{e:MNLS} with $\nu=1$ and NZBC.
First,
substituting \eqref{e:Nupnoreflection} into \eqref{e:Nbarupnoreflection} we have
\[
\label{e:Nbarreflectionless}
\bar{N}_{\text{up}}(x,t,z_n^*)=I_2-iQ_{+}\sum_{j=1}^{J} \frac{e^{2i\theta(x,t,z_j)}C_j}{z_j(z_n^*-z_j)}+\sum_{j=1}^{J} \sum_{l=1}^{J} \frac{e^{2i(\theta(x,t,z_j)-\theta(x,t,z_l^*))}}{(z_n^*-z_j)(z_j-z_l^*)}\bar{N}_{\text{up}}(x,t,z_l^*)\bar{C}_lC_j.
\]
Note that the exponents $i\theta(x,t,z_j)$ and $i\theta(x,t,z_j^*)$ appearing in~\eqref{e:Qpotnoreflection} and~\eqref{e:efs_norefl}
are all real.
For convenience let us take the transpose of the equation \eqref{e:Nbarreflectionless},
\[
\label{e:transposeNbarreflectionless}
\displaystyle
\bar{N}_{\text{up}}^{T}(x,t,z_n^*)=I_2-i\sum_{j=1}^{J} \frac{e^{2i\theta(x,t,z_j)}C_j}{z_j(z_n^*-z_j)}\,Q_{+}+\sum_{j=1}^{J} \sum_{l=1}^{J} \frac{e^{2i(\theta(x,t,z_j)-\theta(x,t,z_l^*))}}{(z_n^*-z_j)(z_j-z_l^*)}C_j\bar{C}_l\bar{N}_{\text{up}}^{T}(x,t,z_l^*).
\]
Introducing $\@X = (Y_1,Y_2,...,Y_J)^{T}$ and $\@B = (V_1,V_2,...,V_J)^{T}$ where
\[
\nonumber
Y_n = \bar{N}_{\text{up}}^{T}(x,t,z_n^*),\qquad
V_n = I_2-i\sum_{j=1}^{J} \frac{e^{2i\theta(x,t,z_j)}C_j}{z_j(z_n^*-z_j)}\,Q_{+},\quad n= 1,..,J,
\]
and defining the $2J \times 2J$ matrix $A = (A_{n,l})$, where
\[
\nonumber
A_{n,l}= \sum_{j=1}^{J}\frac{e^{2i(\theta(x,t,z_j)-\theta(x,t,z_l^*))}}{(z_n^*-z_j)(z_j-z_l^*)}C_j\bar{C}_l\,,\qquad n,l = 1,2,..,J,
\]
the system \eqref{e:transposeNbarreflectionless} becomes simply $R \@X = \@B$, where $R = I_{2J}- A = (R_1,R_2,...,R_{2J})$ and
$\@X,\@B$, and $A$ consists of $2 \times 2$ block matrices.
We can then rewrite this system as
\[
\label{e:decomposesystem}
R X_1 = B_1, \qquad R X_2 = B_2,
\]
with $\@X = (X_1, X_2)$ and $\@B = (B_1,B_2)$, whose solution \eqref{e:decomposesystem} is simply: $X_{n,1} = \det \hat{R}^{\text{ext}}_n / \det R$ and $X_{n,2} = \det \check{R}^{\text{ext}}_n / \det R$ for $n = 1,2,...,2J$, where
\[
\nonumber
\hat{R}^{\text{ext}}_n = (R_1,R_2,\dots,R_{n-1},B_1,R_{n+1},\dots,R_{2J})\,,\qquad
\check{R}^{\text{ext}}_n = (R_1,R_2,\dots,R_{n-1},B_2,R_{n+1},\dots,R_{2J}).
\]
It then follows that
\[
\displaystyle
\bar{N}_{\text{up}}^{T}(x,t,z_n^*) = \begin{pmatrix} X_{2n-1, 1}& X_{2n-1, 2}\\
X_{2n, 1}&X_{2n, 2}
\end{pmatrix}\,,\qquad n = 1,2,\dots,J,
\]
and Eq.~\eqref{e:Qpotnoreflection} yields
\be
\label{e:transposeQpotnoreflection}
Q(x,t)=Q_{+}+i\sum_{j=1}^{J} e^{-2i\theta(x,t,z_j^*)}\bar{C}_j\,\bar{N}_{\text{up}}^{T}(x,t,z_j^*).
\ee
Finally, upon substituting $Y_1,\dots,Y_J$ into the above formula, the resulting expression for the potential can be written compactly as
\[
\displaystyle
Q(x,t) = \frac1{\det R} \begin{pmatrix} \det N_{11}^{\text{aug}}& \det N_{12}^{\text{aug}}\\
\det N_{21}^{\text{aug}}&\det N_{22}^{\text{aug}}
\end{pmatrix},
\label{e:multisoliton}
\]
where the augmented $(2J+1) \times (2J+1)$ matrices are given by
\bse
\[
N_{j k}^{\text{aug}} = \begin{pmatrix} Q_{+,j k} &-i D_j^{T}\\
B_{k}& R
\end{pmatrix}\,,\qquad j,k \in \{1,2\},
\]
and
\[
Q_+ = (Q_{+,ij})\,,\quad i,j \in \{1,2\}\,,\quad
(D_1, D_2)= (E_1,E_2,\dots,E_N)^{T}\,,\quad E_n = e^{-2i\theta(x,t,z_n^*)}\bar{C}_n\,,\quad n = 1,\dots,J\,.
\]
\ese

%%%%%%%%%%%%%%%%%%%%%%%%%%%%%%%%%%%%%%%%%%%%%%%%%%%%%%%%%%%%%%%%%%%%%%%%%%%%%%%%%%%%%%%%%%%%%
\section{One-soliton solutions}
\label{s:direct1}

In this section we discuss and classify the one-soliton solutions,
namely the solutions obtained from~\eqref{e:multisoliton} with $J=1$.

%%%%%%%%%%%%%%%%%%%%%%%%%%%%%%%%%%%%%%%%%%%%%%%%%%%%%%%%%%%%%%%%%%%%%%%%%%%%%%%%%%%%%%%%%%%%%

\subsection{Classification of one-soliton solutions}
\label{s:classification}

Solving
\eqref{e:Qpotnoreflection} when $J=1$  we have
\[
\label{e:mostgeneralpotential}
Q(x,t) = Q_+ + \frac{i e^{2 i \theta(x,t,z_1)}}{z_1 }\Big(I_2 + \dfrac{i e^{2 i \theta(x,t,z_1)}}{(z_1^*-z_1)} \Pi_1 \Big)^{-1}\Pi_1 Q_+,\qquad x\in \Real,\, t\in \Real^+\,,
\]
where $\theta(x,t,z)$ is as in \eqref{e:theta}, $z_1 = \Ko e^{i \varphi}$ with $\varphi \in (0,\pi)$ and
\[
\label{e:Pi}
\Pi_1 = \frac{1}{z_1} Q_+ C_1\,,
\]
with $\Pi_1 = \Pi_1^\dag$ thanks to \eqref{e:symmetryofnormigcnst1}.
Since $\Pi_1$ is Hermitian, there exists a unitary matrix $U$ such that $U\,\Pi_1\, U^\dag = \diag(\gamma_1 , \gamma_{-1}) $ where $\gamma_1$ and $\gamma_{-1}$ are
the (real) eigenvalues of $\Pi_1$, and \eqref{e:mostgeneralpotential} can be written as:
\[
Q(x,t) = U^\dagger\Big[I_2 + \frac{i e^{2 i \theta(x,t,z_1)}}{z_1 }\Big(I_2 + \dfrac{i e^{2 i \theta(x,t,z_1)}}{(z_1^*-z_1)} \diag(\gamma_1 , \gamma_{-1}) \Big)^{-1}\diag(\gamma_1 , \gamma_{-1})\Big] U Q_+\,.
\]
This solution is regular for all $x,t \in \Real$, if and only if $\gamma_1 \leq 0$ and $\gamma_{-1} \leq 0$.
Further simplification yields,
\be
\label{e:simplerpolar}
Q(x,t) = U^\dagger \diag(q_{\text{dark},1}(x,t) , q_{\text{dark},-1}(x,t))\, U Q_+\,,
\ee
whenever $\det \Pi_1 \neq 0$, (i.e. $\gamma_1 < 0$ and $\gamma_{-1}< 0$) and
\[
\label{e:darksoliton}
q_{\text{dark},j}(x,t) = e^{-i \varphi}\{\cos \varphi + i \sin \varphi \tanh [\Ko \sin \varphi (x-x_{j} + 2\Ko\,t \cos\varphi)]\}
\]
with $x_j$ such that $e^{-2x_j\,\Ko\,\sin\varphi} = -2\Ko \,\sin \varphi / \gamma_j$, for $j = 1, -1$. Furthermore, if $\det \Pi_1 = 0 $ then without loss of generality one can assume $\gamma_{-1} = 0$, which yields,
\[
\label{e:simplerferro}
Q(x,t) = U^\dagger \diag(q_{\text{dark},1}(x,t) , 1)\,U Q_+\,.
\]
\iffalse
Now recall that the MNLS equation is invariant under unitary transformations.
Namely, if $Q(x,t)$ is a solution of~\eqref{e:MNLS}, then $\tilde{Q}(x,t)=U\,Q(x,t)V$
is also a solution, for arbitrary unitary matrices $U$ and $V$. These transformations
are admissible provided $\tilde{Q}(x,t)$ is symmetric whenever $Q(x,t)$ is. \fi
Exploiting the invariance of MNLS under unitary transformations, we can consider
\[\label{e:unitaryU}
\tilde{Q}(x,t)=U\, Q(x,t)\, U^\dagger \equiv \diag(q_{\text{dark},1}(x,t) , q_{\text{dark},-1}(x,t)) \,U Q_+U^\dagger\,,
\]
and if the quantization axes have been chosen so that $Q_+=I_2$ (cf. Sec.~\ref{s:canonical}), then
\[
\label{e:diagonalpolar}
\tilde{Q}(x,t)= \diag(q_{\text{dark},1}(x,t) , q_{\text{dark},-1}(x,t))\,.
\]
(The unitary transformations \eqref{e:unitaryU} are obviously
admissible, since $\tilde{Q}$ is diagonal.) The same obviously holds for \eqref{e:simplerferro}, in which case
\[
\tilde{Q}(x,t)= \diag(q_{\text{dark},1}(x,t) , 1)\,.
\label{e:diagonalferro}
\]
Specifically, to derive \eqref{e:diagonalpolar} we have to use two subsequent unitary transformations, first reducing $Q_+$ to identity and secondly, diagonalizing $\Pi_1$. Notice that when we reduce \eqref{e:mostgeneralpotential} to its canonical form (i.e., when $Q_+= I_2$), $\Pi_1$ becomes a real symmetric matrix, therefore we can find
an orthogonal matrix $W$ that diagonalizes $\Pi_1$. Thus using the discussion in section \ref{s:canonical} one can show that the $\~Q(x,t)$ defined in \eqref{e:diagonalpolar} can be written as:
\be
\label{e:twounitarytransformtion}
\~Q(x,t) = \~U Q(x,t) \~{U}^T\,,
\ee
with the unitary matrix $\~U := W \Delta^{-1/2}\,V$, where $\Delta$ and $V$ are defined in equation \eqref{e:eigenmatrixforQ+} and $W$ is the orthogonal matrix which diagonalizes $\Pi_1$.

As shown in Appendix~\ref{s:symmetries}, the transformation~\eqref{e:twounitarytransformtion}
is equivalent to a complex rotation of the quantization axes.
Thus, without loss of generality (i.e., up to admissible unitary transformations), any one-soliton solution can be reduced to a superposition
of two oppositely polarized shifted dark solitons of the scalar NLS equation.

It should be noted that, even though any one-soliton solution is unitarily equivalent to the simpler, diagonal solutions \eqref{e:diagonalpolar} or \eqref{e:diagonalferro}, when more than one soliton are present in general it is not possible to simultaneously reduce both solitons to diagonal forms via unitary transformations.
In particular, $\Pi_1$ and $\Pi_2$ can be simultaneously diagonalized if and only if they commute, which obviously is a very special case.
For this reason, it is important to discuss the form of the one-soliton also in the generic case in which $\Pi_1$ is an arbitrary Hermitian matrix, not necessarily diagonal.
And for similar reasons, i.e., in order to elucidate soliton interactions, it is also important to characterize one-soliton solutions that are not in canonical form.
This will be done in the next subsections, where we will distinguish between ferromagnetic solitons and polar solitons.
From the mathematical point of view, we will refer to a ferromagnetic soliton when the associated norming constant is such that $\det \Pi_1=0$, and to a polar soliton when $\det \Pi_1>0$ which is the full rank case. As we will explain below, the terminology corresponds to the standard one in the physical literature, where
a ferromagnetic soliton has nonzero total spin, while the total spin of a polar soliton is zero.

\iffalse
It is worth mentioning that the MNLS equation is scale invariant. Namely, if $Q(x,t)$ is a solution of \eqref{e:MNLS}
so is $\tilde{Q}(x,t)= c\,Q(cx,c^2t)$ for any constant $c\in \Real$.
Therefore, without loss of generality we can take $\Ko = 1$, and a solution with arbitrary $\Ko$ can be reconstructed from the former using the scale transformation above.
For future reference, we note that we can then reduce equation \eqref{e:mostgeneralpotential} to the following canonical form
\[
\label{e:canonicalform}
Q(x,t) = I_2 + \frac{i e^{2 i \theta(x,t,z_1)}}{z_1 }\Big(I_2 + \dfrac{i e^{2 i \theta(x,t,z_1)}}{(z_1^*-z_1)} \Pi_1 \Big)^{-1}\Pi_1,\qquad x\in \Real,\, t\in \Real^+\,,
\]
where $\Pi_1 = (c_{ij})$ is a real, symmetric $2\times2$ matrix,
and now $z_1 = e^{i \varphi}$ with $\varphi \in (0,\pi)$.
\fi

Finally, using \eqref{e:mostgeneralpotential} and the fact that $Q_+ = I_2$, one can obtain an explicit expression for the
particle number density \eqref{e:ndensity}  and the spin density \eqref{e:fdensity}  for a one-soliton solution:
\bse
\begin{align}
\displaystyle
\label{e:nbar}
\=n(x,t) &= \frac{e^{2i\theta(x,t,z_1) - \varrho}}{D^2}\Big(
- e^{2(2i\theta(x,t,z_1) + \varrho)}\det \Pi_1\,\tr \Pi_1 +4\,e^{2i\theta(x,t,z_1) + \varrho} \det \Pi_1 -\tr \Pi_1
\Big)\,,\\
\label{e:fbar}
\@f(x,t) &= \frac{e^{2i\theta(x,t,z_1) - \varrho}\,\tr (\boldsymbol{\sigma} \Pi_1)}{D^2}\Big(1-e^{2(2i\theta(x,t,z_1) + \varrho)}\det \Pi_1\Big)\,,
\end{align}
\ese
where $D = e^{2(2i\theta(x,t,z_1) + \varrho)}\det \Pi_1 - e^{2i\theta + \varrho}\tr \Pi_1 + 1$, and $e^{-\varrho} = 2 \sin \varphi$ with $\boldsymbol{\sigma}$ as in \eqref{e:densities}.

%\bb Of course, $\det %\Pi_1=\gamma_{1}\gamma_{-1}$ and $\tr %\Pi_1=\gamma_1+\gamma_{-1}$ where %$\gamma_{\pm 1}$ are the real %eigenvalues of $\Pi_1$. \eb

%one-soliton solution admit a  different form when the norming constant $C_1$ is full rank or rank 1.
%In following sections we discuss these soliton solutions in detail, separately.

One can simplify the spin densities~\eqref{e:fbar} further for polar and ferromagnetic states separately.
Specifically, for ferromagnetic solitons (i.e., $\det \Pi_1 = 0$) one has
\bse
\be
\label{e:spinferro}
\@f(x,t)= \frac{1}{4}\tr(\boldsymbol{\sigma} \Pi_1)\,e^{-(\rho+2\alpha_0\,\sin \varphi)} \sech^2[\sin\varphi (x-\alpha_0 + 2t \cos \varphi)]\,,
\ee
where $\alpha_0 = (\rho + \ln(-\tr \Pi_1))/2 \sin \varphi$ and we will show in later sections that $\tr \Pi_1 < 0$. This shows that  the spin density is an even function of $x$,
and therefore the total spin of a ferromagnetic soliton in canonical form is nonzero.

On the other hand, the spin density of a polar soliton in canonical form can be written as
\be
\displaystyle
\label{e:spinpolar}
\@f(x,t)= -2\,\tr(\boldsymbol{\sigma} \Pi_1)\,e^{-(\rho+2\beta_0\,\sin \varphi)} \dfrac{\sinh[2\sin\varphi (x-\beta_0 + 2t \cos \varphi)]}{\{2\cosh[2\sin\varphi (x-\beta_0 + 2t \cos \varphi)]- \tr(\Pi_1)e^{\rho-2\beta_0\,\sin\varphi}\}^2}\,,
\ee
with $\beta_0 = (2\rho + \ln(\det \Pi_1))/4 \sin \varphi$, and one can see that all three components of the spin density are odd functions of $x$. As a consequence,
the total spin of a polar soliton in canonical form is always zero.
\ese
%Using the symmetries of \eqref{e:spinferro}, \eqref{e:spinpolar} and, the total spin one can distinguish polar and ferromagnetic solitons, specifically, if the total spin is zero we have a polar soliton and ferromagnetic soliton corresponds to non-zero total spin.
Finally, we note that in light of \eqref{e:fbarunitrayrelation} and \eqref{e:rotationmatrix} the transformation to canonical form, while changing the total spin, does not change
the nature of the soliton as being polar or ferromagnetic.

\begin{figure}[t!]
\centerline{
  \includegraphics[width=0.275\textwidth]{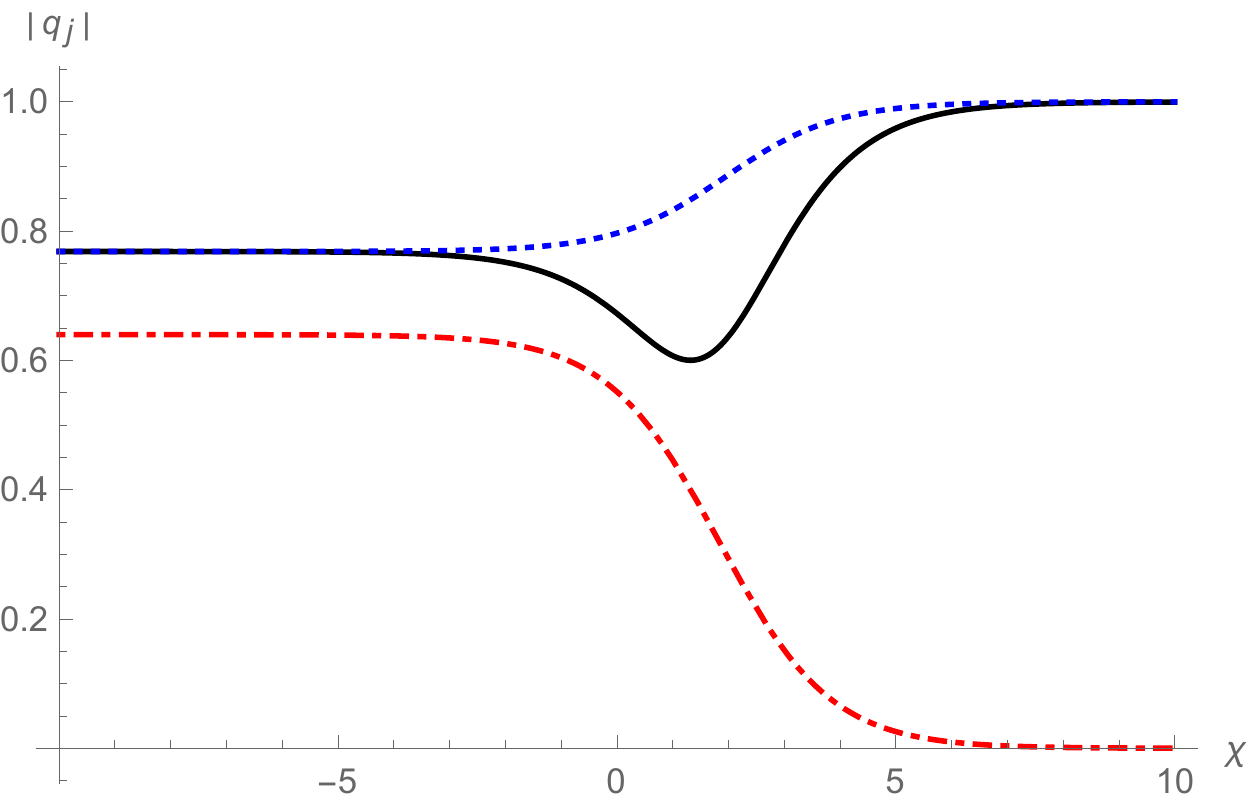}
  \includegraphics[width=0.275\textwidth]{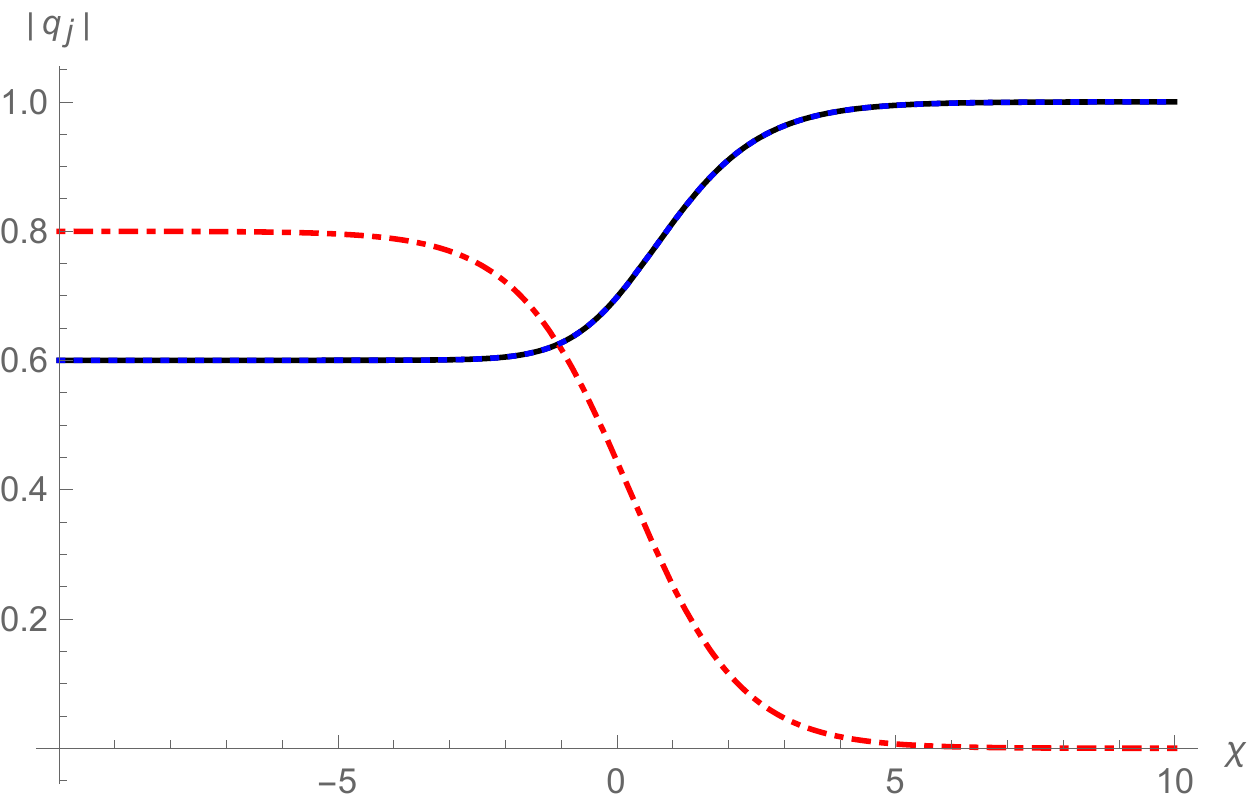}
  \includegraphics[width=0.275\textwidth]{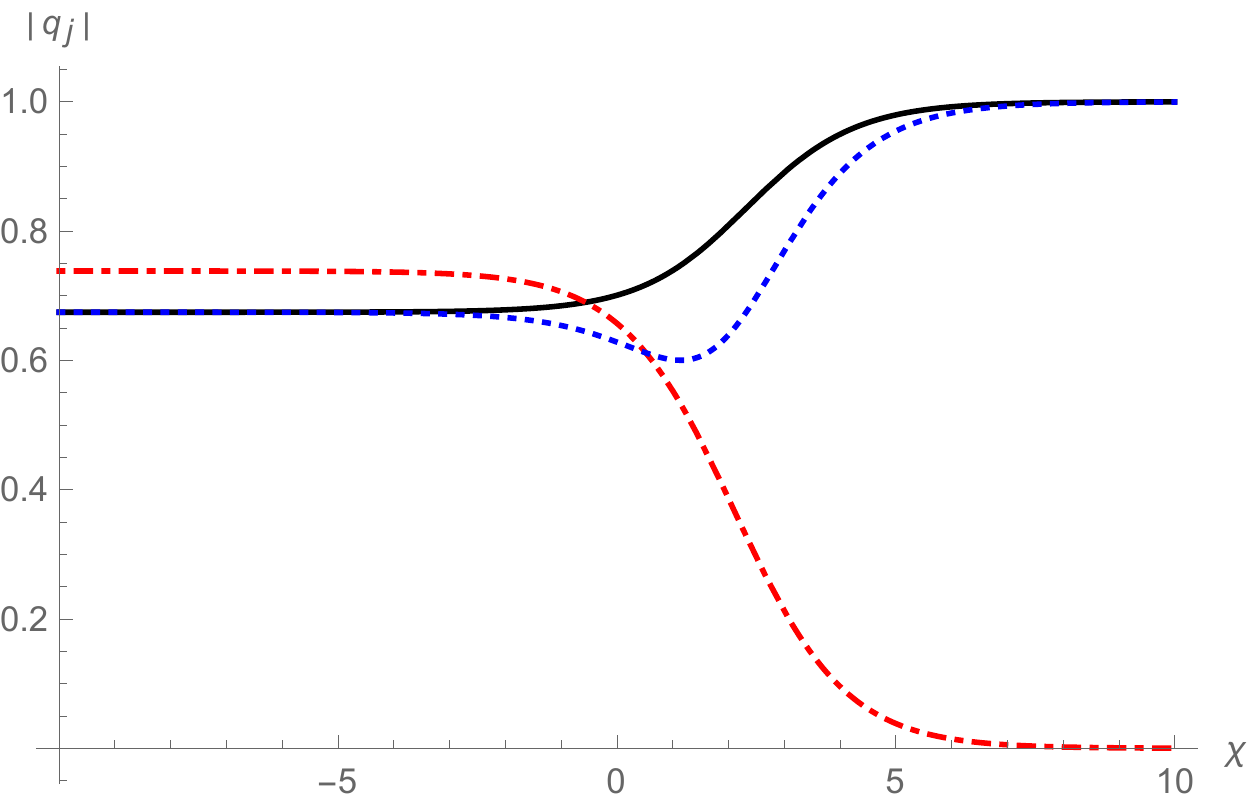}
}
\medskip
\caption{One-soliton solution profiles for ferromagnetic states in canonical form
generated by a discrete eigenvalue $z_1 = e^{0.927i}$.
From left to right:
$\rho = 4$,
~$\rho = 1$
and $\rho = 4/9$.
%where $\chi = -2i\theta(x,t,z_1),\,\varphi = 0.927$.
In each plot, the three components
$|Q_{11}|$ (black solid line),
$|Q_{12}|$ (red dot-dashed line) and
$|Q_{22}|$ (blue dotted line) are shown.}
\label{f:canonicalferro}
\bigskip\bigskip
\centerline{
\includegraphics[width=0.2875\textwidth]{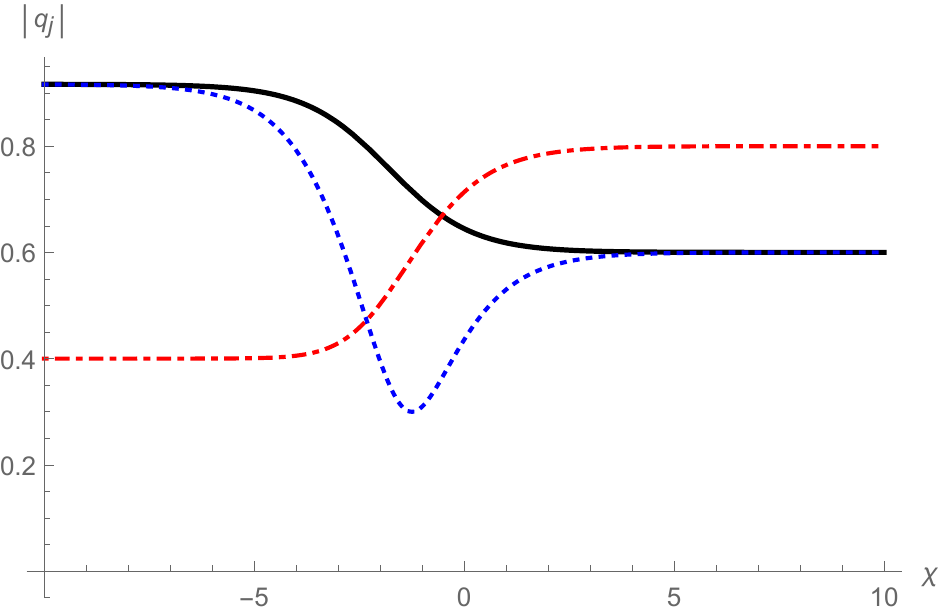}
\includegraphics[width=0.2875\textwidth]{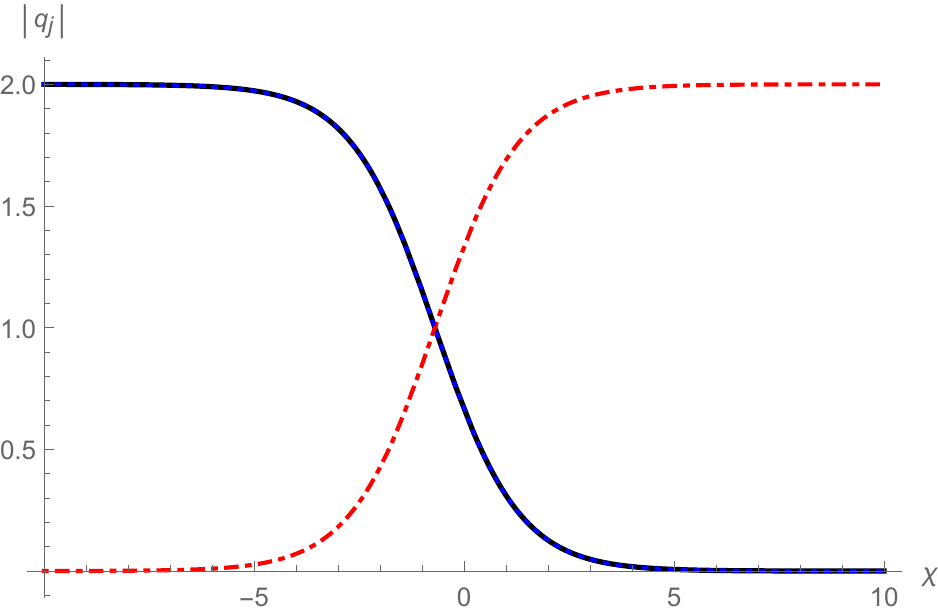}
\includegraphics[width=0.2875\textwidth]{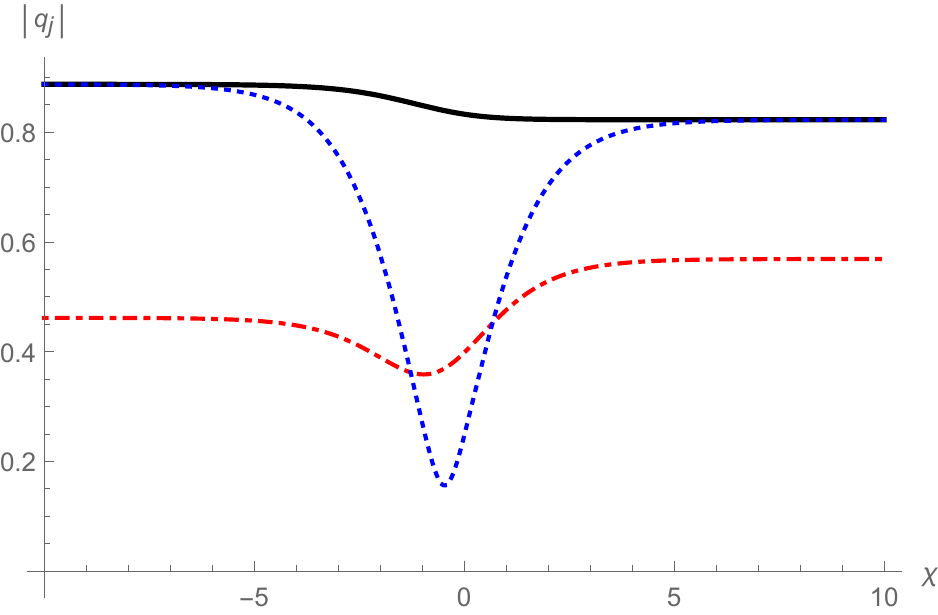}
}
\medskip
\caption{One-soliton solution profiles for ferromagnetic states in non-canonical form,
corresponding respectively to each of the three pair of asymptotic matrices $Q_+$ and norming constants $C_1$ in \eqref{e:Q+forfig2},
as well as, respectively,
$z_1 = e^{i\pi/3}$ (left),
$z_1 = 2i$ (center) and
$z_1 = i$ (right).
As in Fig.~\ref{f:canonicalferro}, in each case
%For each case, the components shown are
the black solid line,
red dot-dashed line and
blue dotted line
correspond respectively to
$|Q_{11}|$,
$|Q_{12}|$
and
$|Q_{22}|$.}
\label{f:non-canonicalferro}
%\end{figure}
\bigskip\bigskip
%\begin{figure}[h!]
\centerline{
  \includegraphics[width=0.2875\textwidth]{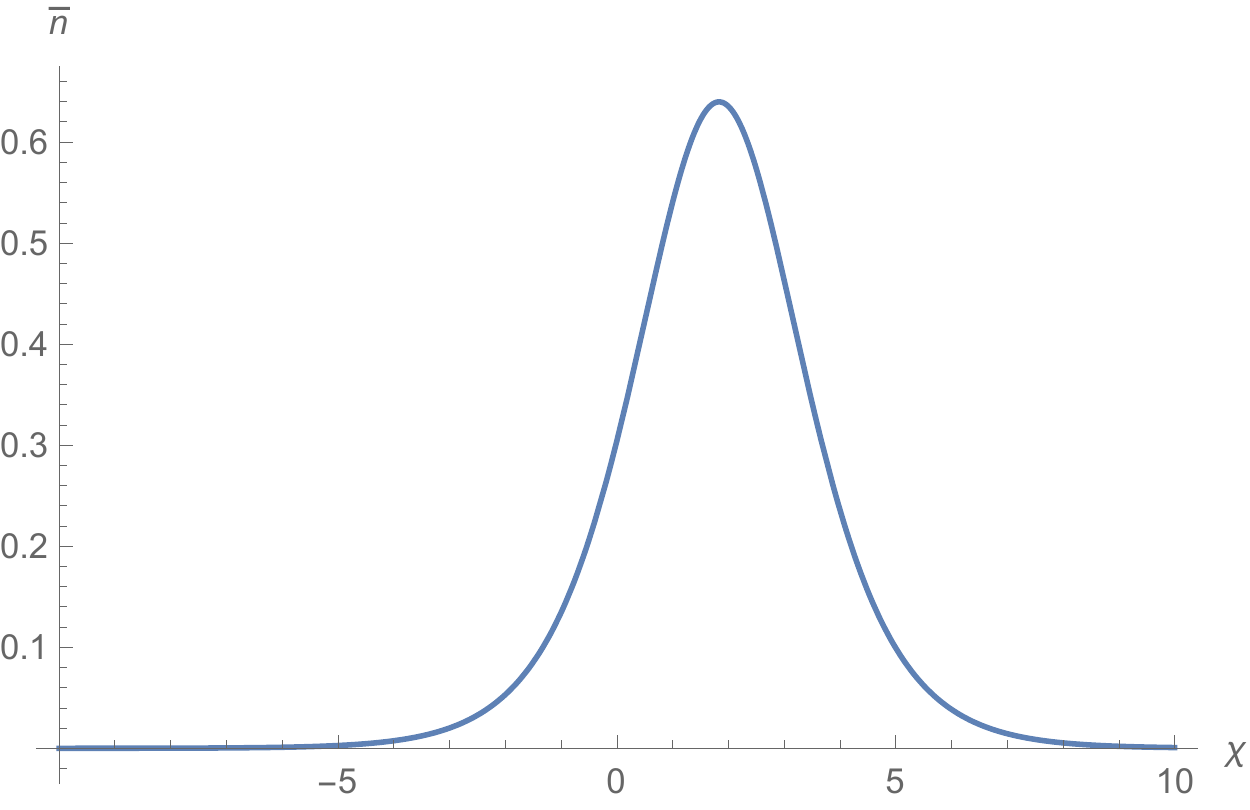}
  \includegraphics[width=0.2875\textwidth]{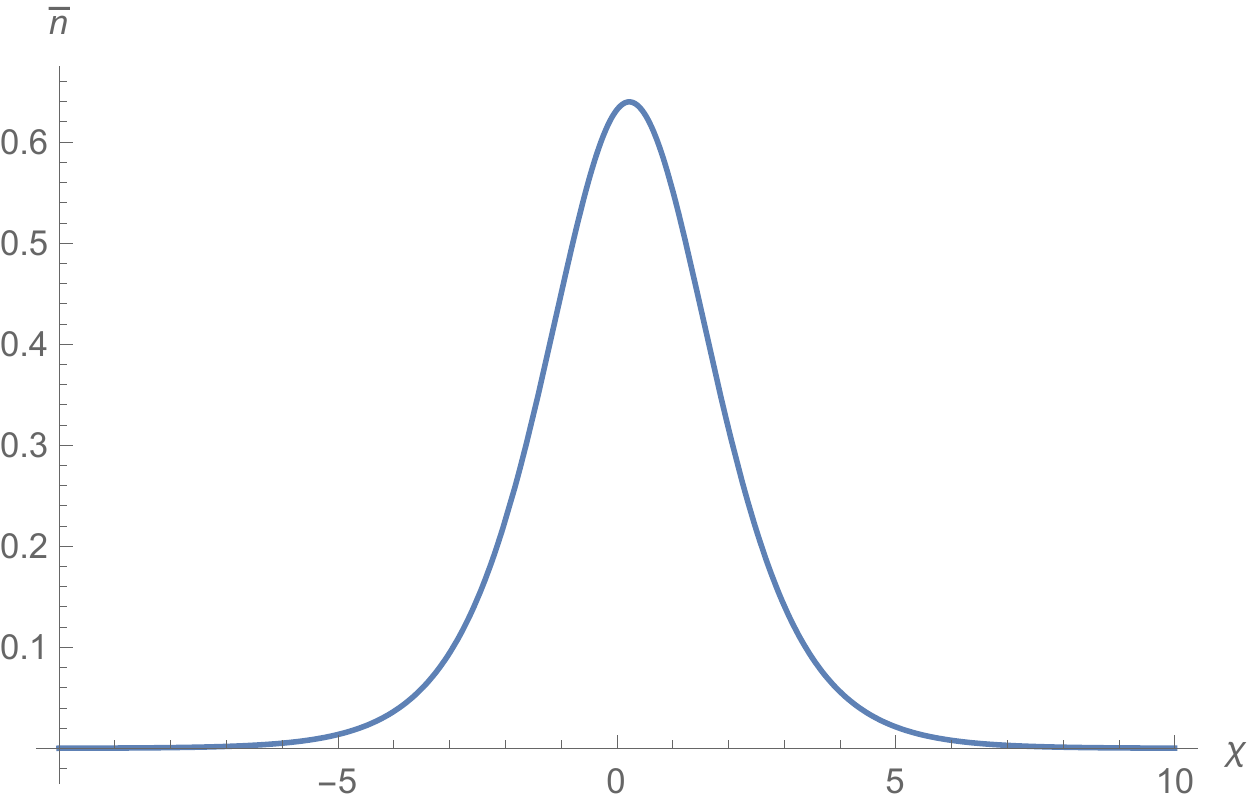}
  \includegraphics[width=0.2875\textwidth]{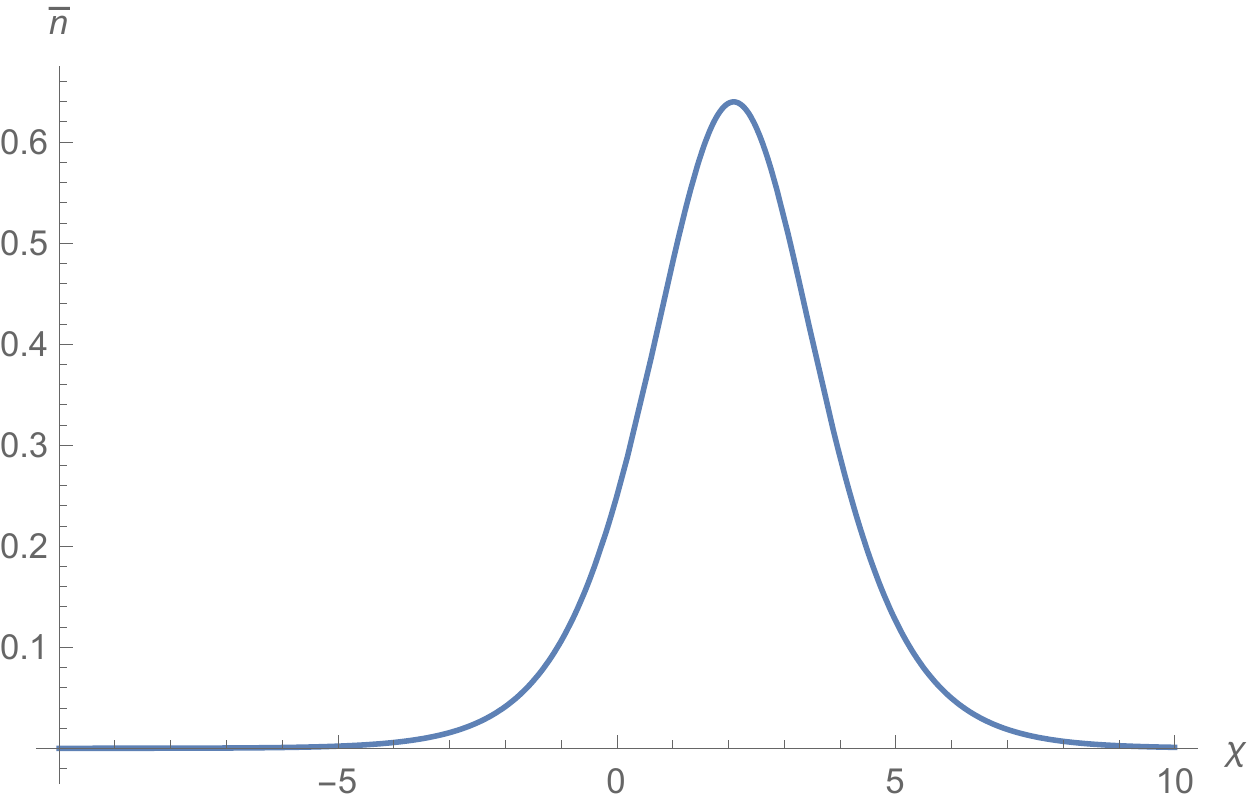}
}
\centerline{
  \includegraphics[width=0.2875\textwidth]{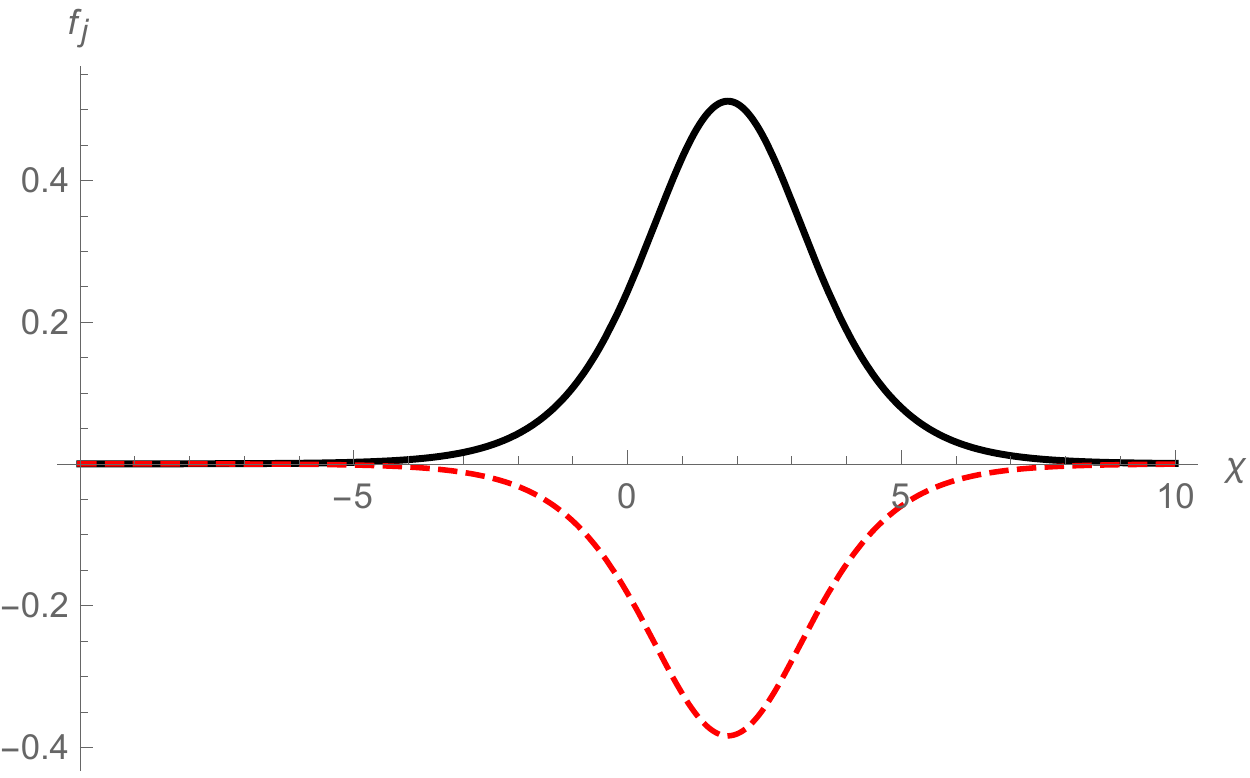}
  \includegraphics[width=0.2875\textwidth]{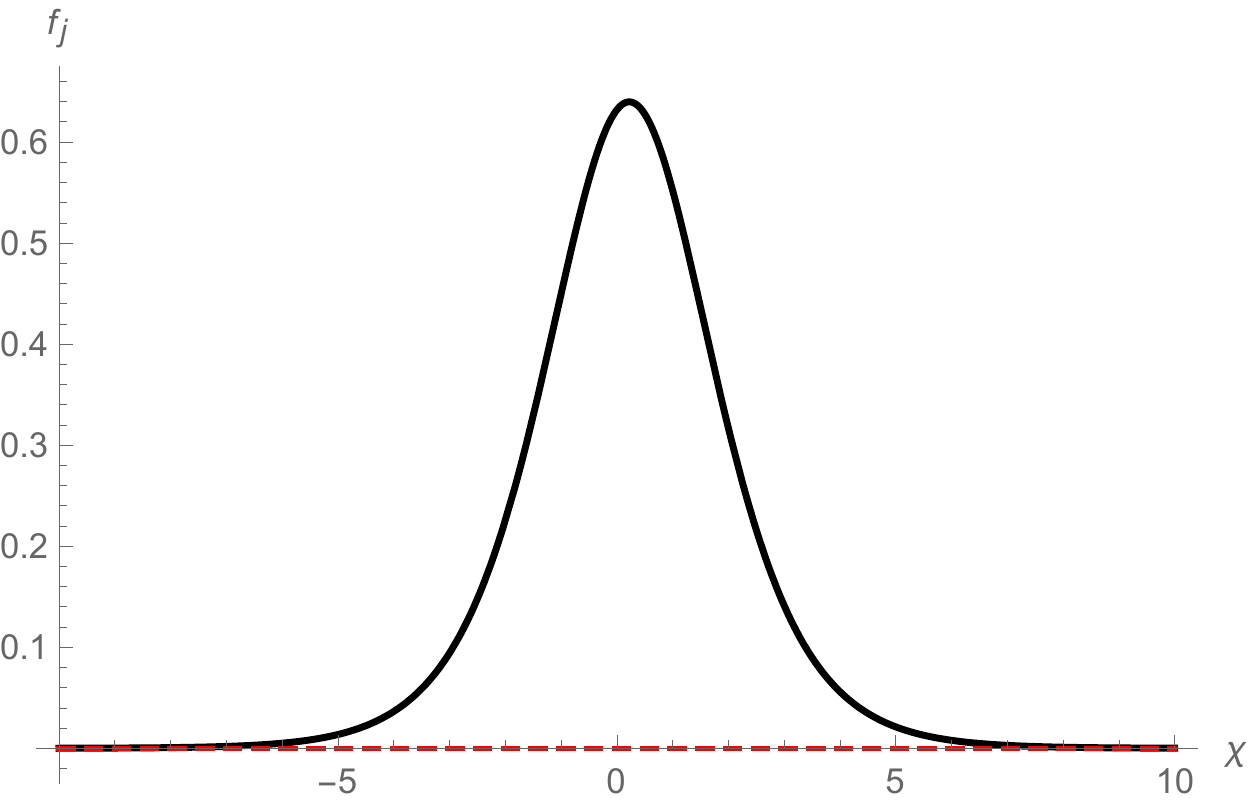}
  \includegraphics[width=0.2875\textwidth]{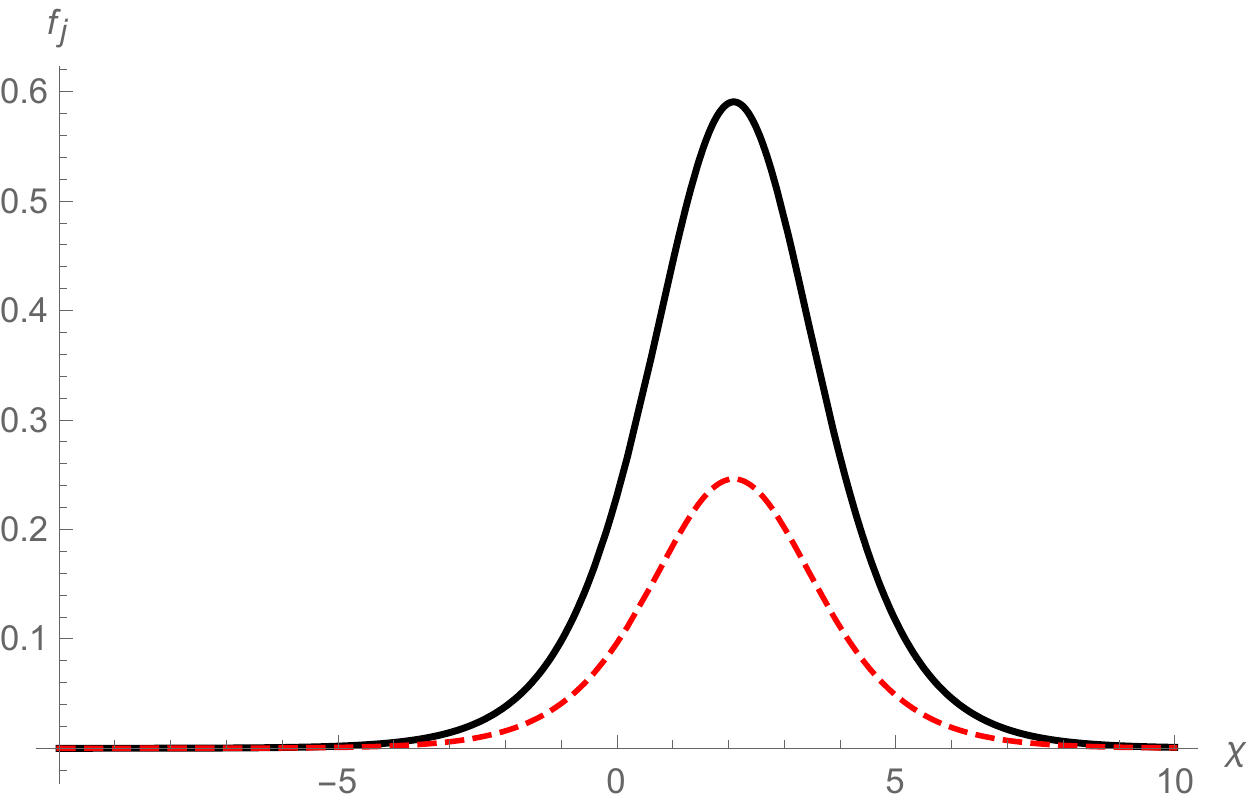}
}
\medskip
\caption{Particle density $\bar{n}$  (top row) and spin densities (bottom row)
corresponding to a ferromagnetic one-soliton solution generated by a discrete eigenvalue $z_1 = e^{0.927i}$.
Left column: $\rho = 4$, middle column: $\rho = 1$, right column: $\rho = 4/9$.
In the bottom row, the black solid lines and the red dashed lines show respectively the
spin density components $f_1$ and $f_{-1}$ (dashed line).
(Note that $f_{0}$ is zero everywhere.)
}
\label{f:densityferro}
\kern-2\bigskipamount
\end{figure}

%%%%%%%%%%%%%%%%%%%%%%%%%%%%%%%%%%%%%%%%%%%%%%%%%%%%%%%%%%%%%%%%%%%%%%%%%%%%%%%%%%%%%%%%%%%%%
\subsection{Ferromagnetic states}

We start by considering the case of a one-soliton solution for which $\Pi_1$ (i.e., its associated norming constant $C_1$)
is rank-1, which corresponds to a ferromagnetic state.
Since in this case  $\det \Pi_1 = 0$, the solution \eqref{e:mostgeneralpotential} simplifies to:
\[
\label{e:generalferropotential}
Q(x,t) = Q_+ + \dfrac{2 i \sin{\varphi}\,e^{-i \varphi}}{2\Kosqr  \sin{\varphi}\,e^{-2i \theta(x,t,z_1)}-\tr( \Pi_1)} \Pi_1 Q_+\,,
\]
where $z_1 = \Ko\,e^{i \varphi}$, with $\varphi \in (0,\pi)$.
Taking the limit $x \to -\infty$ we have
\[
Q_- = \mathcal{V} Q_+,
\]
where $\mathcal{V}  = I_2 - \dfrac{2 i \sin{\varphi}\,e^{-i \varphi}}{\tr( \Pi_1)} \Pi_1 $ and $\mathcal{V}$ is a unitary matrix.
The above equation shows that $Q_-\ne Q_+$ and that in fact, generically, the energy distribution among the components
as $x\to\infty$ and $x\to-\infty$ is different.
As we will see, this is different from what happens in polar states (i.e., when $\det \Pi_1\ne 0$).
When $Q_+ = I_2$, \eqref{e:generalferropotential} yields the canonical form for the ferromagnetic state as
\[
\label{e:ferropotential}
Q(x,t) = I_2 + \dfrac{2 i \sin{\varphi}\,e^{-i \varphi}}{2 \sin{\varphi}\,e^{-2i \theta(x,t,z_1)}-\tr\,\Pi_1} \,\Pi_1\,,
\]
where $\Pi_1 = (c_{ij})$ is now real, symmetric and with zero determinant, and is therefore completely determined by its diagonal entries.
As we said, $\Pi_1$ can always be reduced to a diagonal form (in this case, $\Pi_1=\mathrm{diag}(\gamma_1,0)$ since $\det \Pi_1=0$) via rotations of the quantization axes.
And the above solution in this case in simply given by \eqref{e:diagonalferro}.
However, as we also clarified before, it is important to understand the properties of the solutions when $\Pi_1$ is not diagonal. Therefore, we will consider below
a general (i.e., non-diagonal) rank-1 matrix $\Pi_1$.
In order for \eqref{e:ferropotential} to be regular for all $x,t\in \mathbb{R}$, it is necessary and sufficient that $\tr\,\Pi_1 < 0$. Since $\det \Pi_1 = c_{11}c_{22} - c_{12}^2 = 0$, one can show that $c_{11} <0$ and $c_{22}< 0 $.
In what follows, it will be convenient to express $\Pi_1$ in terms of the ratio of its diagonal entries, $\rho = c_{11}/c_{22}$, and the quantity
$x_o = \ln[-c_{22}/(2\sin\varphi)]/(2\sin\varphi)$
(which amounts to expressing $c_{22}$ as $c_{22} = - 2e^{2x_o\sin\varphi}\sin\varphi$).
That is,
\be
\Pi_1 = - 2e^{2x_o\sin\varphi}\sin\varphi\, \begin{pmatrix} \rho & -\sqrt\rho \\ - \sqrt\rho & 1 \end{pmatrix}\,.
\ee
The reason why the above parametrization is convenient is that
the shape of the solution is controlled only by~$\rho$, whereas
$x_o$ corresponds to an overall translation of the solution.
\iffalse
Indeed, the above parametrization yields
\be
Q(x,t) = I - 2i\sin\varphi \frac{1}{1 + (1+\rho)\,\e^{2\chi}}
  \begin{pmatrix} \rho & - \sqrt\rho \\ - \sqrt\rho & 1 \end{pmatrix},
\label{e:canonical2}
\ee
where $\chi(x,t) = - i\theta(x,t,z_1) + x_o\sin\varphi$.
Explicitly,
\be
\chi(x,t) = \sin\varphi\,(x - x_o + 2t\cos\varphi)\,.
\ee
\fi
Indeed, the above parametrization yields
\be
Q(x,t) = I - 2i\sin\varphi \frac{e^{-i\varphi}}{\e^{2 \chi} + (1+\rho)}
  \begin{pmatrix} \rho & - \sqrt\rho \\ - \sqrt\rho & 1 \end{pmatrix},
\label{e:canonical2}
\ee
where $\chi(x,t) = - i\theta(x,t,z_1) - x_o\sin\varphi$.
Explicitly,
\be
\label{e:chifor1-soliton}
\chi(x,t) = \sin\varphi\,(x - x_o + 2t\cos\varphi)\,.
\ee
The canonical form~\eqref{e:canonical2} of the ferromagnetic one-soliton solutions allows one to characterize their physical properties,
as we show next.
Specifically, one can show that $|Q_{12}(x,t)|$ does not admit minima or maxima, while
exactly one between $|Q_{11}(x,t)|$ and $|Q_{22}(x,t)|$ has a minimum for any choice of $\rho\ne 1$.
%when $c_{11}, c_{22} \in \Real \setminus \{0\}$ and $c_{11}\neq c_{22}$ with $\tr \Pi < 0$.
(Note, however, that these properties do not extend to solitons in non-canonical form.)
More precisely, $|Q_{11}(x,t)|$ has a minimum when $\rho>1$, while $|Q_{22}(x,t)|$ has a minimum when $\rho<1$.
Moreover, the minimum in either case is located on the line
$\chi(x,t) = \frac{1}{2}\ln|1-\rho|$, i.e.:
\[
\label{e:ferrominlocation}
(x - x_o)\,\sin\varphi + t\,\sin(2\varphi) = \frac12\ln{|1-\rho|}\,,
%x + 2 \cos \varphi\,t = \frac{1}{2 \sin\varphi} \big( \ln|c_{22}-c_{11}| - \ln2 \sin \varphi) \big)\,.
%\begin{cases}
%\frac{1}{2 \sin\varphi}\ln(\frac{c_{22}-c_{11}}{2 \sin \varphi}), \qquad c_{22}>c_{11} \\
%\frac{1}{2 \sin\varphi}\ln(\frac{c_{11}-c_{22}}{2 \sin \varphi}), \qquad c_{22}<c_{11}
%\end{cases}
\]
Note that Eq.~\eqref{e:chifor1-soliton} gives $v= -2 \cos \varphi = -2 \Re(z_1)$ for the soliton velocity.
%, and the volocity is maximum when $\varphi = \pi$ and minimum when $\varphi = 0$.

Finally, the depth of the minimum in both cases is given by:
\begin{equation}
1 - |Q_{jj,\mathrm{min}}(x,t)|  = 1 - |\cos\varphi|\,,\quad j = 1,2\,.
\end{equation}
Importantly, note that the depth of the minimum is independent of the norming constant $C_1$.
On the other hand, the ``soliton center'', i.e., the location of the minimum (as well as and the information about the component of $Q(x,t)$ which exhibits the minimum, if it exists), depends on the diagonal entries of the norming constant.

Figure~\ref{f:canonicalferro} shows the profile of ferromagnetic one-soliton solutions in canonical form for which: only $|Q_{11}(x,t)|$ has a minimum ($\rho> 1$); only $|Q_{22}(x,t)|$ has a minimum ($\rho < 1$); none of the components of $Q(x,t)$ has a minimum ($\rho =1$).
For comparison purposes, Fig.~\ref{f:non-canonicalferro} shows examples of different max/min patterns in ferromagnetic one-soliton solutions in non-canonical form,
corresponding respectively to the following pair of asymptotic
matrices and norming constants:
We want to point out that solutions in non-canonical form(i.e., $Q_+
\neq I_2$) have different max/min patterns. Just to provide some
representative
examples of the phenomenology that can arise,
we use the following choices of boundary conditions and norming constants:
\begin{align}
\nonumber
Q_+ &= \frac{1}{5}\begin{pmatrix}1+4 e^{i\pi/3}& -2+2 e^{i\pi/3}\\-2+ 2e^{i\pi/3}& 4+e^{i\pi/3} \end{pmatrix}\,,
\quad
C_1 = \begin{pmatrix}-\sqrt{3}-3i& 2\sqrt{3}+6i\\2\sqrt{3}+6i& -4\sqrt{3}-12i\end{pmatrix}\,,\\
\nonumber
Q_+ &= 2 e^{i \pi/4} \begin{pmatrix}0&1\\1&0\end{pmatrix}\,,
\quad
C_1 = \begin{pmatrix}-1&-e^{i \pi/4}\\-e^{i \pi/4}&-i\end{pmatrix}\,,\\
Q_+ &= \frac{1}{13}\begin{pmatrix}9+4 e^{i\pi/3}& 6-6 e^{i\pi/3}\\6-6 e^{i\pi/3}& 4+9e^{i\pi/3} \end{pmatrix}\,,
\quad
C_1 = \frac{4}{507} \begin{pmatrix}-80\sqrt{3}-11i & 2(86\sqrt{3} + 57i) \\2(86\sqrt{3} + 57i) & -4(28\sqrt{3}+37i) \end{pmatrix}\,,
%\quad
%C_1 = \frac{4}{169} \begin{pmatrix}-47.905+ 58.51 i & 171.742 - %27.623 i \\171.742 - 27.623 i & -337.795-214.51 i %\end{pmatrix}\,.
%C_1 &= \frac{4}{169} \begin{pmatrix}(1071-646\sqrt{3})+ (204-84\sqrt{3}) i & %(669\sqrt{3}-987) + (33-35\sqrt{3}) i \\(669\sqrt{3}-987) + (33-35\sqrt{3}) i & %4(210-170\sqrt{3})+4(21\sqrt{3}-90) i \end{pmatrix}\,.
\label{e:Q+forfig2}
\end{align}
%the profiles of each component are more complicated in general, and more complicated combinations of maxima and minima can arise.
However, one can always reduce the solution to canonical form and characterize the solution in that simpler form, as discussed above.
Note that in all the above cases $Q_+ C_1$ is not a symmetric matrix.

{A key observation is that these ferromagnetic solitary waves arise in the form
  of domain walls between the $\pm 1$ components and the
  $0$-component. These domain
  walls ``harbor'' a structure reminiscent of a dark soliton in one (Fig.~\ref{f:canonicalferro})
  or more (Fig. ~\ref{f:non-canonicalferro}) of the components.}
For completeness, in Fig.~\ref{f:densityferro} we show
the particle number {(this involves the physical particle density
subtracted from that of the background per Eq.~(\ref{e:densities}))} and spin densities for a ferromagnetic one-soliton solution,
illustrating that all of them have a single-hump shape.

%%%%%%%%%%%%%%%%%%%%%%%%%%%%%%%%%%%%%%%%%%%%%%%%%%%%%%%%%%%%%%%%%%%%%%%%%%%%%%%%%%%%%%%%%%%%%
\subsection{Polar states}

We now consider one-soliton solutions with $\det \Pi_1\ne 0$, i.e.,  whose associated norming constant $C_1$ is full rank,
which gives rise to a polar state.
In this case, simplifying \eqref{e:mostgeneralpotential} we obtain:
\[
\label{e:generalpolarpotential}
Q(x,t) =
\dfrac{(e^{2(2i\theta(x,t,z_1) + \varrho- i\varphi)}\,\det (\Pi_1) - e^{2i\theta(x,t,z_1) + \varrho}\,\tr(\Pi_1) + 1) \Kosqr  Q_+ + i e^{i(2\theta(x,t,z_1) - \varphi)} \Pi_1 Q_+}{\Kosqr (e^{2(2i\theta(x,t,z_1) + \varrho)}\,\det (\Pi_1) - e^{2i\theta(x,t,z_1) + \varrho}\,\tr(\Pi_1)+1)}
\]
where $e^{-\varrho} = 2 \Kosqr  \sin \varphi$, $z_1 = \Ko\,e^{i \varphi}$ with $\varphi \in (0,\pi)$. When $x \to -\infty$ we have
\[
Q_- = e^{-2i \varphi} Q_+\,,
\]
showing that for polar solitons the asymptotic states $Q_+$ and $Q_-$ always coincide up to an overall phase factor, determined by
the phase of the discrete eigenvalue.
Choosing $Q_+ = I_2$ and $\Ko = 1$, \eqref{e:generalpolarpotential}
gives the canonical form of the polar state as
\[
\label{e:polarpotential}
Q(x,t) =
\dfrac{(e^{2(2i\theta(x,t,z_1) + \varrho- i\varphi)}\,\det \Pi_1 - e^{2i\theta(x,t,z_1) + \varrho}\,\tr\Pi_1+1) I_2 + i e^{i(2\theta(x,t,z_1) - \varphi)}\Pi_1}{e^{2(2i\theta(x,t,z_1) + \varrho)}\,\det \Pi_1 - e^{2i\theta(x,t,z_1) + \varrho}\,\tr\Pi_1+1}\,.
\]
Note that this solution is regular for any real, symmetric matrix $\Pi_1 = (c_{ij})$ with $\tr \Pi_1 <0$ and $\det \Pi_1 >0$. For future convenience, let us express $\Pi_1$ in terms of $\rho_1 = c_{11}/c_{22}$, $\rho_2 = c_{12}/c_{22}$ and the quantity
$x_o = \ln[-c_{22} \sin \varphi/ (c_{11}c_{22} - c_{12}^2)]/(2 \sin \varphi)$ as
\be
\Pi_1 = \frac{-e^{2x_o\sin\varphi} (c_{11}c_{22} - c_{12}^2)}{\sin\varphi}\, \begin{pmatrix} \rho_1 & \rho_2 \\ \rho_2 & 1 \end{pmatrix}\,.
\ee

\begin{figure}[b!]
\centerline{
  \includegraphics[width=0.2875\textwidth]{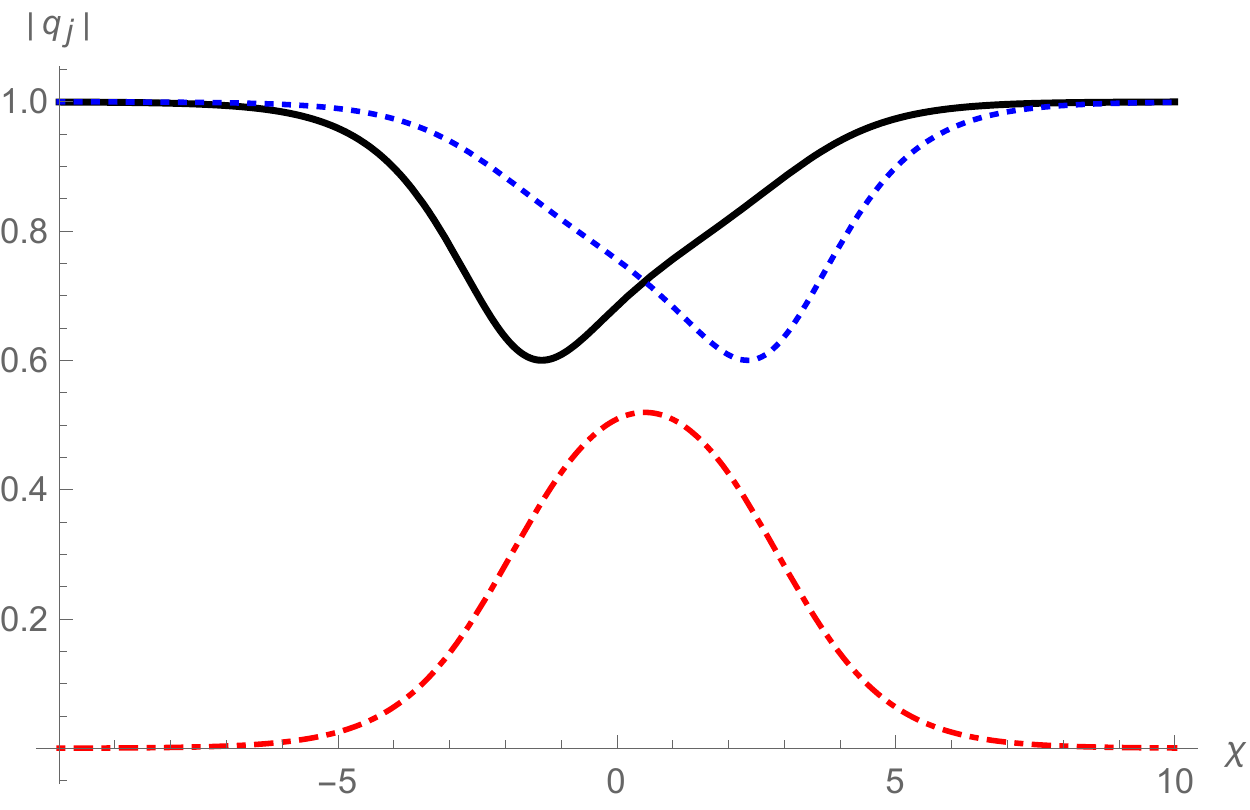}
  \includegraphics[width=0.2875\textwidth]{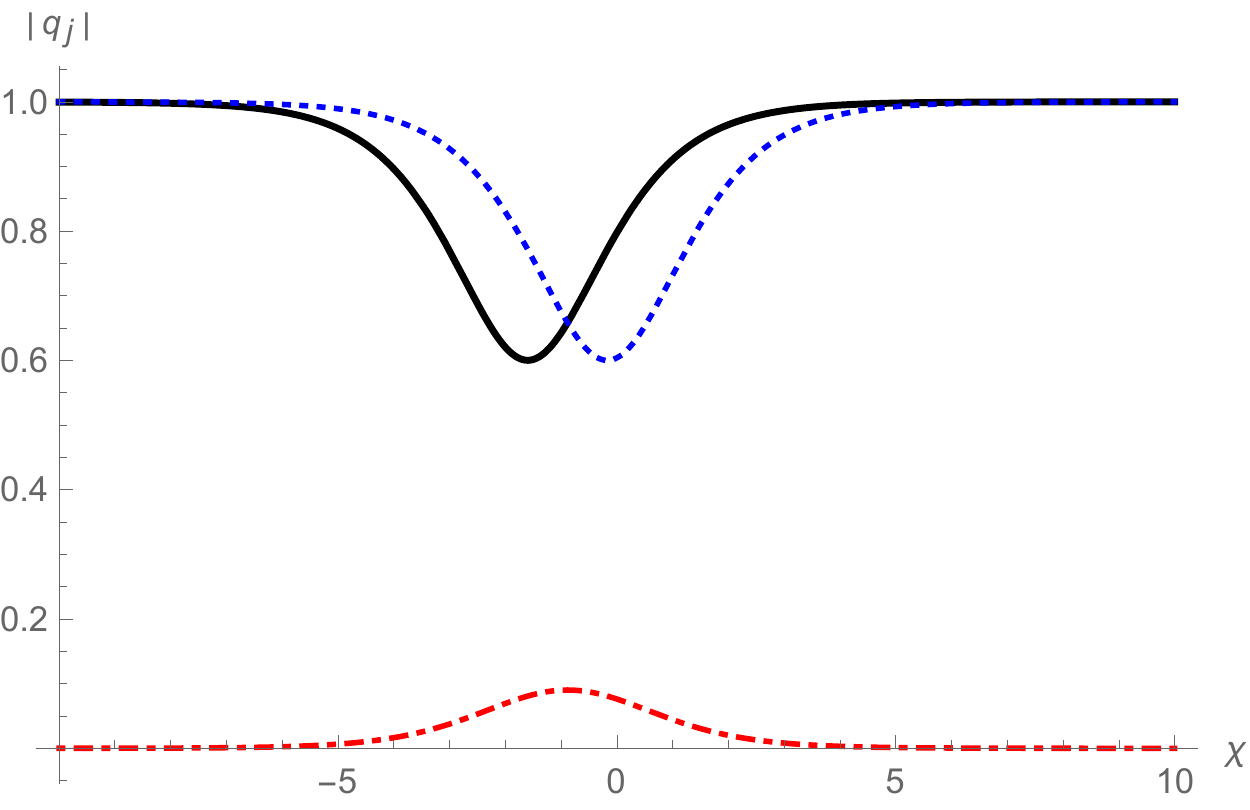}
  \includegraphics[width=0.2875\textwidth]{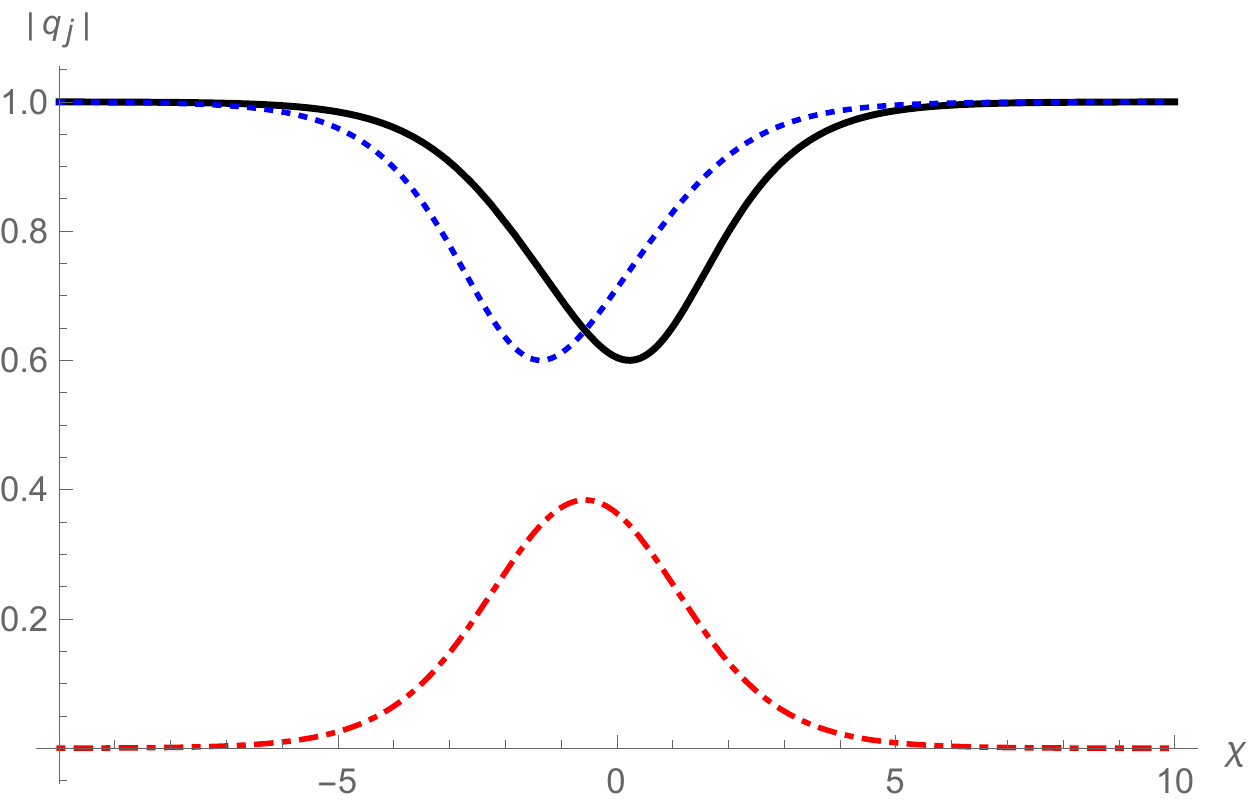}
}
\caption{One-soliton solution profiles for polar states in canonical form
with $\varphi = 0.927$.
~$\rho_1 = 4$, $\rho_2 = -1.94$ (left);
~$\rho_1 = 4$, $\rho_2 = -1/2$ (middle);
~$\rho_1 = 3/8$, $\rho_2 = -1/2$ (right).
%where $\chi = -2i\theta(x,t,z_1),\,\varphi = 0.927$.
For each case, the components shown are
$|Q_{11}|$ (black solid line),
$|Q_{12}|$ (red dot-dashed line),
$|Q_{22}|$ (blue dotted line).}
\label{f:canonicalpolar}
\bigskip\bigskip
\centerline{
\includegraphics[width=0.2875\textwidth]{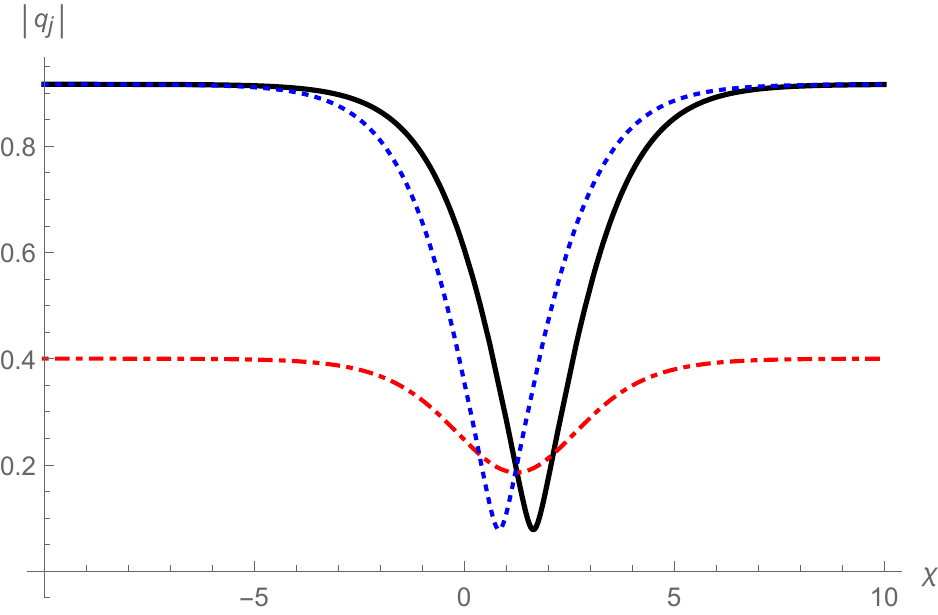}
\includegraphics[width=0.2875\textwidth]{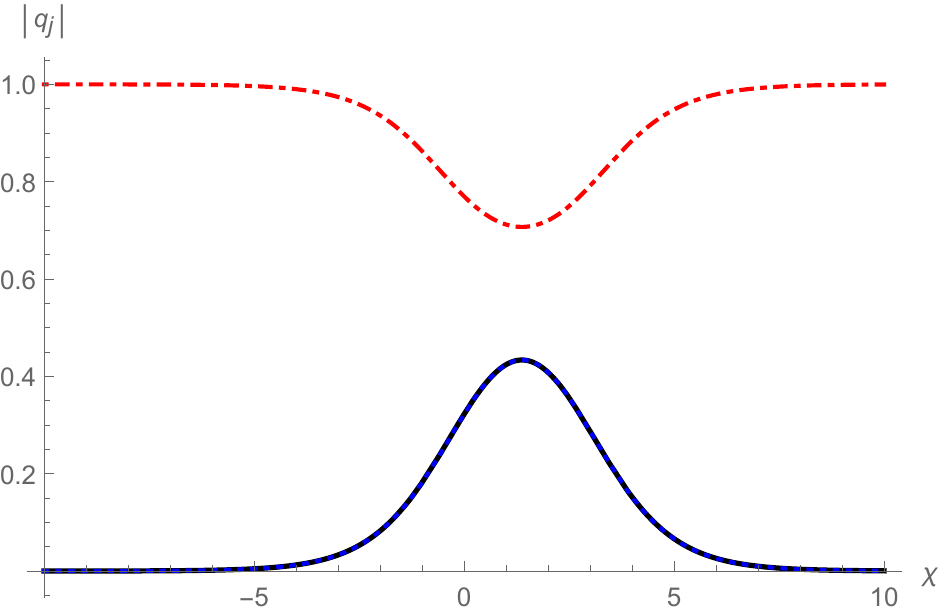}
\includegraphics[width=0.2875\textwidth]{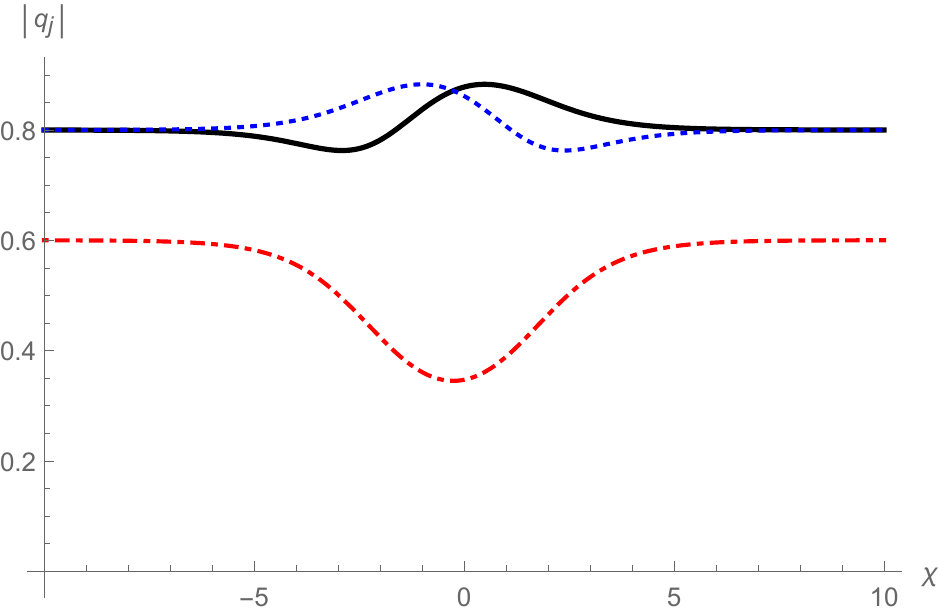}
}
\caption{One-soliton solution profiles for polar states in non-canonical form,
corresponding respectively to each of the three pair of asymptotic matrices $Q_+$ and norming constants $C_1$ in \eqref{e:Q+forfig5},
as well as, respectively,
$z_1 = i$ (left),
$z_1 = e^{i\pi/4}$ (center) and
$z_1 = e^{i\pi/6}$ (right).
For each case,
the black solid line,
red dot-dashed line and
blue dotted line
correspond respectively to
$|Q_{11}|$,
$|Q_{12}|$
and
$|Q_{22}|$.
}
\label{f:non-canonicalpolar}
\bigskip\bigskip
\centerline{
  \includegraphics[width=0.2875\textwidth]{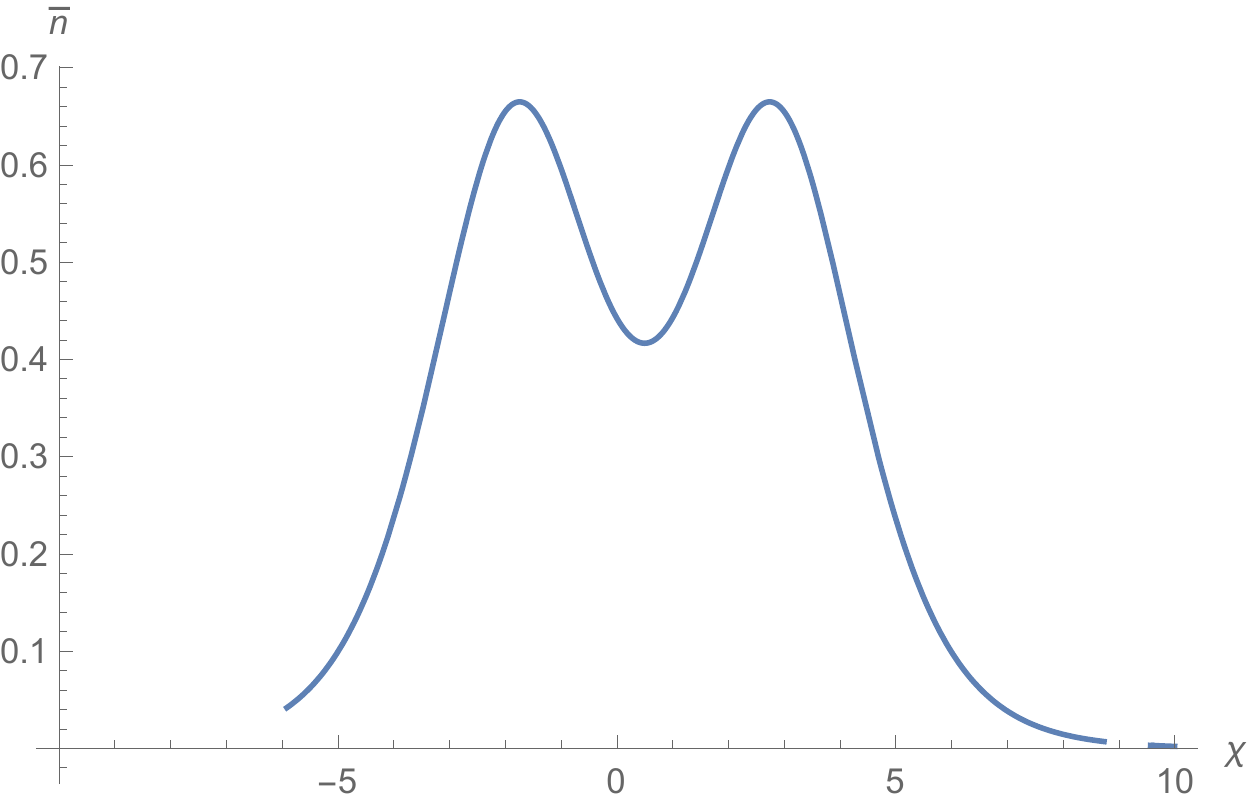}
  \includegraphics[width=0.2875\textwidth]{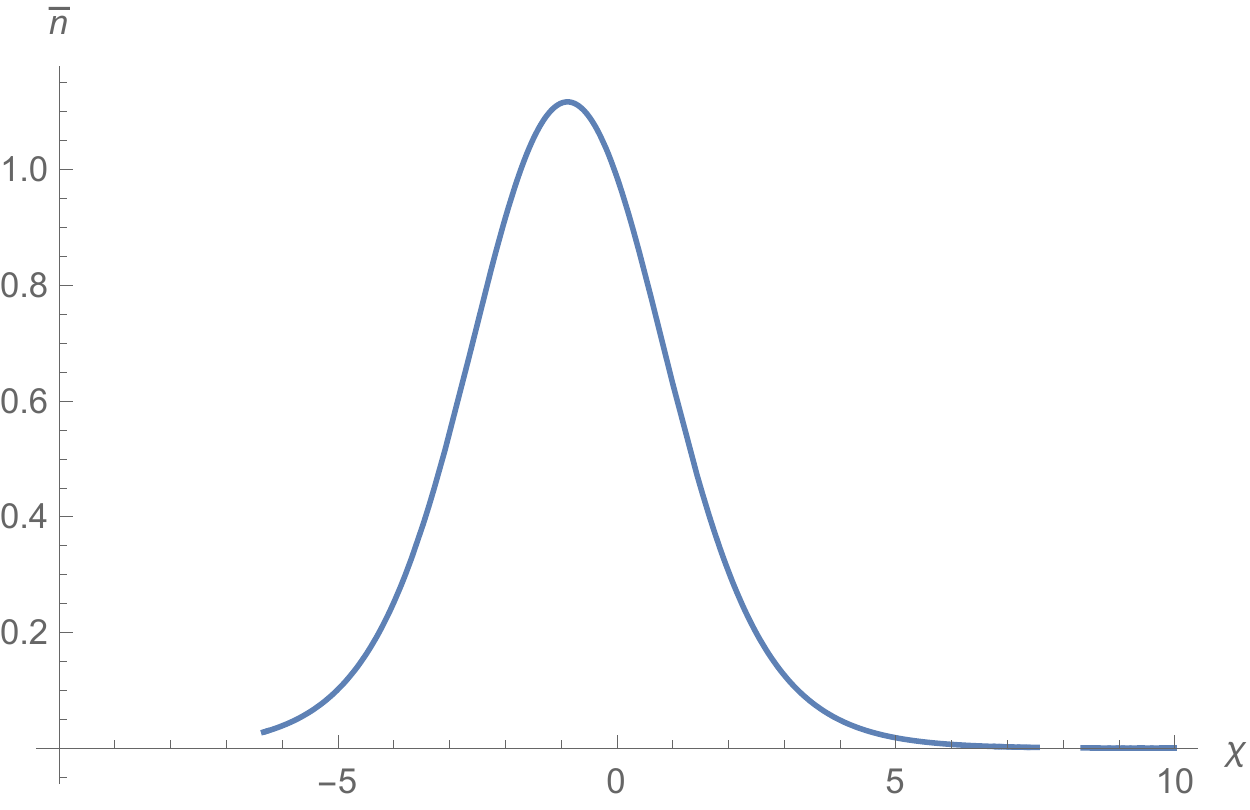}
  \includegraphics[width=0.2875\textwidth]{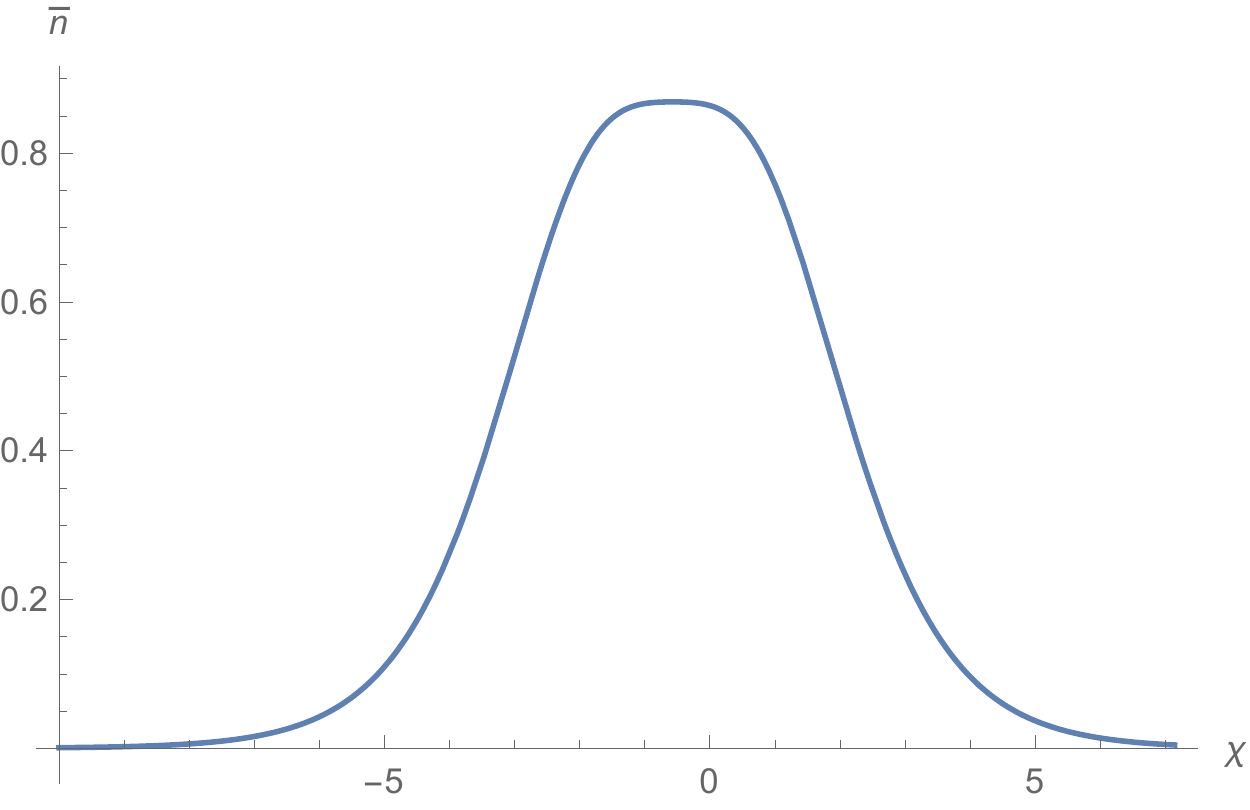}
}
\medskip
\centerline{
  \includegraphics[width=0.2875\textwidth]{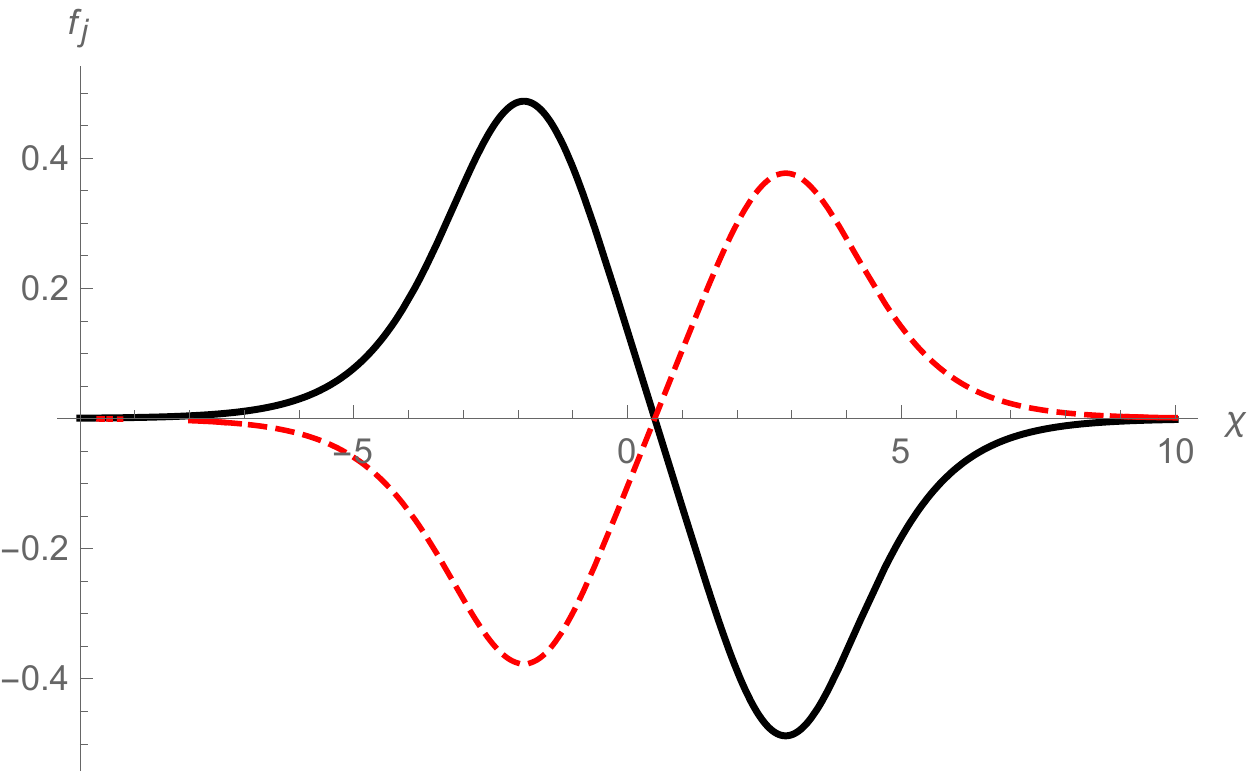}
  \includegraphics[width=0.2875\textwidth]{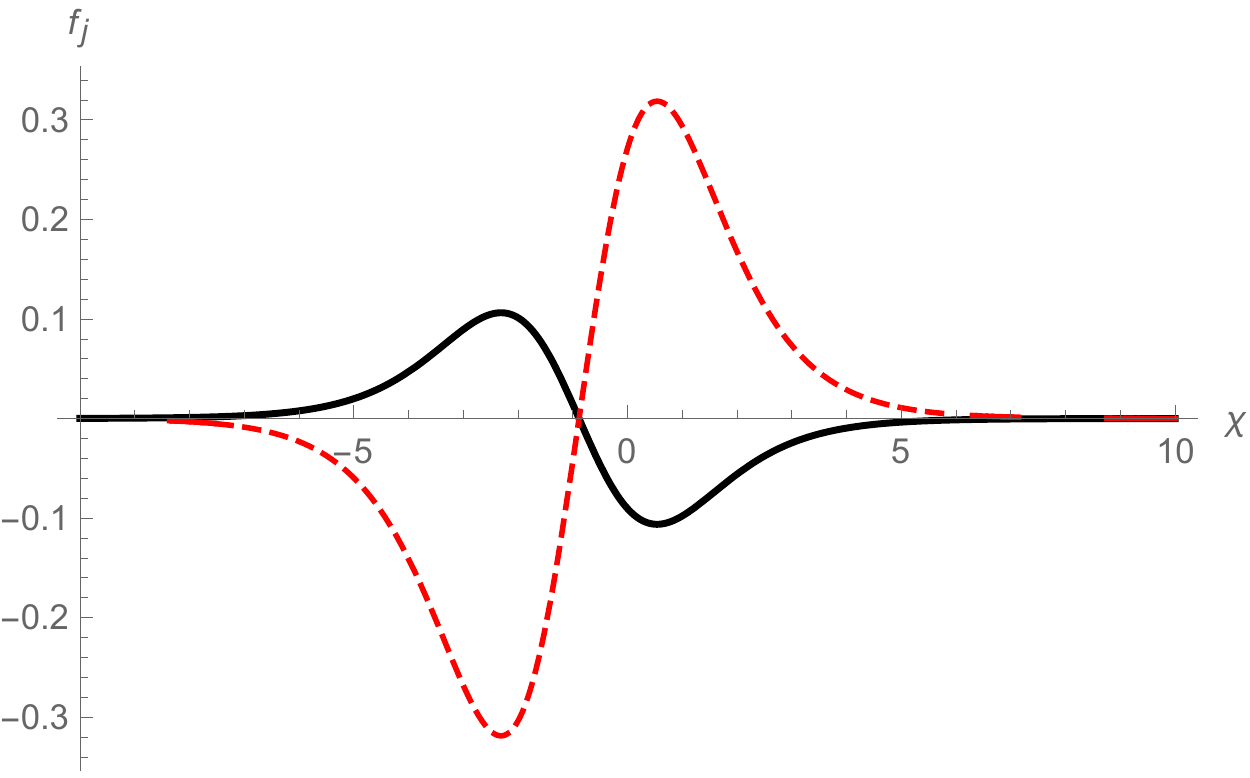}
  \includegraphics[width=0.2875\textwidth]{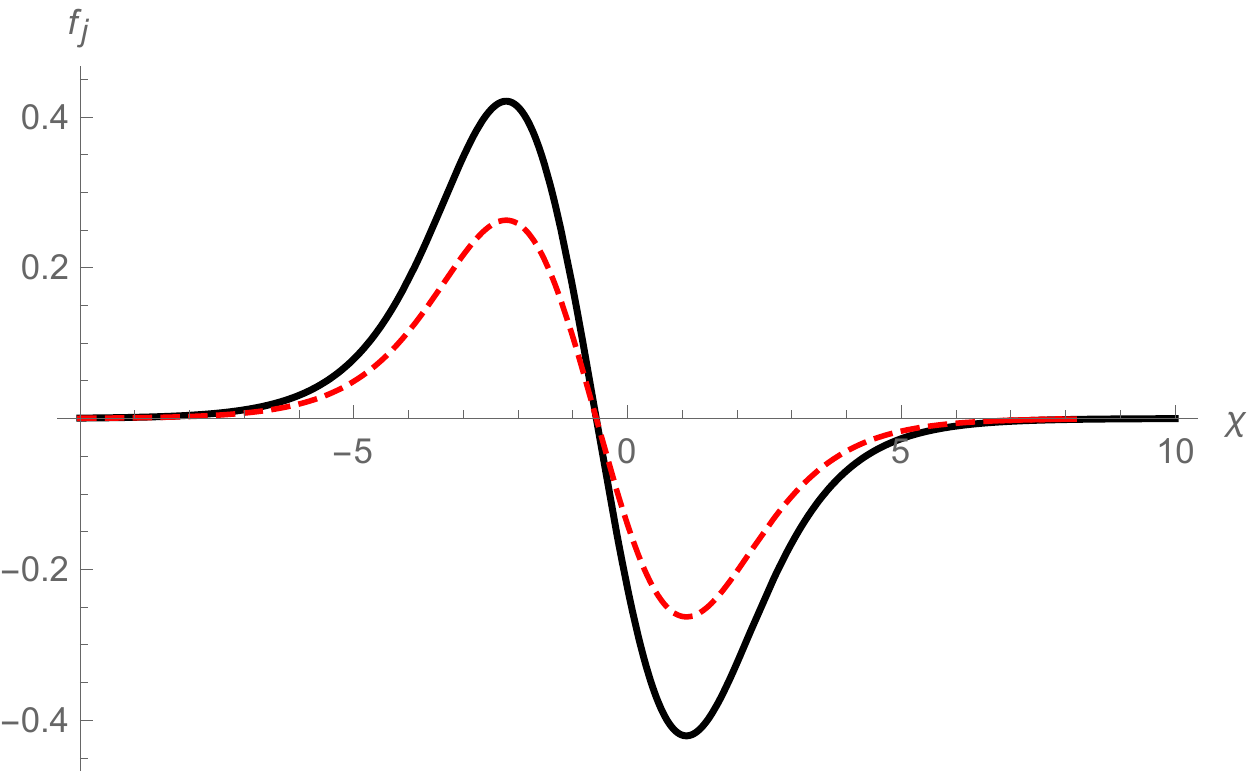}
}
\caption{Particle density $\bar{n}$  (top row) and spin densities (bottom row) for polar state with $\varphi = 0.927$. Left column: $\rho_1 = 4$, $\rho_2 = -1.94$, middle column:$ \rho_1 = 4$, $\rho_2 = -1/2$, right column: $\rho_1 = 3/8$, $\rho_2 = -1/2$. Spin density components $f_1$ (solid line), $f_{-1}$ (dashed line), $f_0$ is zero everywhere}
\label{f:densitypolar}
\kern-\bigskipamount
\end{figure}

Next, we characterize the physical properties of polar states in canonical form, but with $\Pi_1$ non-diagonal.
Unlike the ferromagnetic case, from \eqref{e:polarpotential} one can see that for any choice of $z_1$ and $\Pi_1$, $|Q_{11}(x,t)|,\,|Q_{22}(x,t)|$ both have a minimum,  and the minimum of $|Q_{11}(x,t)|$ (if $\rho>1$) or $|Q_{22}(x,t)|$ (if $\rho<1$) is located on the line
\[
\label{e:polarminlocation}
(x+ x_0) \sin\varphi + \sin(2\varphi)\,t = - \frac12\ln\Big[|\rho_1 - 1|+ \sqrt{(1+\rho_1)^2- 4 \rho_2^2}\Big]\,.
\]
Moreover, $|Q_{12}(x,t)|$ reaches a maximum when $c_{12}\neq 0$, and the maximum is located on the line
\[
\label{e:polarmaxlocation}
(x-x_0) \sin\varphi + \sin(2\varphi)\,t = - \frac12\ln\Big[\frac{2 \sin^2 \varphi}{(\rho_1 - \rho_2^2)^{3/2}}\Big]\,,
\]
with height
\[
\label{e:polarmaxheight}
|Q_{12,max}(x,t)|=\dfrac{2 |\rho_2| \sin \varphi}{\rho_1 +1+ 2 \sqrt{\rho_1-\rho_2^2}}\,.
\]
Note that the equation~\eqref{e:polarmaxlocation} implies the velocity
of the polar one-soliton is $v= -2 \cos \varphi = -2 \Re(z_1)$, where
the velocity reaches
its max and min when $\varphi = \pi$ and $\varphi = 0$, respectively.
Fig.~\ref{f:canonicalpolar} shows the profile of polar one-soliton solutions in canonical form and for comparison purposes, Fig.~\ref{f:non-canonicalpolar} shows examples of polar one-soliton solutions in non-canonical form,
corresponding respectively to the following pair of asymptotic
matrices and norming constants.  In particular, the following pairs were chosen to show the significant difference of min/max patterns.
\begin{align}
\nonumber
Q_+ &= \frac{1}{5}\begin{pmatrix}1+4 e^{i\pi/3}& 2-2 e^{i\pi/3}\\2-2 e^{i\pi/3}& 4+ e^{i\pi/3} \end{pmatrix}\,,
\quad
C_1 = \frac{1}{30}\begin{pmatrix}-4\sqrt{3}-10i& 2\sqrt{3}-10i\\2\sqrt{3}-10i& -\sqrt{3}-25i\end{pmatrix}\,,\\
%Q_+ &= \frac{1}{13}\begin{pmatrix}4+9 e^{i\pi/3}& 6-6 %e^{i\pi/3}\\6- 6e^{i\pi/3}& 9+4e^{i\pi/3} \end{pmatrix}\,,\\
%\quad
%C_1 = \begin{pmatrix}-4.422-0.833i& 1.565-0.658i\\1.565-0.658i& %-0.899-0.099i\end{pmatrix}\,,\\
%\nonumber
%C_1 &= \frac{1}{169}\begin{pmatrix}(2214\sqrt{3}-4582)+(126-154\sqr%t{3})i& (5169-2679\sqrt{3})/2 + %(105-189\sqrt{3})i/2\\(5169-2679\sqrt{3})/2 + (105-189\sqrt{3})i/2& %(802\sqrt{3}-1541)+(63\sqrt{3}-126)i\end{pmatrix}\,,\\
\nonumber
Q_+ &= e^{i \pi/6} \begin{pmatrix}0&1\\1&0\end{pmatrix}\,,
\quad
C_1 = e^{i\pi/12}\begin{pmatrix}7/10&-\pi/4\\
-\pi/4&7/10\end{pmatrix}\,,\\
Q_+ &= \frac{1}{5}\begin{pmatrix}4& 3\\3& -4 \end{pmatrix}\,,
\quad
C_1 = \frac{i}{5\sqrt{3}}\begin{pmatrix}8\sqrt{3}+35\sqrt{3}\,e^{2i\pi/3} & -14\sqrt{3}+10i \\-14\sqrt{3}+10i & -8\sqrt{3}-5\sqrt{3}\,e^{2i\pi/3} \end{pmatrix}\,.
%\quad
%C_1 = \begin{pmatrix}-6.978-2.111i & 0.21-2.704i \\0.21-2.704i & %0.915+0.218i \end{pmatrix}\,.
%C_1 &= \frac{1}{20}\begin{pmatrix}(-72-39\sqrt{3})+(12\sqrt{3}-63)i & %(96-53\sqrt{3})+(4\sqrt{3}-61)i \\(96-53\sqrt{3})+(4\sqrt{3}-61)i & %(72-31\sqrt{3})+(233-132\sqrt{3})i \end{pmatrix}\,.
\label{e:Q+forfig5}
\end{align}
One can show that $Q_+ C_1$ is not symmetric in all of the above cases.
{The relevant patterns in this setting are far more reminiscent
  of the dark- or dark-bright solitonic generalizations that have been
  identified in the spinorial
  setting~\cite{Panos3,Panos6,Panos_PRL2020,katsi}, however here, too,
  there
  are differences. In particular, the dark solitons on the $\pm 1$
  components
  are not necessarily collocated (in terms of their density extrema);
  in addition they may contain ``anti-dark'' patterns that may exceed the
  asymptotic density of the respective species.}

However, one can always use canonical form to characterize the solution in non-canonical form, as discussed above.
The particle number with spin densities for polar soliton solutions are plotted in Fig.~\ref{f:densitypolar}
as functions of $\chi:=2i\theta(x,t,z_1)$.

%%%%%%%%%%%%%%%%%%%%%%%%%%%%%%%%%%%%%%%%%%%%%%%%%%%%%%%%%%%%%%%%%%%%%%%%%%%%%%%%%%%%%%%%%%%%%
\section{Soliton interactions}
\label{s:solitoninteraction}

In this section we will discuss the soliton interaction in detail by computing the long-time asymptotics of the two-soliton solutions,
i.e., the solutions obtained from the general expression~\eqref{e:multisoliton} with $J=2$.

%%%%%%%%%%%%%%%%%%%%%%%%%%%%%%%%%%%%%%%%%%%%%%%%%%%%%%%%%%%%%%%%%%%%%%%%%%%%%%%%%%%%%%%%%%%%%
\subsection{General set up}
\label{s:interactionsetup}

\begin{figure}[t!]
\centering
\includegraphics[width=3.4in, height=2.7in]{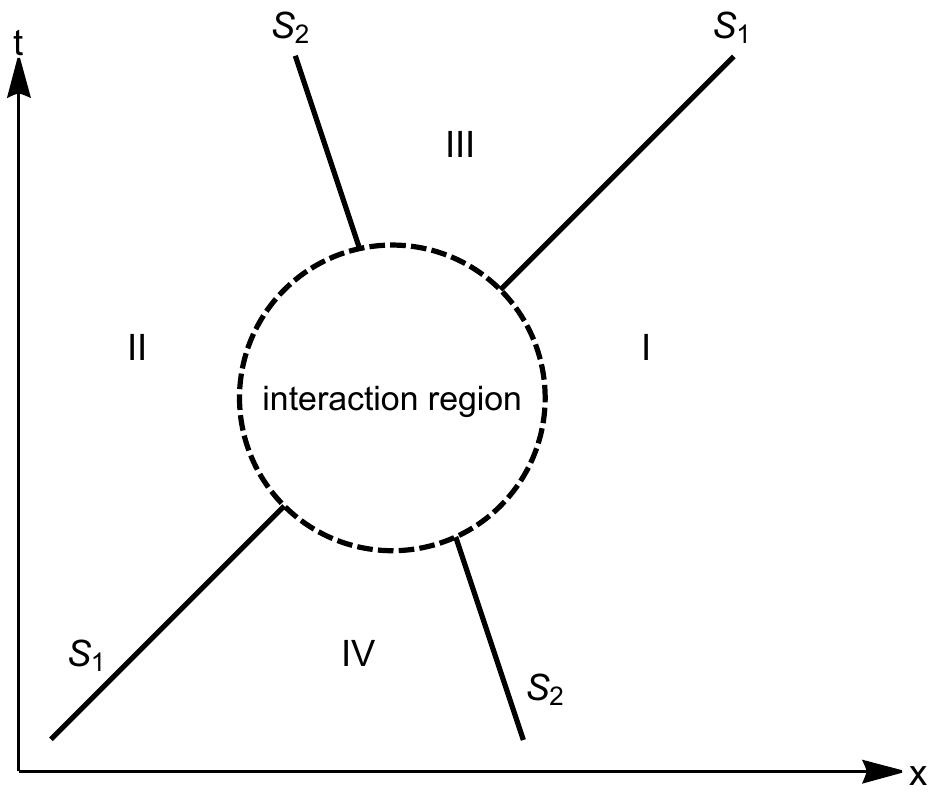}
\caption{Schematic diagram of a two-soliton interaction showing the solitons $s_1$ and $s_2$, the interaction region, and the fundamental domains $\I,\dots,\IV$ for the analysis in the text.}
\label{f:soliton interaction}
\end{figure}

The canonical form of a two-soliton solution is given by \eqref{e:Qpotnoreflection} with $J = 2,\, \Ko = 1$ and $Q_+ = I_2$, namely
\bse
\label{e:2-solitonsolution}
\begin{gather}
Q(x,t)=I_2+i\sum_{j=1}^{2} e^{-2i\theta(x,t,z_j^*)}\bar{N}_{\text{up}}(x,t,z_j^*)\bar{C}_j,\\
\noalign{\noindent where}
\bar{N}_{\text{up}}(x,t,z_n^*)=I_2-i \sum_{j=1}^{2} \frac{e^{2i\theta(x,t,z_j)}C_j}{z_j(z_n^*-z_j)}+\sum_{j=1}^{2} \sum_{l=1}^{2} \frac{e^{2i(\theta(x,t,z_j)-\theta(x,t,z_l^*))}}{(z_n^*-z_j)(z_j-z_l^*)}\bar{N}_{\text{up}}(x,t,z_l^*)\bar{C}_lC_j,\\
\noalign{\noindent with}
\nonumber
C_n = \Pi_{n}\,e^{i \phi_n},\,z_n = e^{i \phi_{n}},\quad n=1,2,\, \phi_n \in (0,\pi),\quad \Pi_{1} = (c_{ij}),\, \Pi_{2} = (d_{ij})\,,\quad i,j\in \{1,2\}.
\end{gather}
\ese
For the rest of the paper, we denote the discrete eigenvalues as follows:
$$
z_j = \zeta_j + i \eta_j, \qquad \eta_j>0, \quad\text{for } j=1,2, \qquad \zeta_1>\zeta_2\,.
$$
The assumption $\zeta_1 > \zeta_2$ is obviously without loss of
generality, and since the velocities of the solitons are
$v_j=-2\zeta_j$ for $j=1,2$, it corresponds to labeling as soliton $1$
the slowest soliton.
Now let $\chi_j = x + 2\,\zeta_j t$ for $j = 1,2$ denote the direction of each soliton.
Note that $\chi_2 = \chi_1 + 2 (\zeta_2 - \zeta_1)t$. Since $z_1,\,z_2 \in C_0^+$, we have $\zeta_j = k_j$ and $i \eta_j = \lambda_j $ for $j = 1,2$. Recalling \eqref{e:theta} we have
\bse
\begin{gather}
\label{e:chi1}
e^{2i \theta(x,t,z_1)} = e^{-2 \eta_1\,\chi_1}= e^{-2 \eta_1\,\chi_2} e^{4\eta_1 (\zeta_2 - \zeta_1)t}\,,\\
\label{e:chi2}
e^{2i \theta(x,t,z_2)} = e^{-2 \eta_2\,\chi_2} = e^{-2 \eta_2\,\chi_1} e^{4\eta_2 (\zeta_1 - \zeta_2)t}\,.
\end{gather}
\ese
Next, we compute the long-time asymptotics as $t \to \pm \infty$ along the direction of each soliton, i.e., keeping $\chi_j$ fixed, first for $j=1$ and then for $j=2$. When the direction $\chi_1$ is fixed, using \eqref{e:chi2} we have
\bse
\begin{gather}
e^{2i \theta(x,t,z_2)} =
\begin{cases}
0\,, \qquad t \to - \infty\,,\\
\infty\,, \qquad t \to \infty\,.\\
\end{cases}\\
\noalign{\noindent Conversely, when $\chi_2$ is fixed, using \eqref{e:chi1}}
e^{2i \theta(x,t,z_1)} =
\begin{cases}
\infty\,, \qquad t \to - \infty\,,\\
0\,, \qquad t \to \infty\,.\\
\end{cases}
\end{gather}
\ese
After rewriting equation \eqref{e:2-solitonsolution} in terms of $\chi_1$ and $\chi_2$, we compute the leading order behavior as $t \to \pm \infty$. Using this idea we will analyze the two-soliton solution for the polar-polar, ferromagnetic-ferromagnetic and polar-ferromagnetic cases in following sections.
For future convenience, before going into the detail of the soliton interactions,  we introduce the following notations (cf Fig~\ref{f:soliton interaction}):
\bse
\begin{align}
&\forall (x,t) \in \I:\quad Q(x,t) = Q_+ + o(1),\quad x \to \infty\,,\\
&\forall (x,t) \in \II:\quad Q(x,t) = Q_- + o(1),\qquad x \to -\infty\,,\\
&\forall (x,t) \in \III:\quad Q(x,t) = Q_\III + o(1),\qquad t \to \infty\,,\\
&\forall (x,t) \in \IV:\quad Q(x,t) = Q_\IV + o(1),\qquad t \to -\infty\,.
\end{align}
\ese
Also, from now on we will use subscripts $\pm$ to denote limits as $x \to \pm \infty$, and superscripts $\pm$ to denote limits as $t \to \pm \infty$.

\begin{figure}[t!]
\centering
\includegraphics[width=2in, height=1.5in]{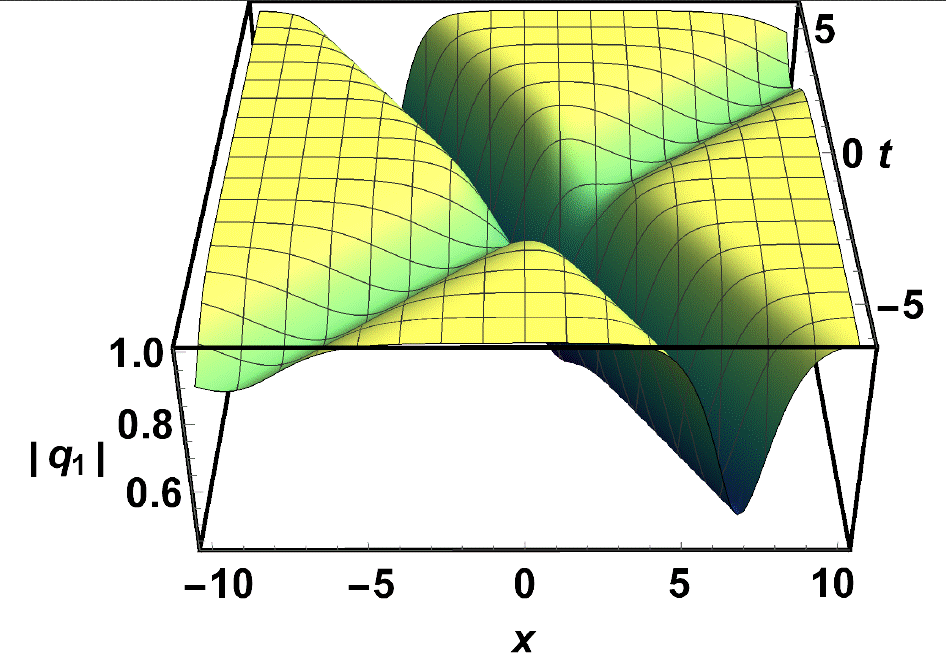}
\includegraphics[width=2in, height=1.5in]{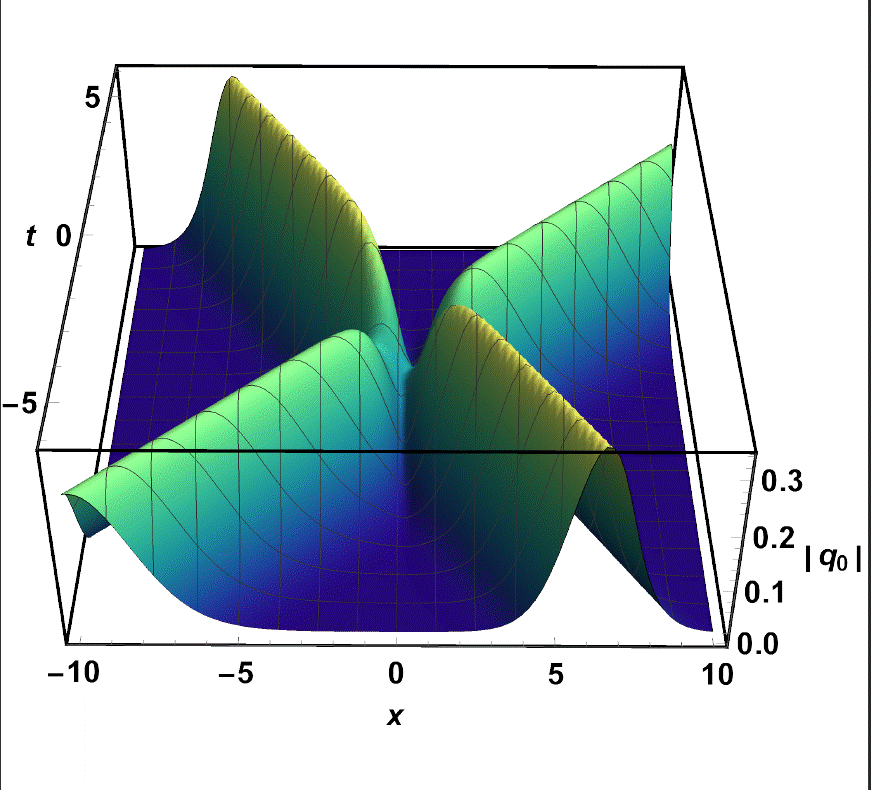}
\includegraphics[width=2in, height=1.5in]{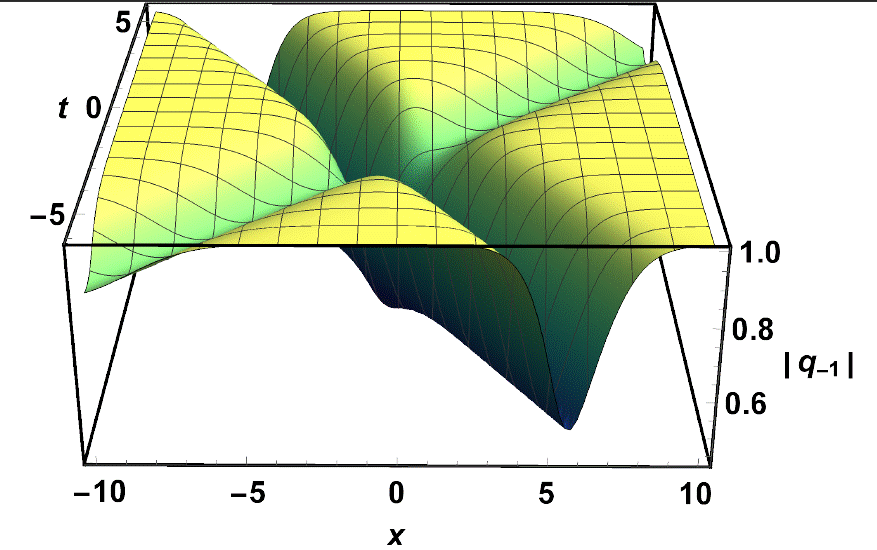}
\medskip
\caption{Plot of a polar-polar soliton interaction. The soliton parameters are as follows: $c_{11}= -8,\,c_{12}= 3,\,c_{22}= -2$ for the norming constant $C_1$, $d_{11}= -9,\,d_{12}= 5,\,d_{22}= -4,$ for the entries of the norming constant $C_2$. Also, $\Ko = 1$, and $\phi_1 = \pi/3,\, \phi_2 = 5\pi/6$ are the phases of the corresponding
discrete eigenvalues.}
%, with $q_{i,\pm}$ denotes the asymptotic expansion of the component $q_i$ as $t \to \pm \infty$ for $i \in \{1,0,-1\}$.}
\label{f:polar-polar}
\end{figure}

%%%%%%%%%%%%%%%%%%%%%%%%%%%%%%%%%%%%%%%%%%%%%%%%%%%%%%%%%%%%%%%%%%%%%%%%%%%%%%%%%%%%%%%%%%%%%
\subsection{Polar-polar soliton interaction}
\label{s:polarpolar}

We start by considering the interaction between two polar solitons since, as we shall see, their interaction is trivial and
the computation is straightforward. In this case, we assume both $\det C_1 \neq 0$ and $\det C_2 \neq 0$. The long-time asymptotic expansion of the two-soliton solution as $t \to -\infty$ with $\chi_1 = x + 2 \zeta_1 t$ fixed gives:
\[
\label{e:LTApolarpolar-infty}
Q(x,t) \sim
\dfrac{(e^{2(-2\eta_1 \chi_{1} + \varrho_1- i\varphi_1)}\,\det \Pi_{1,\text{eff}}^- - e^{-2\eta_1 \chi_1 + \varrho_1}\,\tr \Pi_{1,\text{eff}}^- + 1) I_2 + i e^{-2\eta_1 \chi_1 - i\varphi_1} \Pi_{1,\text{eff}}^-}{e^{2(-2\eta_1 \chi_1 + \varrho_1)}\,\det \Pi_{1,\text{eff}}^- - e^{-2\eta_1 \chi_1 + \varrho_1}\,\tr \Pi_{1,\text{eff}}^- + 1}\,,
\]
where
%$\chi = 2 i \theta(x,t,z_1) \in \Real$ and
$e^{-\varrho_1} = 2 \sin\varphi_1$ and $\Pi_{1,\text{eff}}^- =\Pi_1$.
On the other hand, the limit as $t \to \infty$ with $\chi_1 = x + 2 \zeta_1 t$ fixed yields
\begin{gather}
\label{e:LTApolarpolar+infty}
Q(x,t) \sim  e^{-2i \varphi_2}
\dfrac{(e^{2(-2\eta_1 \chi_1 + \varrho_1 - i\varphi_1)}\,\det \Pi_{1,\text{eff}}^+ - e^{-2\eta_1 \chi_1 + \varrho_1}\,\tr \Pi_{1,\text{eff}}^+ + 1) I_2 + i e^{-2\eta_1 \chi_1- i\varphi_1} \Pi_{1,\text{eff}}^+}{e^{2(-2\eta_1 \chi_1 + \varrho_1 )}\,\det \Pi_{1,\text{eff}}^+ - e^{-2\eta_1 \chi_1 + \varrho_1}\,\tr \Pi_{1,\text{eff}}^+ +1}
\\
\noalign{\noindent where}
\Pi_{1,\text{eff}}^+ = \dfrac{(z_1-z_2)(z_2^*-z_1^*)}{(z_1^*-z_2)(z_2^*-z_1)} \Pi_1 = \Big|\dfrac{z_1-z_2}{z_1^*-z_2}\Big|^2 \Pi_{1,\text{eff}}^-
\end{gather}
and $e^{-\varrho_1} = 2 \sin\varphi_1$ as before. Fig.~\ref{f:polar-polar} gives the 2 polar soliton interaction for a specific choice of the soliton parameters,
while Fig.~\ref{f:polar-polartest} in Appendix~\ref{s:appendix2}
shows the differences between the solution and the long-time asymptotics in each direction derived above.
Hereafter, $\Pi_{i,\text{eff}}^\pm$ denote the polarization matrix along the direction of soliton $i$ for $i =1$ or $2$, as $t\to -\infty$ $(-)$ and $t\to \infty$ $(+)$.
It can be easily seen that the same result holds along the direction of the second soliton. Specifically, the long-time asymptotic behavior can be obtained from \eqref{e:LTApolarpolar-infty} and \eqref{e:LTApolarpolar+infty} by switching the indices 1 and 2, and the limits $t \to \pm \infty$, yielding
\[
\Pi_{2,\text{eff}}^- = \Big|\dfrac{z_1-z_2}{z_1^*-z_2}\Big|^2 \Pi_{2,\text{eff}}^+\,.
\]
The above asymptotics show that the interaction of polar solitons is
always trivial, since the polarization matrices of each soliton are
affected by the interaction only by an overall phase factor. {Indeed,
these results for each component are somewhat reminiscent of the
unscathed interaction of 3-component dark-dark-bright Manakov
solitons.}

%%%%%%%%%%%%%%%%%%%%%%%%%%%%%%%%%%%%%%%%%%%%%%%%%%%%%%%%%%%%%%%%%%%%%%%%%%%%%%%%%%%%%%%%%%%%%
\subsection{Ferromagnetic-ferromagnetic soliton interaction}
\label{s:ferroferro}

\begin{figure}[t!]
\centering
\includegraphics[width=2in, height=1.5in]{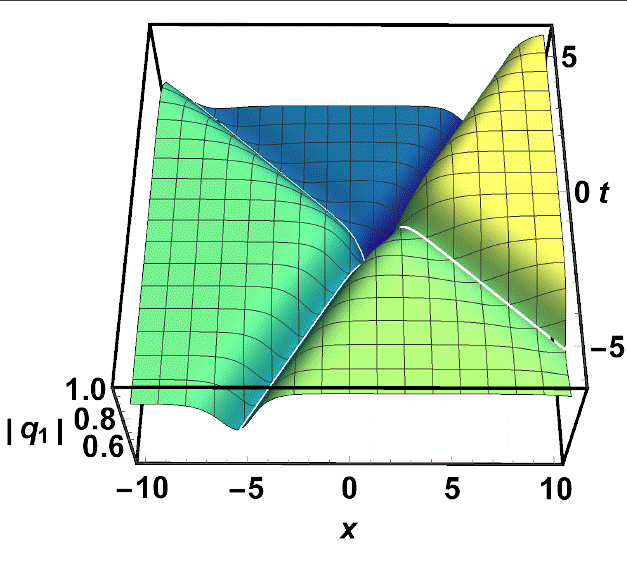}
\includegraphics[width=2in, height=1.5in]{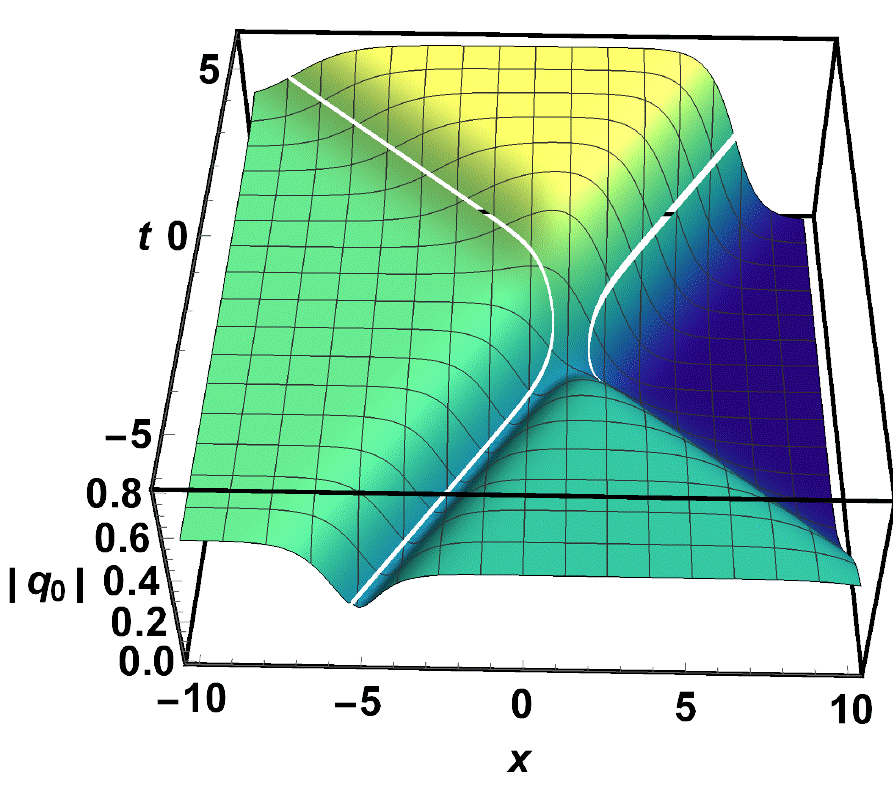}
\includegraphics[width=2in, height=1.5in]{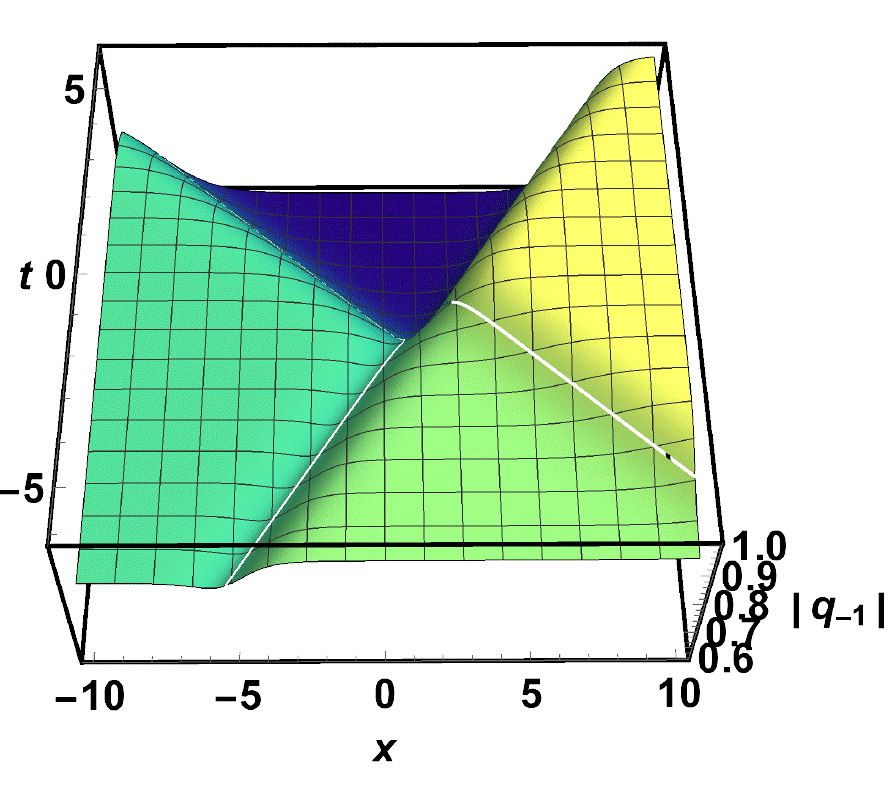}
\medskip
\caption{Plot of a ferromagnetic-ferromagnetic soliton interaction.
The soliton parameters are: $c_{11}= -8,\,c_{12}= 4,\,c_{22}= -2$ for the norming constant $C_1$, and $d_{11}= -9,\,d_{12}= 6,\,d_{22}= -4$ for the entries of the norming constant $C_2$. Also, $\Ko = 1$, and $\phi_1 = \pi/6,\, \phi_2 = \pi-\pi/3$ are the phases of the two discrete eigenvalues.}
%, with $q_{i,\pm}$ denotes the asymptotic expansion of the component $q_i$ as $t \to \pm \infty$ for $i \in \{1,0,-1\}$.}
\label{f:ferro-ferro}
\end{figure}

Next, we consider the interaction between two ferromagnetic solitons. In this case, we assume both $\det C_1 = 0$ and $\det C_2 = 0$, and we consider
a solution in canonical form, i.e., with $Q_+=I_2$.
The long-time asymptotic expansion of the two-soliton solution as $t\to -\infty$ with $\chi_1 = x + 2 \zeta_1 t$ fixed gives:
\[
\label{e:LTAferroferro-infty}
Q(x,t) \sim  I_2 +
\dfrac{2 i \sin{\varphi_1}\,e^{-i \varphi_1}}{2 \sin{\varphi_1}\,e^{2i \eta_1 \chi_1}-\tr \Pi_{1,\text{eff}}^-} \Pi_{1,\text{eff}}^-\,,%\quad t \to -\infty
\]
where $\Pi_{1,\text{eff}}^- = \Pi_1$.
On the other hand, the limit as $t\to +\infty$ with $\chi_1 = x + 2 \zeta_1 t$ fixed yields
\bse
\label{e:finalLTAferroferro+infty}
\begin{gather}
\displaystyle
Q(x,t) \sim Q_1^+ + \dfrac{2 i \sin{\varphi_1}\,e^{-i \varphi_1}}{2 \sin{\varphi_1}\,e^{2i \eta_1 \chi_1}-\tr( \Pi_{1,\text{eff}}^+)} \Pi_{1,\text{eff}}^+ Q_1^+\\
\noalign{\noindent where }
%$C_j = \Pi_{j}\,e^{i \varphi_j},\,\Pi_j \in Mat_{2\times 2}(\Real)$ and $z_j = e^{i \varphi_j}$ with $\varphi_j \in (0,\pi)$ for $j\in \{1,2\}$ and }
Q_1^+ = I_2 - \dfrac{2i \sin \varphi_2 e^{-i \varphi_2}}{\tr(\Pi_2^2)}\Pi_2^2\,,
\end{gather}
\begin{multline}
\Pi_{1,\text{eff}}^+ = \frac{(z_1^*-z_1)}{z_2\,\tr(\Pi_2^2)\sqrt{m\,\tr(\Pi_2^2)}}
\Big\{ \frac{z_1(z_2^*-z_2) }{ (z_1^*-z_2)(z_2^*-z_1)} \Big[\frac{2\,\tr(\Pi_1^2\Pi_2^2)}{(z_1^*-z_1)(z_2^*-z_2)} - \frac{\tr^2(\Pi_1\Pi_2)}{(z_1^*-z_2)(z_2^*-z_1)} \Big] \Pi_2^2 -
\\
\tr(\Pi_2^2)\,\Big[(z_2\,B_1+z_1\,B_2)\Pi_1 + z_2\,B_3\,\Pi_2 + \frac{z_1\,\Pi_2\Pi_1^2\Pi_2}{(z_1^*-z_1)(z_1^*-z_2)(z_2^*-z_1)}\Big]
 \Big\}(Q_1^+)^\dag,
\end{multline}
\begin{gather}
m =\frac{\tr^2(\Pi_1 \Pi_2)}{|z_1^*-z_2|^4} + \frac{\tr(\Pi_1^2)\tr(\Pi_2^2)}{16 \sin^2\varphi_1 \sin^2\varphi_2} - \frac{2\,\tr(\Pi_1^2 \Pi_2^2)}{4\,|z_1^*-z_2|^2 \sin\varphi_1 \sin\varphi_2},\\
B_1 = \frac{\Pi_1}{(z_2^*-z_2)}\Big(\frac{\Pi_2^2}{(z_2^*-z_1)(z_1^*-z_2)} -\frac{\tr(\Pi_2^2)I_2}{(z_1^*-z_1)(z_2^*-z_2)} \Big),\\
B_2 = \frac{\Pi_2}{(z_2^*-z_1)}\Big(\frac{\Pi_2\Pi_1}{(z_1^*-z_1)(z_2^*-z_2)} -\frac{\tr(\Pi_2\Pi_1)I_2}{(z_1^*-z_2)(z_2^*-z_1)} \Big),\\
B_3 = \frac{\Pi_1}{(z_1^*-z_2)}\Big(\frac{\Pi_1\Pi_2}{(z_1^*-z_1)(z_2^*-z_2)} -\frac{\tr(\Pi_1\Pi_2)I_2}{(z_2^*-z_1)(z_1^*-z_2)} \Big).
\end{gather}
\ese
For soliton 2, the above expressions hold, as before, with indices 1 and 2 switched, and with the limits $t \to \pm \infty$ also interchanged. The above asymptotics show that the interaction of ferromagnetic solitons is nontrivial, as generically the polarization matrices of the solitons change due to the interaction
according to \eqref{e:finalLTAferroferro+infty}, which result in a
redistribution of energy among the spin components of each soliton.
Indeed, the domain wall character of the ferromagnetic solitons plays
a central role in this interaction.
Note that this is true even if for one of the solitons one assumes the associated norming constant is diagonal (say, either $\Pi_1$ or $\Pi_2$ is diagonal), and the corresponding solution in one of the directions is simply a dark soliton as given by \eqref{e:diagonalferro}.
%Note also that the two-soliton solution in \eqref{e:LTAferroferro-infty} was already assumed in canonical form, with $Q_+=I_2$.
Figure~\ref{f:ferro-ferro} gives the 2 ferromagnetic soliton solution for a specific choice of the soliton parameters,
while Fig.~\ref{f:ferro-ferrotest} in Appendix~\ref{s:appendix2}
shows the differences between the solution and the long-time asymptotics in each direction derived above.

As a special case, assume now both $\Pi_1$ and $\Pi_2$ are diagonal. Note that there are two possible choices for $\Pi_1$ and $\Pi_2$. First, consider the case when $\Pi_1 = \diag (\gamma_1,0)$ and $\Pi_2 = \diag (0, \delta_{-1})$ with $\gamma_1<0$ and $\delta_{-1} <0$. Note that in this case $\Pi_1 \Pi_2$ vanishes identically. Then the asymptotic expansion of the two-soliton solutions as $t \to -\infty$, namely  \eqref{e:LTAferroferro-infty}, gives
\[
\label{e:LTAferrodiagonal1}
Q(x,t) \sim \diag (q_{\text{dark},1-}(x,t),1)
\]
where $\diag (q_{\text{dark},1-}(x,t),1)$ is as defined in \eqref{e:darksoliton}, with soliton center $x_{1,-}$ given by $e^{-2x_{1,-}\sin \varphi_1}=-2\sin \varphi_1/\gamma_1$.
Also, when $t \to \infty$ \eqref{e:finalLTAferroferro+infty} simplifies to
\be
Q(x,t) \sim \diag(q_{\text{dark},1+}(x,t) , e^{-2i\varphi_2})\,,
\ee
where
\[
q_{\text{dark},1+}(x,t) = e^{-i \varphi_1}\{\cos \varphi_1 + i \sin \varphi_1 \tanh [\sin \varphi_1 (x-x_{1,+} + 2t \cos\varphi_1)]\}
\]
with $x_{1,+}$ such that $e^{-2x_{1,+}\,\sin\varphi_1} = -2 \sin \varphi_2 / \gamma_1$. % and $Q_1^+ = \diag (1,e^{-2i\varphi_2})$.
The asymptotic expansion of the two-soliton solutions along the direction of the second soliton can be obtained by switching the indices 1 and 2, and interchanging the the diagonal elements of $Q(x,t)$, with $\gamma_1$ replaced by $\delta_{-1}$.
\iffalse
we have $\Pi_{1,\text{eff}}^- = \diag (\gamma_1,0)$ and
\[
\Pi_{1,\text{eff}}^+ =-\frac{\sin \varphi_1}{\sin \varphi_2} \diag (\gamma_1,0)\,,\qquad
Q_1^+ = \diag (1,e^{-2i\varphi_2})\,,
\]
\fi

Now suppose $\Pi_2 = \diag(\delta_{1},0)$ with $\delta_{1}<0$ and
$\Pi_1=\diag(\gamma_{1},0)$ as before. In this case, the asymptotic
expansion when $t \to -\infty$ \eqref{e:LTAferroferro-infty} remains
the same as in \eqref{e:LTAferrodiagonal1}, but the asymptotic expansion along the soliton 1 direction when $t \to \infty$, Eq.~\eqref{e:finalLTAferroferro+infty}, yields
\be
Q(x,t) \sim \diag(e^{-2i\varphi_2} q_{\text{dark},1+}(x,t) , 1)\,,
\ee
where
\[
q_{\text{dark},1+}(x,t) = e^{-i \varphi_1}\{\cos \varphi_1 + i \sin \varphi_1 \tanh [\sin \varphi_1 (x-x_{1,+} + 2t \cos\varphi_1)]\}
\]
with  $x_{1,+}$ such that
\[
e^{-2x_{1,+}\,\sin\varphi_1} = -2 \sin \varphi_2 / (\gamma_1\,\omega)\,, \qquad \omega = \frac{|z_1-z_2|^4}{|(z_1-z_2)(z_1^*-z_2)|^2+2|z_1^*-z_2|^4}>0\,.
\]
To obtain the asymptotic expansion along the direction of the second soliton, one has to replace $\gamma_1$ by $\delta_{1}$ in addition to switching the indices 1 and 2.

%and with $Q_1^+ = \diag (e^{-2i\varphi_2},1)$.

%%%%%%%%%%%%%%%%%%%%%%%%%%%%%%%%%%%%%%%%%%%%%%%%%%%%%%%%%%%%%%%%%%%%%%%%%%%%%%%%%%%%%%%%%%%%%
\subsection{Polar-ferromagnetic soliton interaction}
\label{s:polarferro}

Finally, we discuss the interaction between a polar and a ferromagnetic soliton, i.e., we take the two norming constants $C_1,C_2$ such that $\det C_1 \neq 0$ and $\det C_2 =0$. As $t \to -\infty$ and when the direction $\chi_1$ is fixed, the asymptotic expansion is given by equation \eqref{e:LTApolarpolar-infty}. The long time asymptotic expansion of the two-soliton solution after a polar-ferromagnetic interaction when the direction $\chi_1$ is fixed and as $t \to \infty$ is given by
\bse
\begin{gather}
\displaystyle
\label{e:LTApolar-ferrochi1fixed+infty}
Q(x,t) \sim
\dfrac{(e^{2(-2\eta_1 \chi_1 + \varrho_1- i\varphi_1)}\det (\Pi_{1,\text{eff}}^+) - e^{-2\eta_1 \chi_1 + \varrho_1}\tr(\Pi_{1,\text{eff}}^+) + 1) Q_1^+ + i e^{-2\eta_1 \chi_1 - i\varphi_1} \Pi_{1,\text{eff}}^+ Q_1^+}{e^{2(-2\eta_1 \chi_1 + \varrho_1)}\,\det (\Pi_{1,\text{eff}}^+) - e^{-2\eta_1 \chi_1 + \varrho_1}\,\tr(\Pi_{1,\text{eff}}^+)+1},\\
\noalign{\noindent where $e^{-\varrho_1} = 2 \sin \varphi_1$ with}
Q_1^+= I_2 + \frac{(z_2^* - z_2)}{z_2 \tr(\Pi_2^2)}\Pi_2^2\,,\\
\Pi_{1,\text{eff}}^+ = \frac{(z_2^* - z_2)^2}{\tr(\Pi_2^2)}\, \Big[ \Big(\tr(\Pi_2^2)I_2 - \frac{(z_2^*-z_2)}{(z_1^*-z_2)}\Pi_2^2\Big)\frac{\Pi_1}{(z_2^*-z_2)^2} + \Big(\frac{\tr(\Pi_1 \Pi_2)}{(z_1^*-z_2)}I_2 - \frac{\Pi_1 \Pi_2}{(z_2^*-z_2)}\Big)\frac{\Pi_2}{(z_1^*-z_2)}\Big]\,(Q_1^+)^\dag\,.
\end{gather}
\ese

\begin{figure}[b!]
\centering
\includegraphics[width=2in, height=1.5in]{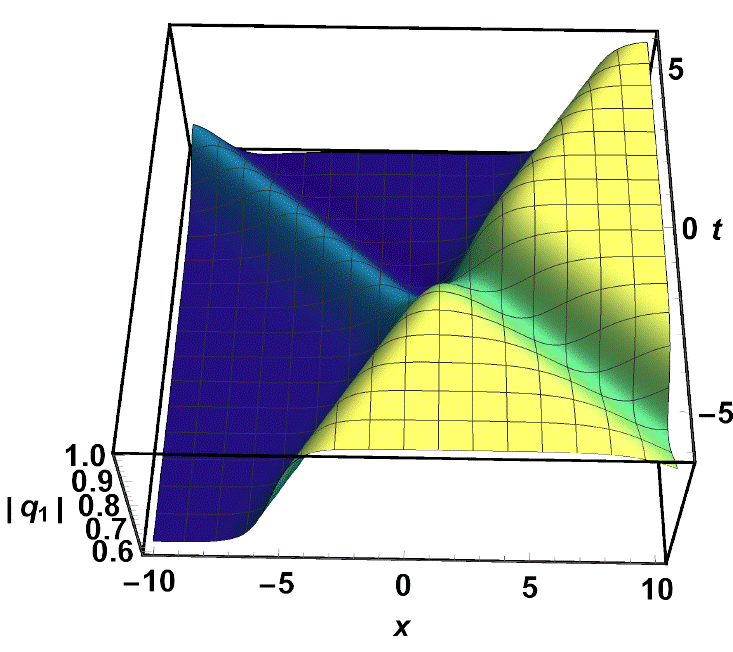}
\includegraphics[width=2in, height=1.5in]{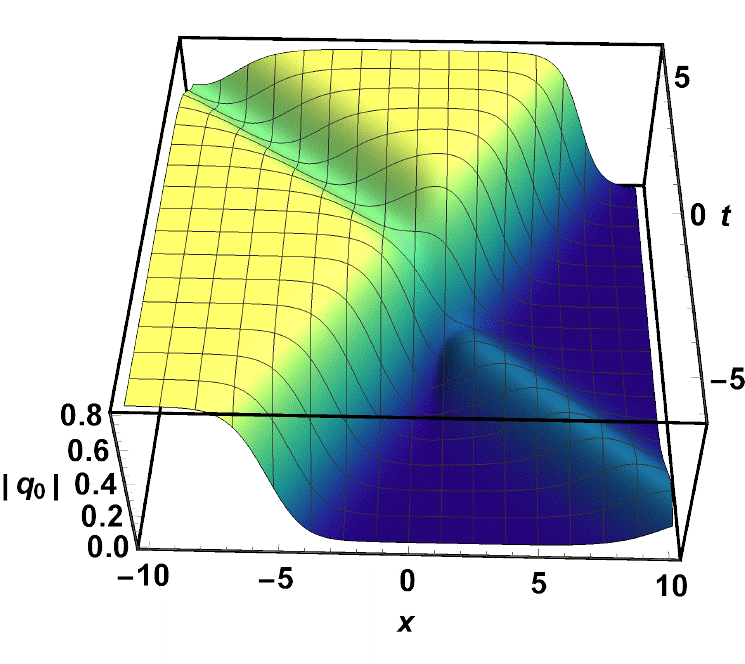}
\includegraphics[width=2in, height=1.5in]{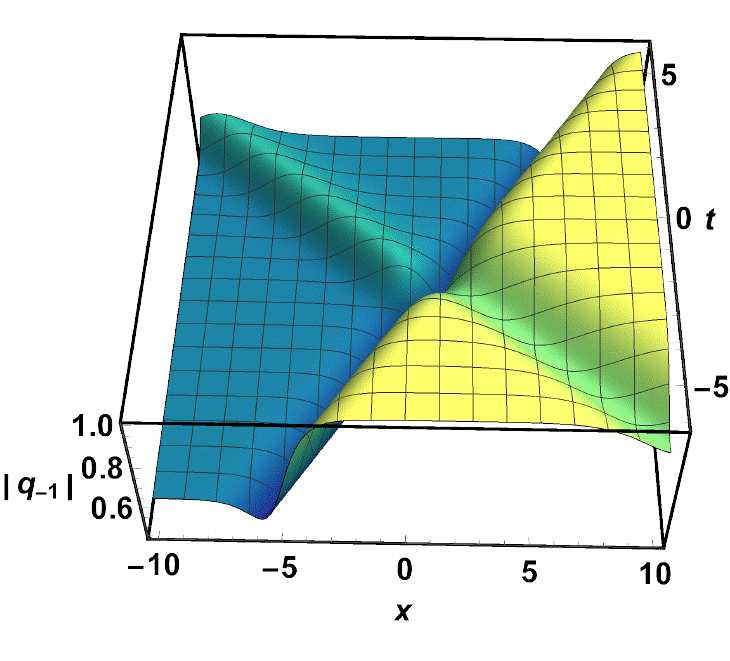}
\medskip
\caption{Plot of a polar-ferromagnetic soliton interaction. The soliton parameters are: $c_{11}= -8,\,c_{12}= 3,\,c_{22}= -2$ for $C_1$, $d_{11}= -4,\,d_{12}= 6,\,d_{22}= -9$ for the entries of $C_2$. Also, $\Ko = 1$, and $\phi_1 = \pi/6,\, \phi_2 = \pi-\pi/3$ are the phases of the two discrete eigenvalues.}
%with $q_{i,\pm}$ denotes the asymptotic expansion of the component $q_i$ as $t \to \pm \infty$ for $i \in \{1,0,-1\}$.}
\label{f:polar-ferro}
\end{figure}

In the case of polar-ferromagnetic interactions, the soliton solution is obviously not symmetric with respect to the interchange of soliton 1 and 2. Therefore the asymptotic behavior as $t \to \pm \infty$ along the direction of second soliton has to be computed independently. In particular, as $t \to \infty$ and when the direction $\chi_2$ is fixed, the asymptotic expansion is
\[
\label{e:LTApolar-ferrochi2fixed+infty}
Q(x,t) \sim I_2 + \dfrac{2 i \sin{\varphi_2}\,e^{-i \varphi_2}}{2 \sin{\varphi_2}\,e^{2i \eta_2 \chi_2}-\tr\,\Pi_{2,\text{eff}}^+} \,\Pi_{2,\text{eff}}^+\,
\]
where $\Pi_{2,\text{eff}}^+ = \Pi_2$.
On the other hand, the long-time asymptotic when $t \to -\infty$ with $\chi_2$ fixed has the form
\bse
\label{e:LTApolar-ferrochi2fixed-infty}
\begin{gather}
\displaystyle
Q(x,t) \sim Q_2^- + \dfrac{2 i \sin{\varphi_2}\,e^{-i \varphi_2}}{2 \sin{\varphi_2}\,e^{2i \eta_2 \chi_2}-\tr( \Pi_{2,\text{eff}}^-)} \Pi_{2,\text{eff}}^- Q_2^-\,,\\
\noalign{\noindent where}
Q_2^- = e^{-2i \varphi_1} I_2,
\end{gather}
\begin{multline}
\Pi_{2,\text{eff}}^- =
\\
\dfrac{z_1 (z_1^*-z_2)(z_2^*-z_2)(z_1^*-z_1)^2}{(z_1-z_2)(z_1^*-z_2^*)\tr(\Pi_1\Pi_2)\sqrt{\tr(\Pi_2^2)}}\Bigg\{\dfrac{(z_1-z_2)^2(z_2^*-z_1^*) \tr(\Pi_1 \Pi_2)}{(z_1^*-z_2)^2(z_1^*-z_1)(z_2^*-z_2)}
\Bigg(\dfrac{z_2(z_2^*-z_1^*)\tr(\Pi_2^2)}{(z_2^*-z_1)(z_2^*-z_2)}I_2 -\dfrac{\Pi_2^2}{z_1 (z_1^*-z_1)} \Bigg)
\\
- (z_2^*-z_1)(z_1^*-z_1)
\Big[ \frac{z_2 \tr(\Pi_2 N_2 \Pi_2)}{(z_1^*-z_2)(z_2^*-z_1)}I_2 - \frac{z_2 \tr(\Pi_1 N_1 \Pi_2)}{4\,\sin \varphi_1 \sin \varphi_2}I_2 +
\frac{(z_1-z_2)}{z_1 (z_2^*-z_1)(z_1^*-z_2)(z_1^*-z_1)} \Pi_2 N_2 \Pi_2\Big]\,
\Bigg\},
\end{multline}
with
\begin{gather}
N_1 = \dfrac{\tr(\Pi_2^2)}{(z_1^*-z_1)(z_2^*-z_2)}I_2 - \dfrac{\Pi_2^2}{(z_1^*-z_2)(z_2^*-z_1)}\,,
\\
N_2 = \dfrac{\tr(\Pi_2 \Pi_1)}{(z_1^*-z_2)(z_2^*-z_1)}I_2 - \dfrac{\Pi_2 \Pi_1}{(z_1^*-z_1)(z_2^*-z_2)}\,.
\end{gather}
\ese
Figure~\ref{f:polar-ferro} gives a two-soliton solution with one polar and one ferromagnetic soliton,
while Figs.~\ref{f:polar-ferrochi1fixed} and \ref{f:polar-ferrochi2fixed} in Appendix~\ref{s:appendix2}
show the differences between the solution and the long-time asymptotics in each direction derived above

As a special case, assume both $\Pi_1$ and $\Pi_2$ are diagonal.  Suppose $\Pi_1 = \diag (\gamma_1,\gamma_{-1})$ and $\Pi_2 = \diag (\delta_{1},0)$ with $\gamma_{\pm1}<0$ and $\delta_{1} <0$. Then the asymptotic expansion of the two-soliton solutions after the interaction along the direction of soliton 1 and as $t \to \infty$, namely, Eq.~\eqref{e:LTApolar-ferrochi1fixed+infty}, gives
\[
\label{e:LTApolarferrodiagonal1}
Q(x,t) \sim \diag (e^{-i \varphi_2} q_{\text{dark},1}^+(x,t),q_{\text{dark},-1}^+(x,t)) %Q_1^+
\]
where $q_{\text{dark},j}^+(x,t) = e^{-i \varphi_1}\{\cos \varphi_1 + i \sin \varphi_1 \tanh [\sin \varphi_1 (x-x_{j} + 2t \cos\varphi_1)]\}$ with $x_j$ such that
\begin{gather*}
e^{-2x_j\,\sin\varphi_1} = -2 \sin \varphi_1 / (\gamma_j \omega_j)  \qquad j = 1,-1 \\
\omega_{-1} = 1\,, \qquad \omega_1 = \Big((z_1-z_2^*+z_1^*-z_2)/|z_1^*-z_2|^2\Big)^2>0\,.
\end{gather*}
% and $Q_1^+ = \diag (e^{-2i\varphi_2},1)$.
In a similar way one can simplify the asymptotic expansions along the direction of the second soliton.
First, the asymptotic behavior as $t \to \infty$ when the direction $\chi_2$ is fixed simplifies to
\[
Q(x,t) \sim \diag (q_{\text{dark},2+}^+(x,t),1)
\]
where $q_{\text{dark},2+}^+(x,t) = e^{-i \varphi_2}\{\cos \varphi_2 + i \sin \varphi_2 \tanh [\sin \varphi_2 (x-x_{2,+} + 2t \cos\varphi_2)]\}$ with $x_{2,+}$ such that $e^{-2x_{2,+}\,\sin\varphi_2} = -2 \sin \varphi_2 / \delta_1$.
On the other hand, the asymptotic expansion when $t \to -\infty$, Eq.~\eqref{e:LTApolar-ferrochi2fixed-infty} reduces to
\be
Q(x,t) \sim \diag(e^{-2i\varphi_1}q_{\text{dark},2-}^-(x,t) , 1)\,, %\,Q_2^-\,,
\ee
where $q_{\text{dark},2-}^-(x,t) = e^{-i \varphi_2}\{\cos \varphi_2 + i \sin \varphi_2 \tanh [\sin \varphi_2 (x-x_{2-} + 2t \cos\varphi_2)]\}$ with $x_{2,-}$ such that $e^{-2x_{2,-}\,\sin\varphi_2} = -2 \sin \varphi_2 / \delta_1 \omega$ with $\omega = (|z_1-z_2|/|z_1^*-z_2|)^2>0$. % and $Q_2^- = e^{-2i\varphi_1}I_2$.
%
\iffalse
we have $\Pi_{1,\text{eff}}^- = \diag (\gamma_1,0)$ and
\[
\Pi_{1,\text{eff}}^+ =-\frac{\sin \varphi_1}{\sin \varphi_2} \diag (\gamma_1,0)\,,\qquad
Q_1^+ = \diag (1,e^{-2i\varphi_2})\,,
\]
\fi
%
Indeed, in this case too, we observe that the dynamics leads to
  nontrivial changes in the profiles of the relevant waveforms. While
  the domain wall of the ferromagnetic soliton seems to maintain its
  profile, the dark-bright pattern of the polar soliton seems to
  change to a dark-antidark one~\cite{schmied2}.
That is, it contains a bright structure on top of a non-vanishing background.

%%%%%%%%%%%%%%%%%%%%%%%%%%%%%%%%%%%%%%%%%%%%%%%%%%%%%%%%%%%%%%%%%%%%%%%%%%%%%%%%%%%%%%%%%%%%%
\section{Concluding remarks}

In the present work we have revisited the defocusing version of the integrable spinor model initiated by the work of~\cite{Wadati1,Wadati1.1}.
We have highlighted the relevance as well as the differences of the present model from the
3-component coupled NLS system, in which
solely density-dependent (i.e., spin-independent)
interactions are accounted for.
Indeed, this opposite yet still integrable limit involves the case of equal spin-dependent
and spin-independent interactions.
The recent experimental manipulation~\cite{Panos_PRL2020} of the spin-dependent interactions
to achieve the Manakov model holds some promise towards varying the
relevant ratio of interactions. Perhaps even more importantly, the availability of gases
such as the strongly ferromagnetic $F=1$ $^7$Li~\cite{jaeyoon} creates
a platform  where the spin-dependent part of the interaction is nearly
half that of the spin-independent one. In light of this, it becomes
progressively relevant to explore analytically tractable mathematical limits that may
yield novel waveforms that may emerge as being relevant for potential
observation in experiments.

It is in this vein that the present work has explored the possible waveforms
in the defocusing variant of the MNLS equation. We have leveraged
the earlier integrable formulation of~\cite{BP4} to identify the
prototypical soliton solutions which we classified into two major
categories. Polar solitons correspond to waveforms reminiscent
of dark- and dark-bright solitons, although with some key distinguishing
features regarding the location of their centers or their potential
to elevate above their asymptotic background. On the other hand,
the ferromagnetic waveforms presented structures that had a
fundamentally distinct pattern involving domain walls asymptoting to different
background values between $x \to -\infty$ and $x \to\infty$.
Going beyond the single soliton states, we explored also multi-soliton
collisions. These were more straightforward in preserving the nature
of the waveforms when same types of solitons (e.g. polar-polar or
ferromagnetic-ferromagnetic) collided. Yet, the scenario was clearly
richer and could involve an apparent change of the density distribution
of the profile when a polar and a ferromagnetic soliton might collide.

There results are a clear basis for numerous further studies at the level
of numerical computation and theoretical analysis and are even suggestive of novel
physical experiments. It would be especially relevant to continue parametrically
the solutions identified herein to explore their range of persistence
as the spin-dependent interaction is varied. If these states could
be continued even down to a ration of $1/2$, the recent experiments
of~\cite{jaeyoon} might enable their observation. Another relevant
possibility might be to compare these waveforms with the
magnetic waves recently explored for 3-component spinor systems
in~\cite{raman1,raman2}.  A comparison of the latter with dark-bright
waves in two-component systems has recently taken place
in~\cite{raman3}. An additional direction of interest concerns
the generalization of the patterns considered herein in higher-dimensional
systems. While identifying integrable generalizations in the higher-dimensional
realm would be a major challenge in its own right, it is certainly
plausible that vortical (i.e., topologically charged) generalizations
of the states presented herein may exist in the two-dimensional
analogue of the present system. Considering such domain-wall
and vortex-bright soliton structures is a numerical and experimental challenge in its own right.

%%%%%%%%%%%%%%%%%%%%%%%%%%%%%%%%%%%%%%%%%%%%%%%%%%%%%%%%%%%%%%%%%%%%%%%%%%%%%%%%%%%%%%%%%%%%%
\subsection*{Acknowledgement}
%This work was partially supported by the National Science Foundation under grant number
%DMS-2009487.
This material is based upon work
supported by the US National Science Foundation under Grants No.
DMS-2009487 (GB),
PHY-1602994 and DMS-1809074 (PGK).

%%%%%%%%%%%%%%%%%%%%%%%%%%%%%%%%%%%%%%%%%%%%%%%%%%%%%%%%%%%%%%%%%%%%%%%%%%%%%%%%%%%%%%%%%%%%%
\section*{Appendix}

\setcounter{section}1
\def\thesection{\Alph{section}}
\def\thesubsection{\Alph{section}.\arabic{subsection}}
\setcounter{equation}0
\def\theequation{\Alph{section}.\arabic{equation}}

%%%%%%%%%%%%%%%%%%%%%%%%%%%%%%%%%%%%%%%%%%%%%%%%%%%%%%%%%%%%%%%%%%%%%%%%%%%%%%%%%%%%%%%%%%%%%
\subsection{Unitary transformations, symmetric matrices and rotation of the quantization axes}
\label{s:symmetries}

Recall that the matrix NLS equation~\eqref{e:MNLS} is invariant under unitary transformations from the left or from the right.
That is, if $Q(x,t)$ solves~\eqref{e:MNLS}, so does
\be
\~Q(x,t) = U Q(x,t) V\,,
\label{e:unitarytransf}
\ee
for all constant $U$ and $V$ such that $U^\dag = U^{-1}$ and $V^\dag = V^{-1}$.
On the other hand, in order for $\~Q(x,t)$ to also represent a spinor wave function, the transformation~\eqref{e:unitarytransf}
must preserve matrix symmetry.
That is, one must have $\~Q^T(x,t) = \~Q(x,t)$ whenever $Q^T(x,t) = Q(x,t)$.
In this appendix we characterize the set of unitary transformations that preserve the symmetry constraint.
We also show that all such transformations correspond to a rotation of the quantization axes.

We begin by representing arbitrary unitary matrices $U$ and $V$
without loss of generality in terms of the Pauli matrices as
\bse
\label{e:UVdef}
\begin{gather}
U = \exp[ iu_0 I_2 + i\@u\cdot\bfsigma ] =
  e^{iu_0} \begin{pmatrix} \cos u + i\^u_3\,\sin u &  i(\^u_1 - i\^u_2)\,\sin u \\
    i(\^u_1 + i\^u_2)\,\sin u & \cos u - i\^u_3\sin u
\end{pmatrix},
\\
V = \exp[ iv_0 I_2 + i\@v\cdot\bfsigma ] =
  e^{iv_0} \begin{pmatrix} \cos v + i\^v_3\,\sin v &  i(\^v_1 - i\^v_2)\,\sin v \\
    i(\^v_1 + i\^v_2)\,\sin v & \cos v - i\^v_3\sin v
\end{pmatrix},
\end{gather}
\ese
where $\bfsigma = (\sigma_1,\sigma_2,\sigma_3)^T$ is the vector of
Pauli matrices, here chosen as
\be
\sigma_1 = \begin{pmatrix} 0 & 1 \\ 1 & 0 \end{pmatrix},\quad
\sigma_1 = \begin{pmatrix} 0 & -i \\ i & 0 \end{pmatrix},\quad
\sigma_1 = \begin{pmatrix} 1 & 0 \\ 0 & -1 \end{pmatrix},
\label{e:Paulidef}
\ee
where $u_0$, $v_0$, $\@u = (u_1,u_2,u_3)^T$ and $\@v = (v_1,v_2,v_3)^T$ are all real,
with
$\^{\@u} = \@u/u$
and
$\^{\@v} = \@v/v$,
and where
\be
u = \sqrt{\@u\cdot\@u} = \sqrt{u_1^2 + u_2^2 + u_3^2}\,,\qquad
v = \sqrt{\@v\cdot\@v} = \sqrt{v_1^2 + v_2^2 + v_3^2}\,.
\ee
Since $u_0$ and $v_0$ just produce overall phase rotations, without loss of generality we can set $u_0=v_0=0$
owing to the phase invariance of the MNLS equation.
Without loss of generality, we can also take $u$ and $v$ in $[0,2\pi]$.

Inserting~\eqref{e:UVdef}
in~\eqref{e:unitarytransf} and requiring the equality of the off-diagonal entries of $\~Q(x,t)$ then yields the following three real constraints:
\bse
\label{e:symmetryconstraint}
\begin{gather}
[ (\^u_2\^v_3 + \^u_3\^v_2)\sin u + \^v_1 \cos u ]\,\sin v - \^u_1\,\sin u \,\cos v = 0\,,
\\
[ (\^u_1\^v_3 - \^u_3\^v_1)\sin u + \^v_2 \cos u ]\,\sin v + \^u_2\,\sin u\,\cos v = 0\,.
\\
[ (\^u_1\^v_2 + \^u_2\^v_1)\sin u - \^v_3 \cos u ]\,\sin v + \^u_3\,\sin u\,\cos v = 0\,.
\end{gather}
\ese
It is relatively straightforward to see that~\eqref{e:symmetryconstraint} are solved by
\be
(\^v_1,\^v_2,\^v_3)\,\tan v  = (\^u_1,-\^u_2,\^u_3)\,\tan u\,.
\label{e:uvsoln}
\ee
In turn, \eqref{e:uvsoln} implies that \eqref{e:symmetryconstraint}
admit the following inequivalent
classes of solutions, obtained respectively when $v = u$ and $v = 2\pi - u$:
\be
S_+:~(\^v_1,\^v_2,\^v_3) = (\^u_1, - \^u_2, \^u_3)\,,\qquad
S_-:~(\^v_1,\^v_2,\^v_3) = (- \^u_1, \^u_2, - \^u_3)\,.
\ee
One can now check that $S_+$ implies $V=U^T$ while $S_-$ implies $V = -U^T$.
Since an overall minus sign can always be rescaled using the phase invariance of the MNLS equation, however,
without loss of generality we can limit ourselves to considering only those transformations produced by $S_+$.

Next we show that the unitary transformation~\eqref{e:unitarytransf} is equivalent to
a complex rotation of the quantization axes.
Let
$\@q(x,t) = (q_1, \sqrt{2}\,q_0 , q_{-1})^T$
be the vector wave functions associated with $Q(x,t)$,
and let $\~{\@q}(x,t) = (\~q_1, \sqrt{2}\,\~q_0 , \~q_{-1})^T$ be the one associated with
$\~Q(x,t)$.
Observe that a sign change of $Q(x,t)$ obviously translates into a sign change in $\@q(x,t)$
and recall that, in the quantum-mechanical context, an overall phase of the wave function is immaterial.
Therefore, we can again  limit ourselves to considering transformations produced by $S_+$.
It is straightforward to show that
\bse
\begin{gather}
\~{\@q}(x,t) = R\,\@q(x,t)\,,
\\
\noalign{\noindent where}
R = \begin{pmatrix}
 c_+^2
 & \sqrt{2}i(\^u_1-i\^u_2)c_+\sin u
 & - (\^u_1-i \^u_2)^2\sin^2u
\\
 \sqrt{2}i(\^u_1+i\^u_2)c_+\sin u
 & \cos^2u-(1-2\^u_3^2)\sin^2u
 & \sqrt{2}i(\^u_1-i\^u_2)c_-\sin u
\\
 - (\^u_1+i\^u_2)^2\sin^2u
 & \sqrt{2}i(\^u_1+i \^u_2)c_-\sin u
 & c_-^2
\end{pmatrix},
\\
\noalign{\noindent and where for brevity we defined}
  c_\pm = \cos u \pm i \^u_3\sin u\,.
\end{gather}
\ese
It is also straightforward to check that $R$ is a unitary matrix, i.e., $RR^\dag = R^\dag R = I_3$,
and that $\det R = 1$, implying $R\in\mathrm{SU}(3)$.
Finally, it is also important to realize that $R$ corresponds to a rotation of the quantization axes.
Consider again the transformation \eqref{e:unitarytransf} with $V = U^T$, and
again let $u_0 = 0$ without loss of generality.
It is straightforward to show that
\be
\label{e:rotationmatrix1}
\displaystyle
R = e^{2i\@u\cdot\@f}\,,
\ee
where $\@f = (f_1,f_2,f_3)^T$, and $f_1,f_2,f_3$ are representation of the angular momentum operators in $\mathrm{SU}(3)$,  namely:
\be
f_1 = \frac{1}{\sqrt{2}}\begin{pmatrix}0&1&0\\
1&0&1\\ 0&1&0
\end{pmatrix},\qquad
f_2 = \frac{i}{\sqrt{2}}\begin{pmatrix}0&-1&0\\
1&0&-1\\ 0&1&0
\end{pmatrix},\qquad
f_3 = \begin{pmatrix}1&0&0\\
0&0&0\\ 0&0&-1
\end{pmatrix}.
\ee
In closing, we also point out that the above relations are purely local symmetries, and are therefore
completely independent of the boundary conditions
satisfied by $Q(x,t)$ as $x\to\pm\infty$.

%%%%%%%%%%%%%%%%%%%%%%%%%%%%%%%%%%%%%%%%%%%%%%%%%%%%%%%%%%%%%%%%%%%%%%%%%%%%%%%%%%%%%%%%%%%%%
\subsection{Asymptotics of two-soliton interactions}
\label{s:appendix2}

\begin{figure}[t!]
\centering
\begin{subfigure}{.27\textwidth}
  \centering
  \includegraphics[width=0.875\linewidth]{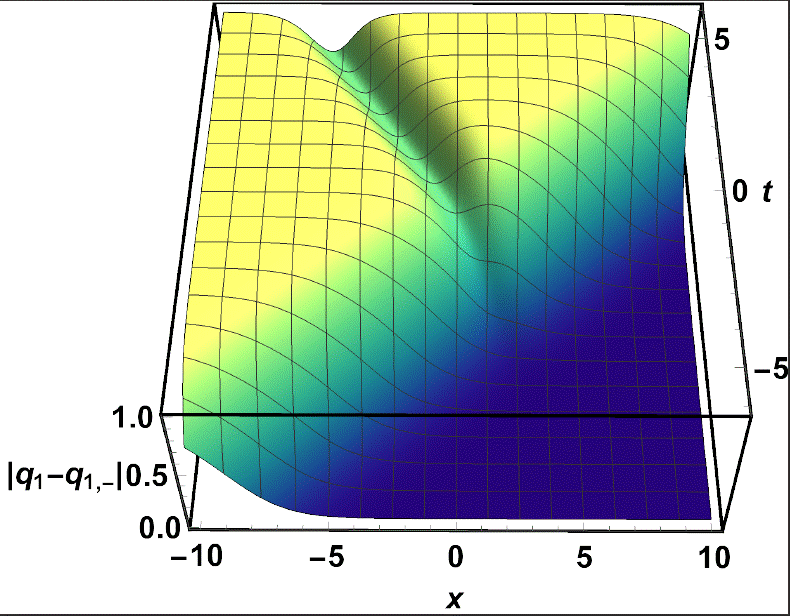}
  \caption{$|q_1-q_{1,-}|$}
  \label{fig8:sub-second}
\end{subfigure}
\begin{subfigure}{.27\textwidth}
  \centering
  \includegraphics[width=0.875\linewidth]{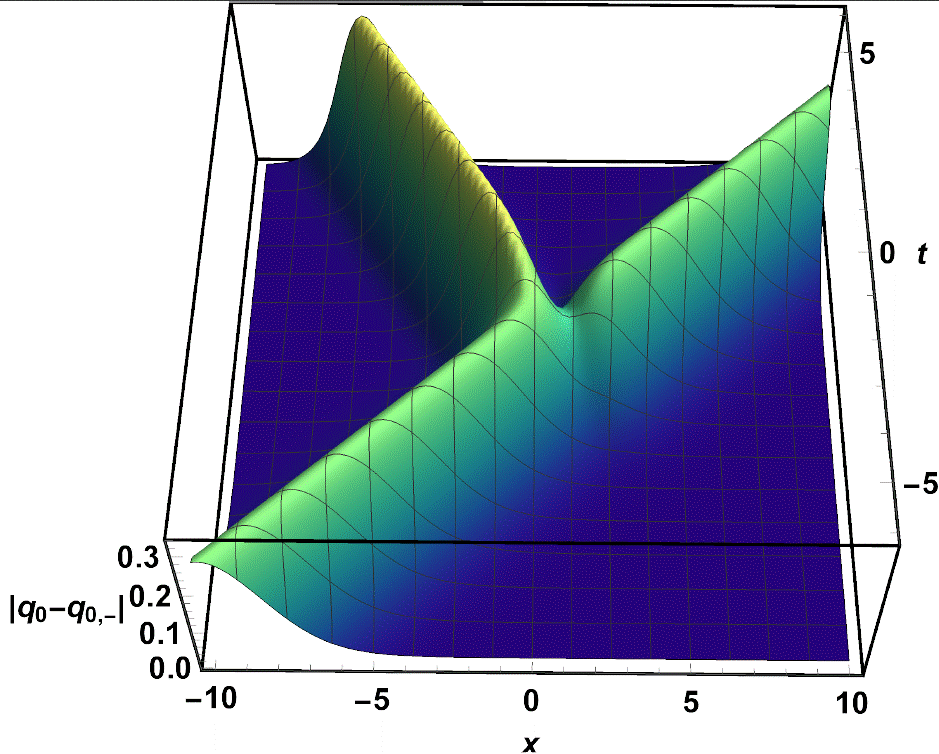}
  \caption{$|q_0-q_{0,-}|$}
  \label{fig8:sub-fifth}
\end{subfigure}
\begin{subfigure}{.27\textwidth}
  \centering
  \includegraphics[width=0.875\linewidth]{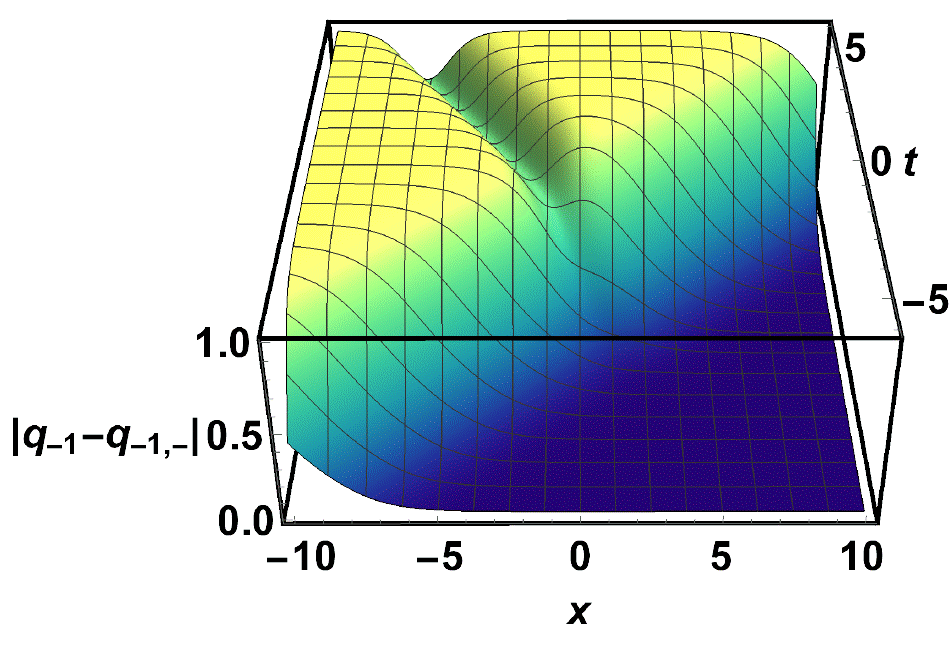}
  \caption{$|q_{-1}-q_{-1,-}|$}
  \label{fig8:sub-eighth}
\end{subfigure}
\vglue\bigskipamount
\begin{subfigure}{.27\textwidth}
  \centering
  \includegraphics[width=0.875\linewidth]{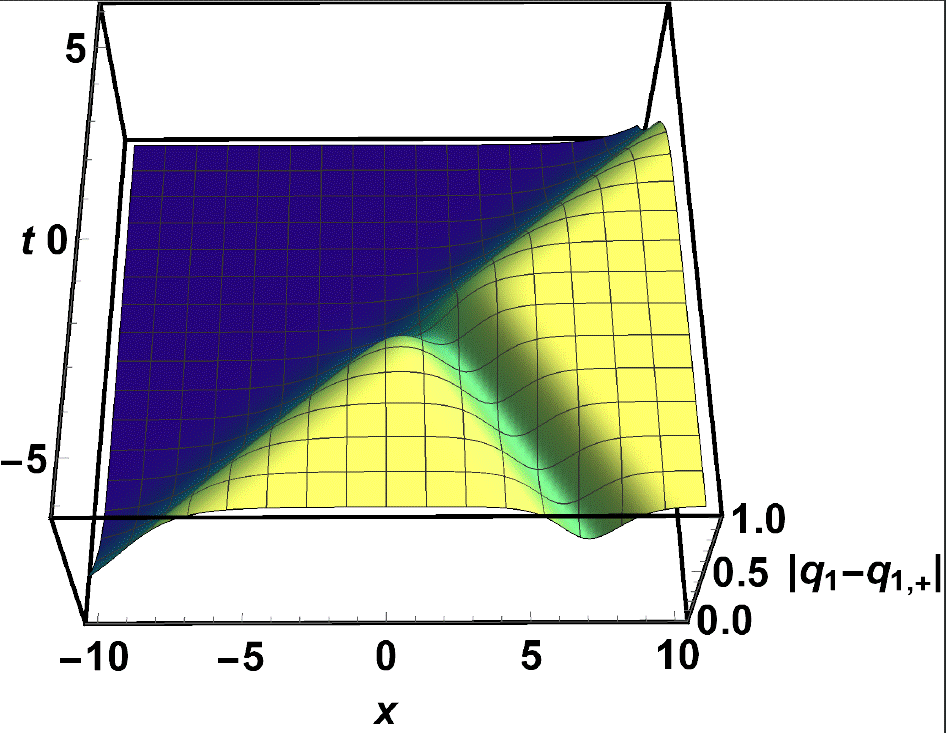}
  \caption{$|q_1-q_{1,+}|$}
  \label{fig8:sub-third}
\end{subfigure}
\begin{subfigure}{.27\textwidth}
  \centering
  \includegraphics[width=0.875\linewidth]{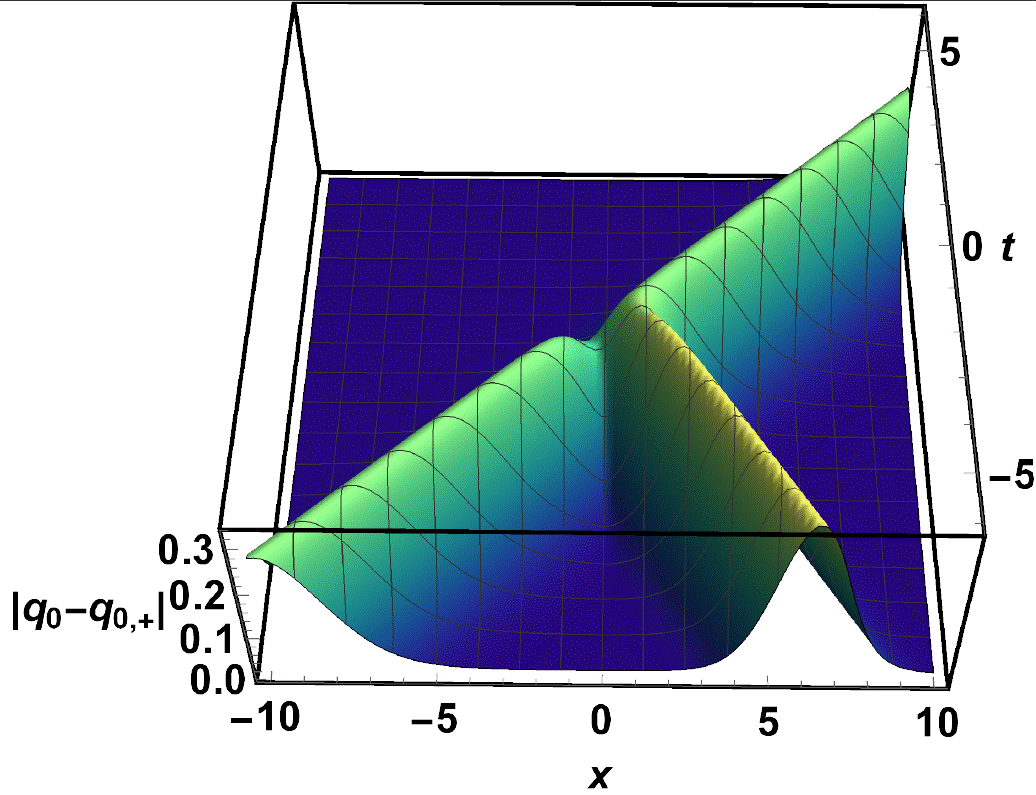}
  \caption{$|q_0-q_{0,+}|$}
  \label{fig8:sub-sixth}
\end{subfigure}
\qquad
\begin{subfigure}{.27\textwidth}
  \includegraphics[width=0.875\linewidth]{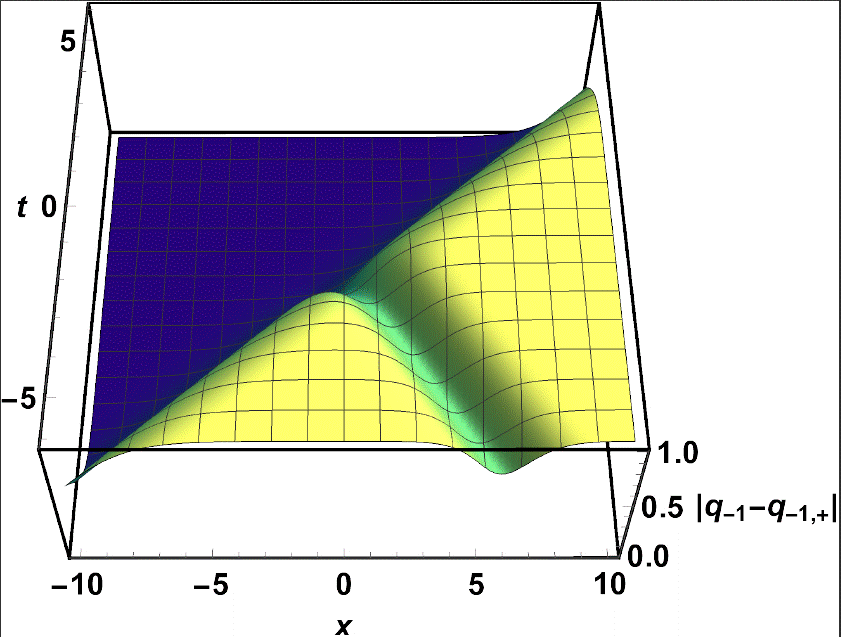}
 \caption{$|q_{-1}-q_{-1,+}|$}
\label{fig8:sub-nineth}
\end{subfigure}
\caption{Plot of the difference between
a polar-polar two-soliton solution and its asymptotic expression along the direction of soliton 1, presented in section~\ref{s:polarpolar}.
Top row: $t\to -\infty$. Bottom row: $t\to \infty$.
The soliton parameters are the same as in figure \ref{f:polar-polar}.}
%, with $q_{i,\pm}$ denotes the asymptotic expansion of the component $q_i$ as $t \to \pm \infty$ for $i \in \{1,0,-1\}$.}
\label{f:polar-polartest}
%\end{figure}
\bigskip
\bigskip
%\begin{figure}[t!]
\centering
\begin{subfigure}{.27\textwidth}
  \centering
  \includegraphics[width=0.875\linewidth]{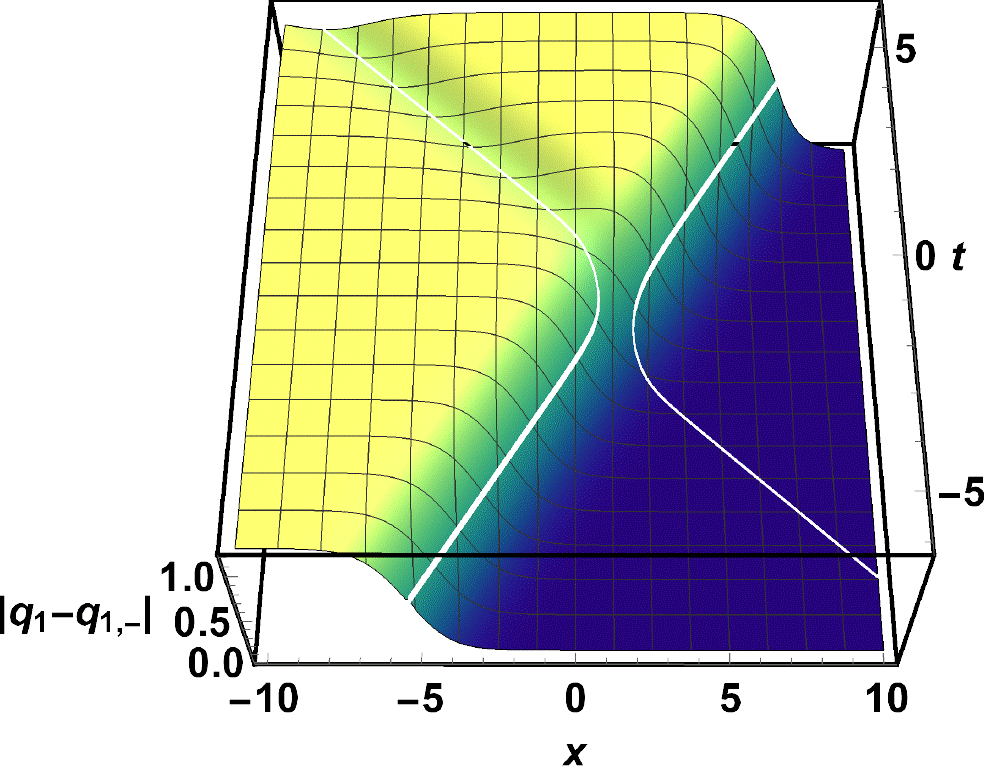}
  \caption{$|q_1-q_{1,-}|$}
  \label{fig9:sub-second}
\end{subfigure}
\begin{subfigure}{.27\textwidth}
  \centering
  \includegraphics[width=0.875\linewidth]{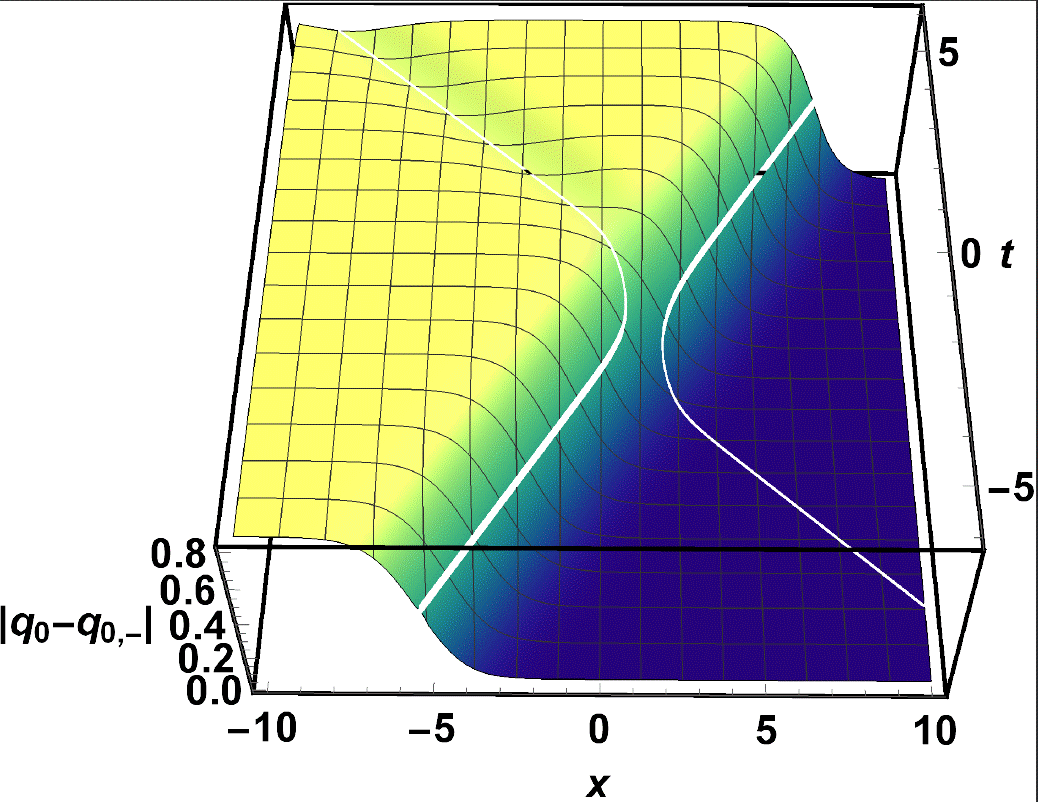}
  \caption{$|q_0-q_{0,-}|$}
  \label{fig9:sub-fifth}
\end{subfigure}
\begin{subfigure}{.27\textwidth}
  \centering
  \includegraphics[width=0.875\linewidth]{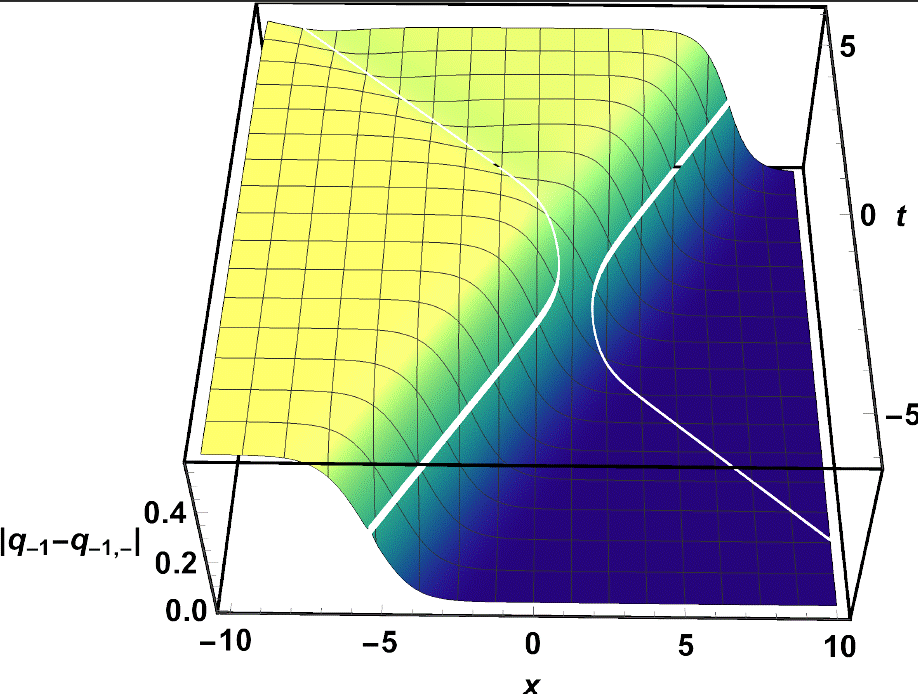}
  \caption{$|q_{-1}-q_{-1,-}|$}
  \label{fig9:sub-eighth}
\end{subfigure}
\vglue\bigskipamount
\begin{subfigure}{.27\textwidth}
  \centering
  \includegraphics[width=0.875\linewidth]{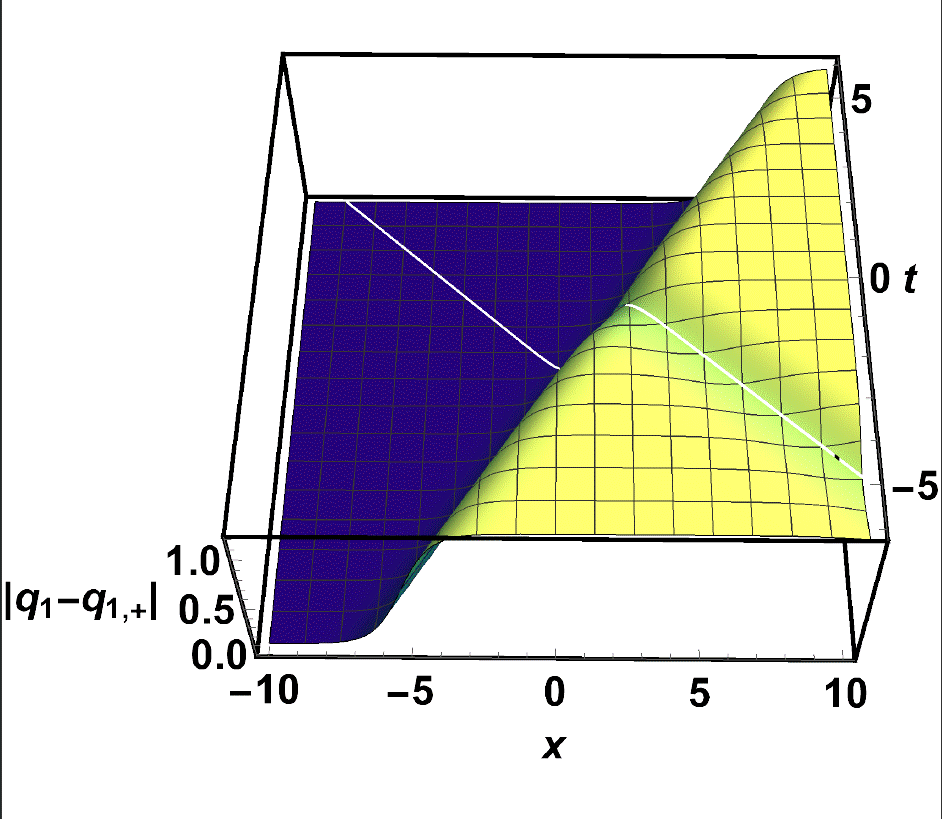}
  \caption{$|q_1-q_{1,+}|$}
  \label{fig9:sub-third}
\end{subfigure}
\begin{subfigure}{.27\textwidth}
  \centering
  \includegraphics[width=0.875\linewidth]{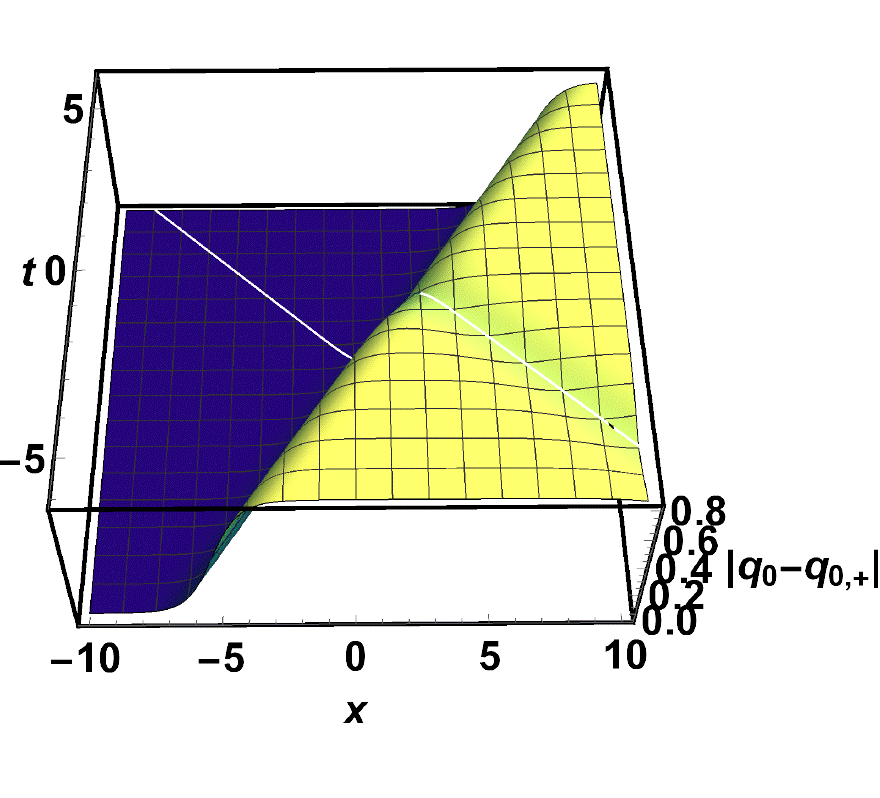}
  \caption{$|q_0-q_{0,+}|$}
  \label{fig9:sub-sixth}
\end{subfigure}
\begin{subfigure}{.27\textwidth}
  \centering
  \includegraphics[width=0.875\linewidth]{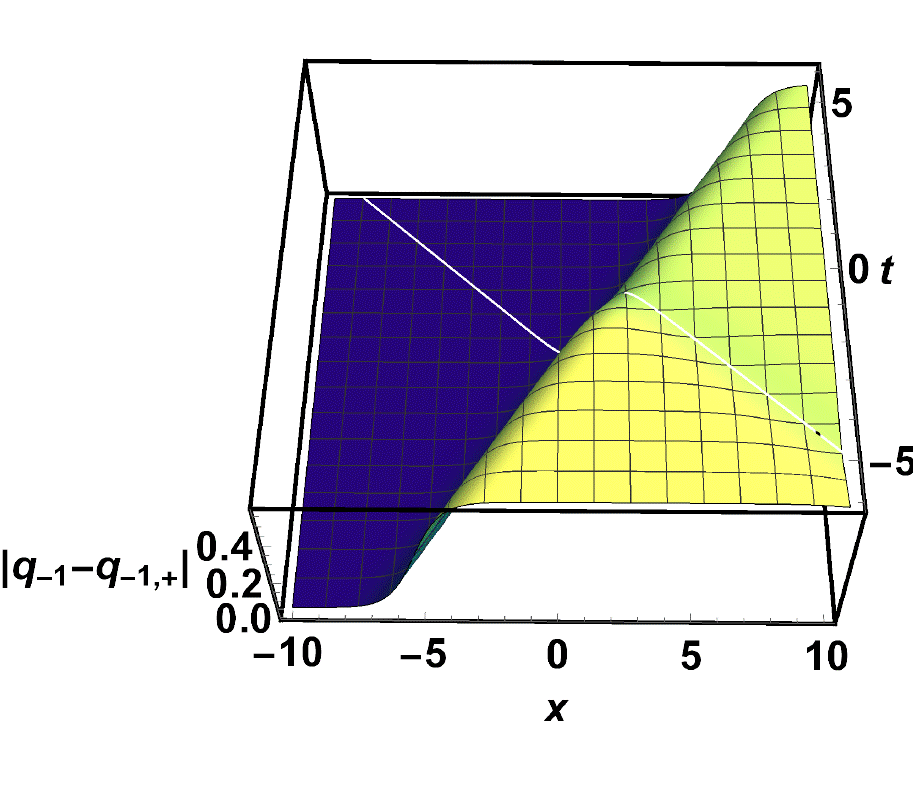}
  \caption{$|q_{-1}-q_{-1,+}|$}
  \label{fig9:sub-nineth}
\end{subfigure}
\caption{Same as Fig.~\ref{f:polar-polartest}, but for a ferromagnetic-ferromagnetic two-soliton solution,
whose asymptotics was presented in section~\ref{s:ferroferro}.
The soliton parameters are as in figure \ref{f:ferro-ferro}.}
%, with $q_{i,\pm}$ denotes the asymptotic expansion of the component $q_i$ as $t \to \pm \infty$ for $i \in \{1,0,-1\}$.}
\label{f:ferro-ferrotest}
\end{figure}

\begin{figure}[t!]
\centering
\begin{subfigure}{.27\textwidth}
  \centering
  \includegraphics[width=0.875\linewidth]{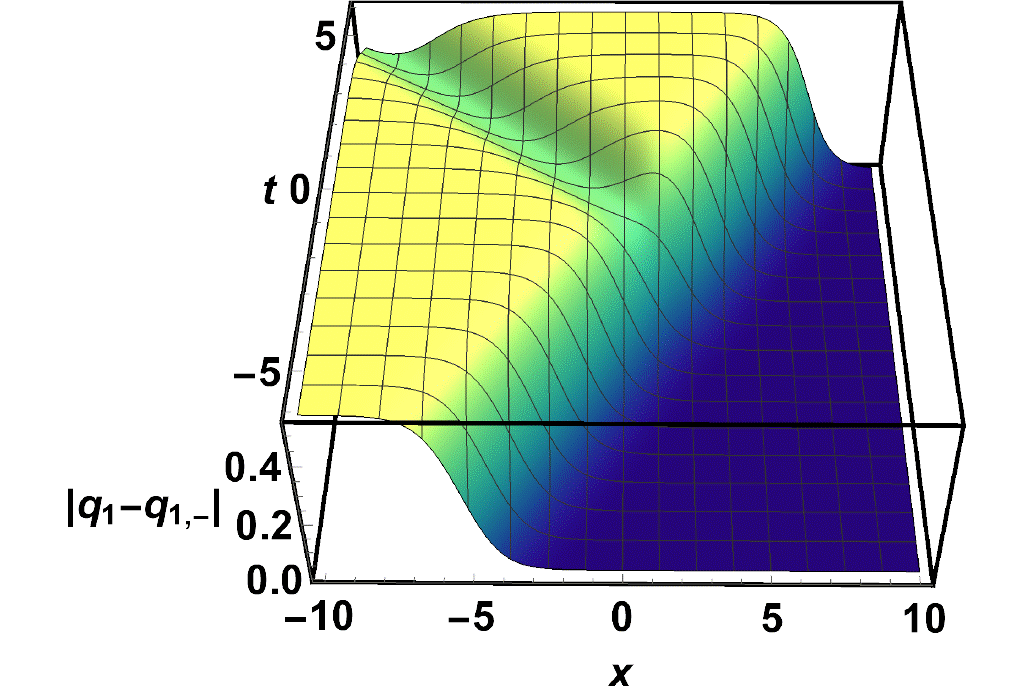}
  \caption{$|q_1-q_{1,-}|$}
  \label{fig8:-second}
\end{subfigure}
\begin{subfigure}{.27\textwidth}
  \centering
  \includegraphics[width=0.875\linewidth]{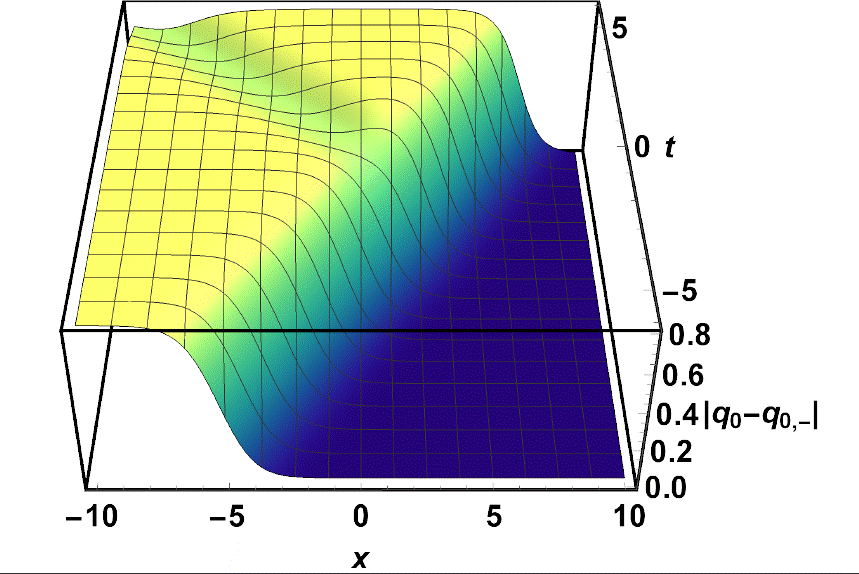}
  \caption{$|q_0-q_{0,-}|$}
  \label{fig8:-fifth}
\end{subfigure}
\begin{subfigure}{.27\textwidth}
  \centering
  \includegraphics[width=0.875\linewidth]{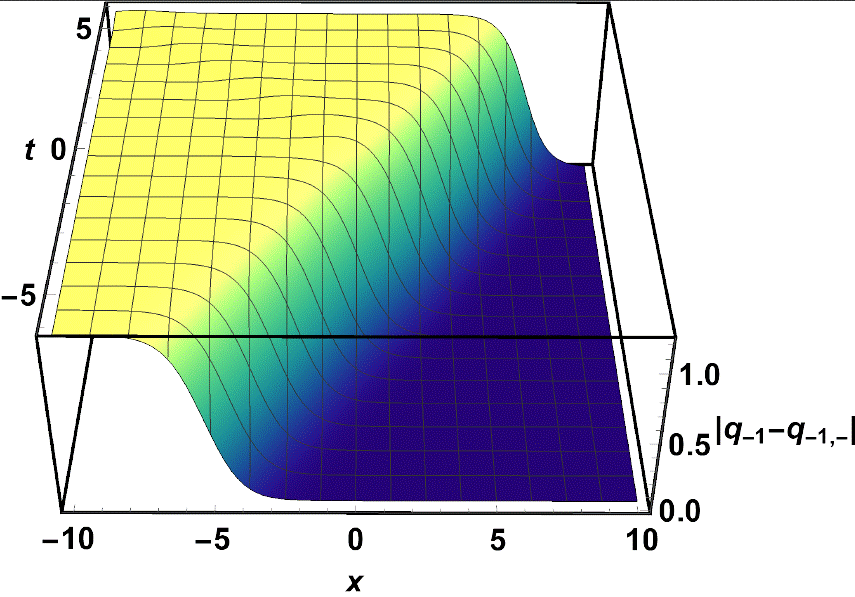}
  \caption{$|q_{-1}-q_{-1,-}|$}
  \label{fig8:-eighth}
\end{subfigure}
\vglue\bigskipamount
\begin{subfigure}{.27\textwidth}
  \centering
  \includegraphics[width=0.875\linewidth]{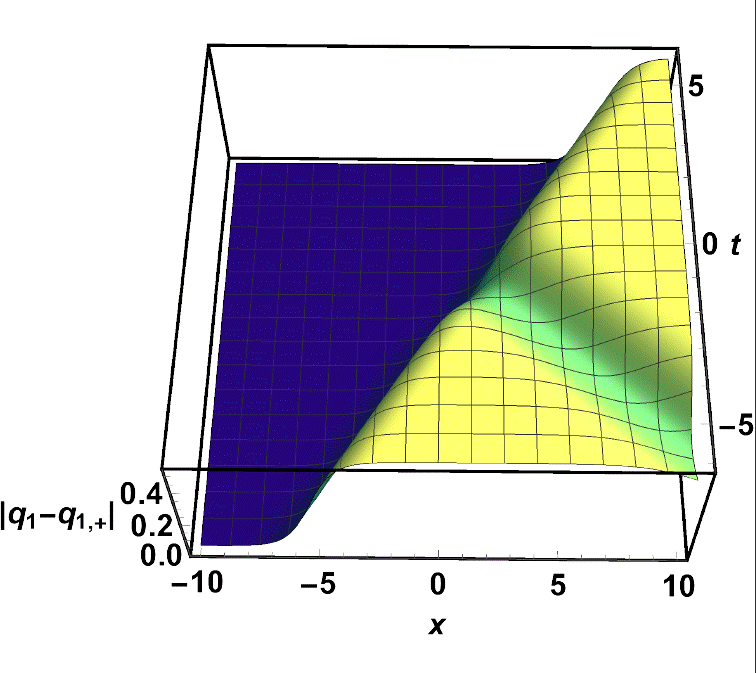}
  \caption{$|q_1-q_{1,+}|$}
  \label{fig8:-third}
\end{subfigure}
\begin{subfigure}{.27\textwidth}
  \centering
  \includegraphics[width=0.875\linewidth]{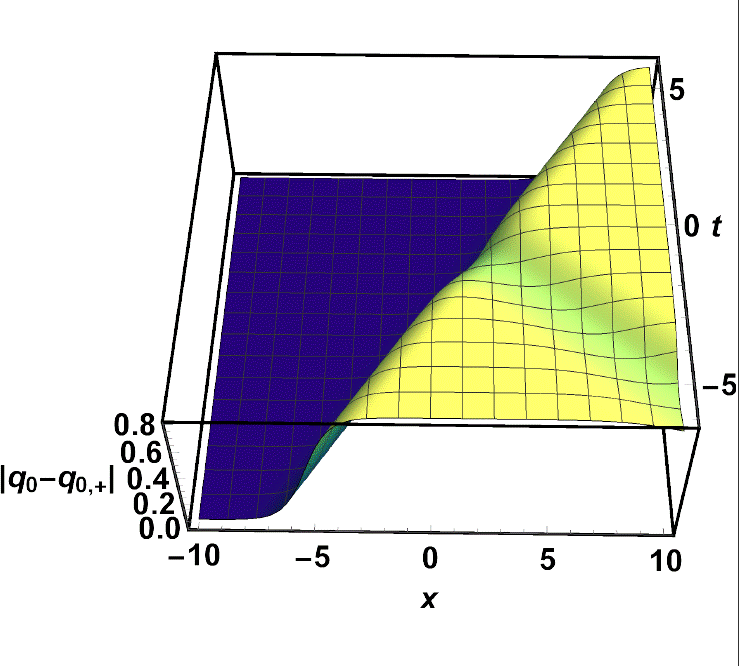}
  \caption{$|q_0-q_{0,+}|$}
  \label{fig8:-sixth}
\end{subfigure}
\begin{subfigure}{.27\textwidth}
  \centering
  \includegraphics[width=0.875\linewidth]{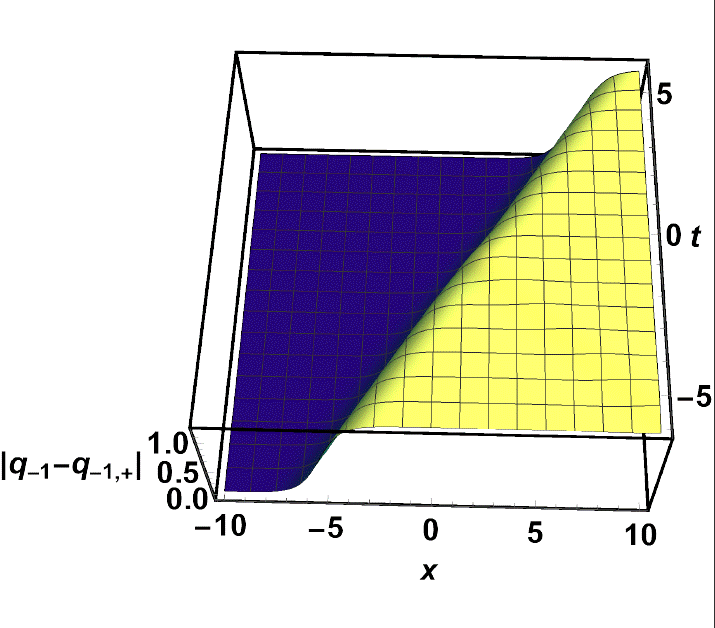}
  \caption{$|q_{-1}-q_{-1,+}|$}
  \label{fig8:-nineth}
\end{subfigure}
\medskip
\caption{Same as Fig.~\ref{f:polar-polartest}, but for a polar-ferromagnetic two-soliton solution,
whose asymptotics was presented in section~\ref{s:polarferro}.
The soliton parameters are as in fig~\ref{f:polar-ferro}.} %with $q_{i,\pm}$ denotes the asymptotic expansion of the component $q_i$ as $t \to \pm \infty$ for $i \in \{1,0,-1\}$.}
\label{f:polar-ferrochi1fixed}
%\end{figure}
\bigskip
\bigskip
%\begin{figure}[ht!]
\centering
\begin{subfigure}{.27\textwidth}
  \centering
  \includegraphics[width=0.875\linewidth]{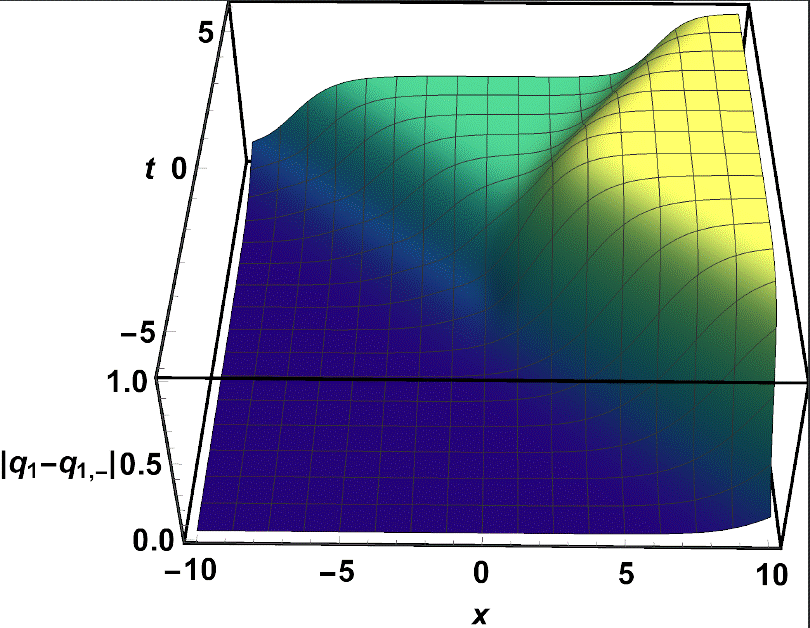}
  \caption{$|q_1-q_{1,-}|$}
  \label{fig9:-second}
\end{subfigure}
\begin{subfigure}{.27\textwidth}
  \centering
  \includegraphics[width=0.875\linewidth]{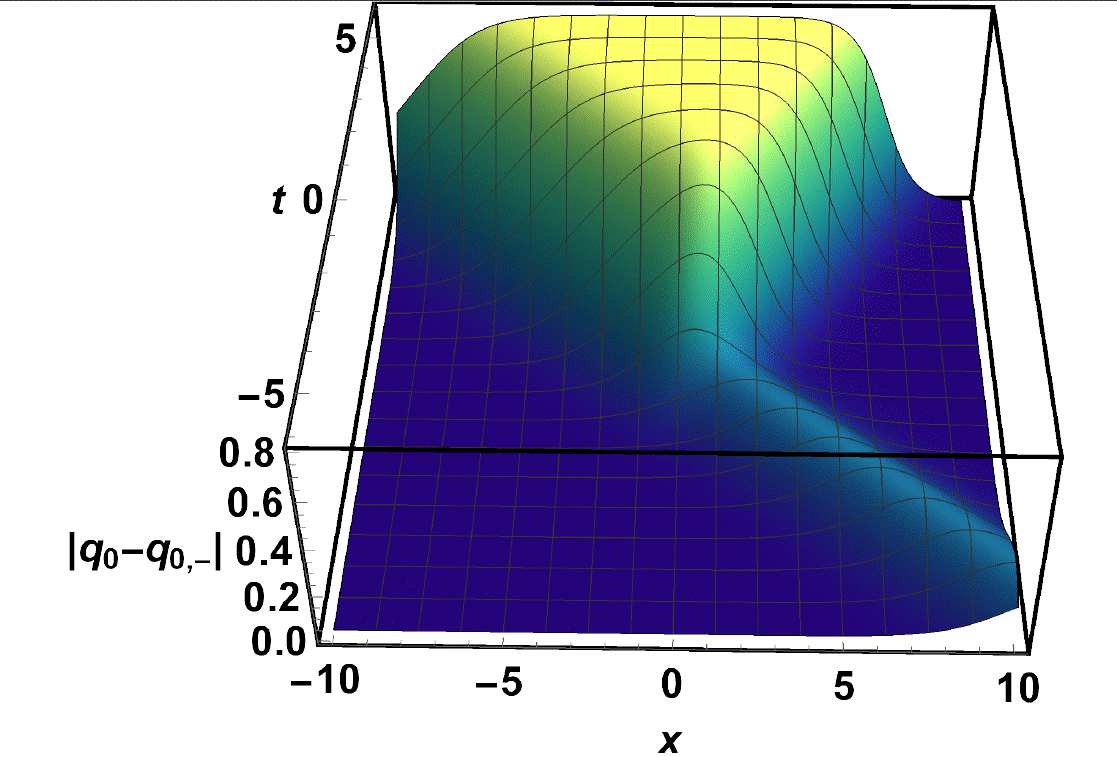}
  \caption{$|q_0-q_{0,-}|$}
  \label{fig9:-fifth}
\end{subfigure}
\begin{subfigure}{.27\textwidth}
  \centering
  \includegraphics[width=0.875\linewidth]{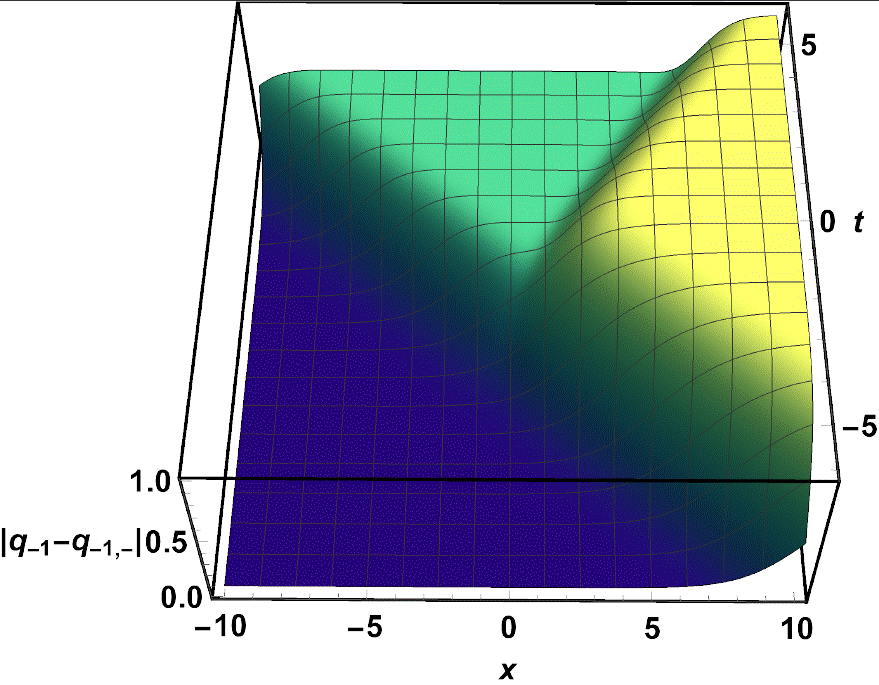}
  \caption{$|q_{-1}-q_{-1,-}|$}
  \label{fig9:-eighth}
\end{subfigure}
\vglue\bigskipamount
\begin{subfigure}{.27\textwidth}
  \centering
  \includegraphics[width=0.875\linewidth]{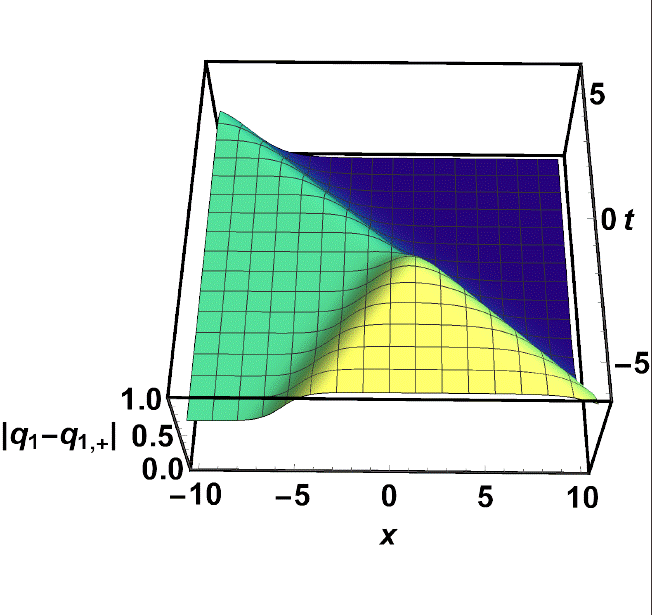}
  \caption{$|q_1-q_{1,+}|$}
  \label{fig9:-third}
\end{subfigure}
\begin{subfigure}{.27\textwidth}
  \centering
  \includegraphics[width=0.875\linewidth]{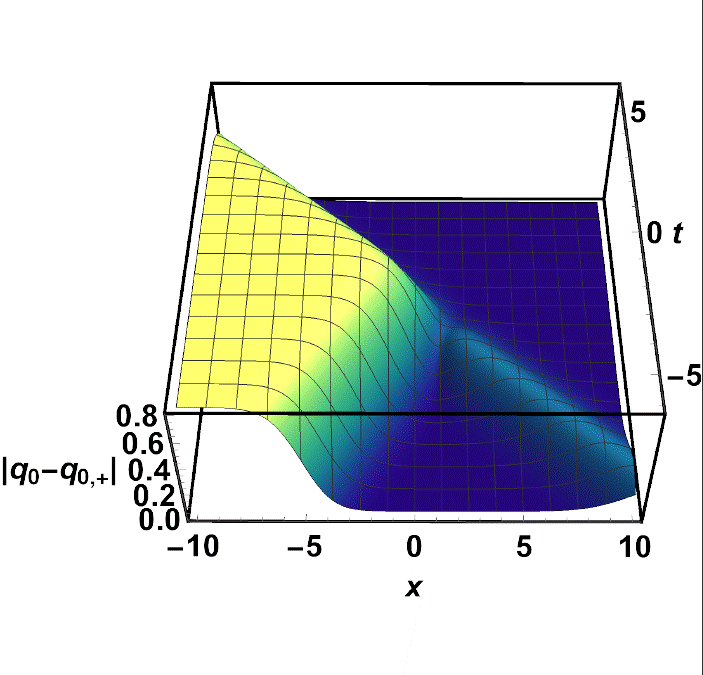}
  \caption{$|q_0-q_{0,+}|$}
  \label{fig9:-sixth}
\end{subfigure}
\begin{subfigure}{.27\textwidth}
  \centering
  \includegraphics[width=0.875\linewidth]{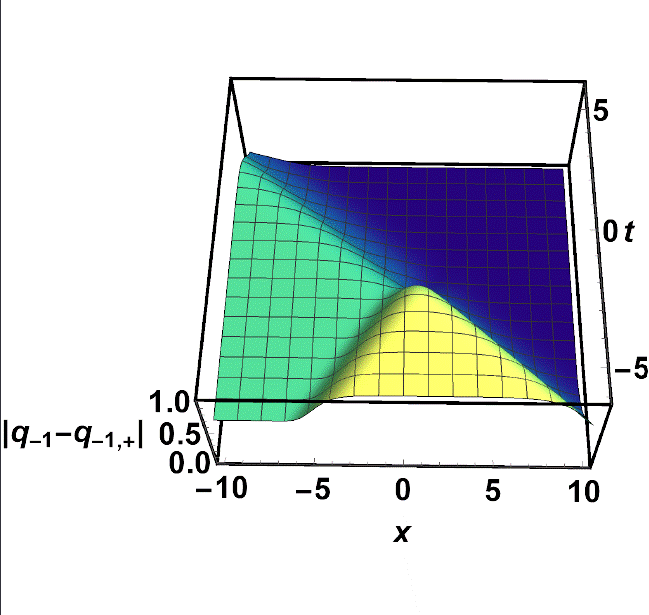}
  \caption{$|q_{-1}-q_{-1,+}|$}
  \label{fig9:-nineth}
\end{subfigure}
\medskip
\caption{Same as Fig.~\ref{f:polar-ferrochi1fixed}, except that the asymptotic is now calculated
along the direction of the ferromagnetic soliton.
The soliton parameters are as in fig~\ref{f:polar-ferro}.}
%with $q_{i,\pm}$ denotes the asymptotic expansion of the component $q_i$ as $t \to \pm \infty$ for $i \in \{1,0,-1\}$.}
\label{f:polar-ferrochi2fixed}
\end{figure}

In this appendix we present a collection of figures to corroborate
the asymptotics analysis of the two-soliton solutions discussed in Section~\ref{s:solitoninteraction}.
Figures \ref{f:polar-polartest}--\ref{f:polar-ferrochi1fixed}
display the difference between the exact two-soliton solution obtained from~\eqref{e:multisoliton} with $J=2$ and the asymptotic expressions,
presented in Section~\ref{s:solitoninteraction},
computed along the direction of soliton~1 as $t\to-\infty$ (top row of each figure) and as $t\to\infty$ (bottom row).
Specifically,
Fig.~\ref{f:polar-polartest} shows the case of a polar-polar two-soliton interaction,
Fig.~\ref{f:ferro-ferrotest} that of a ferromagnetic-ferromagnetic soliton interaction,
and Fig.~\ref{f:polar-ferrochi1fixed} that of a polar-ferromagnetic interaction.
For completeness, Fig.~\ref{f:polar-ferrochi2fixed} also shows the same polar-ferromagnetic interaction but where the asymptotic behavior being subtracted is along the direction of soliton 2,
since in this case the two solitons are of different type.
The fact that the soliton leg vanishes in the appropriate limit in each case serves as a clear visual demonstration of the fact that
the asymptotic expressions do indeed capture the correct behavior of the soliton in both of these limits,
including both the redistribution of mass among the three spin components as well as the position and phase shift.

\makeatletter
\def\@biblabel#1{#1.}
\makeatother
\begingroup
\def\journal#1{\textit{\frenchspacing #1}}
\def\title#1{``#1''}
\def\booktitle#1{\textsl{#1}}
\def\v#1{\textbf{#1}}
\endgroup

\end{document}